\newif\ifdouble
\newif\ifsingle
\newif\ifchange
\doubletrue

\ifdouble
  \documentclass[sigconf, usenames, dvipsnames]{acmart}
\else
  \documentclass[manuscript]{acmart}
  \singletrue
\fi

\usepackage{dblfloatfix} 
\usepackage{booktabs} 
\usepackage{arydshln}
\usepackage{subfig}
\newcommand{\removed}[1]{\iffalse #1 \fi}

\newcommand{\subsub}[1]{\noindent\textit{\textbf{#1:}}}
\newcommand{\width}{0.327\linewidth}

\AtBeginDocument{%
  \providecommand\BibTeX{{%
    \normalfont B\kern-0.5em{\scshape i\kern-0.25em b}\kern-0.8em\TeX}}}

\copyrightyear{2023}
\acmYear{2023}
\setcopyright{acmlicensed}\acmConference[CHI '23]{Proceedings of the 2023 CHI Conference on Human Factors in Computing Systems}{April 23--28, 2023}{Hamburg, Germany}
\acmBooktitle{Proceedings of the 2023 CHI Conference on Human Factors in Computing Systems (CHI '23), April 23--28, 2023, Hamburg, Germany}
\acmPrice{15.00}
\acmDOI{10.1145/3544548.3581449}
\acmISBN{978-1-4503-9421-5/23/04}

\newcommand{\system}{Teachable Reality}

\begin{document}
\pagenumbering{arabic}
\pagestyle{plain}
\title{\system{}: Prototyping Tangible Augmented Reality with Everyday Objects by Leveraging Interactive Machine Teaching}

\author{Kyzyl Monteiro}
\affiliation{%
  \institution{Weave Lab, IIIT-Delhi}
  \city{New Delhi}
  \country{India}}
\affiliation{%
  \institution{University of Calgary}
  \city{Calgary}
  \country{Canada}}  
\email{kyzyl17296@iiitd.ac.in}

\author{Ritik Vatsal}
\affiliation{%
  \institution{Weave Lab, IIIT-Delhi}
  \city{New Delhi}
  \country{India}}
\affiliation{%
  \institution{University of Calgary}
  \city{Calgary}
  \country{Canada}}  
\email{ritik19321@iiitd.ac.in}

 \author{Neil Chulpongsatorn}
\affiliation{%
  \institution{University of Calgary}
  \city{Calgary}
  \country{Canada}}
\email{thobthai.chulpongsat@ucalgary.ca}

 \author{Aman Parnami}
\affiliation{%
  \institution{Weave Lab, IIIT-Delhi}
  \city{New Delhi}
  \country{India}}
 \email{aman@iiitd.ac.in}

\author{Ryo Suzuki}
\affiliation{%
  \institution{University of Calgary}
  \city{Calgary}
  \country{Canada}}
\email{ryo.suzuki@ucalgary.ca}

\renewcommand{\shortauthors}{Monteiro, et al.}
\begin{abstract}
This paper introduces Teachable Reality, an augmented reality (AR) prototyping tool for creating interactive tangible AR applications with arbitrary everyday objects. Teachable Reality leverages vision-based \textit{\textbf{interactive machine teaching}} (e.g., Teachable Machine), which captures  real-world interactions for AR prototyping. It identifies the user-defined tangible and gestural interactions using an on-demand computer vision model. Based on this, the user can easily create functional AR prototypes without programming, enabled by a trigger-action authoring interface. Therefore, our approach allows the flexibility, customizability, and generalizability of tangible AR applications that can address the limitation of current marker-based approaches. We explore the design space and demonstrate various AR prototypes, which include tangible and deformable interfaces, context-aware assistants, and body-driven AR applications. The results of our user study and expert interviews confirm that our approach can lower the barrier to creating functional AR prototypes while also allowing flexible and general-purpose prototyping experiences.
\end{abstract}

\ccsdesc[500]{Human-centered computing~Mixed / augmented reality}

\keywords{Augmented Reality; Mixed Reality; Prototyping Tools; Tangible Interactions; Everyday Objects; Interactive Machine Teaching; Human-Centered Machine Learning;}

\begin{teaserfigure}
\includegraphics[width=\textwidth]{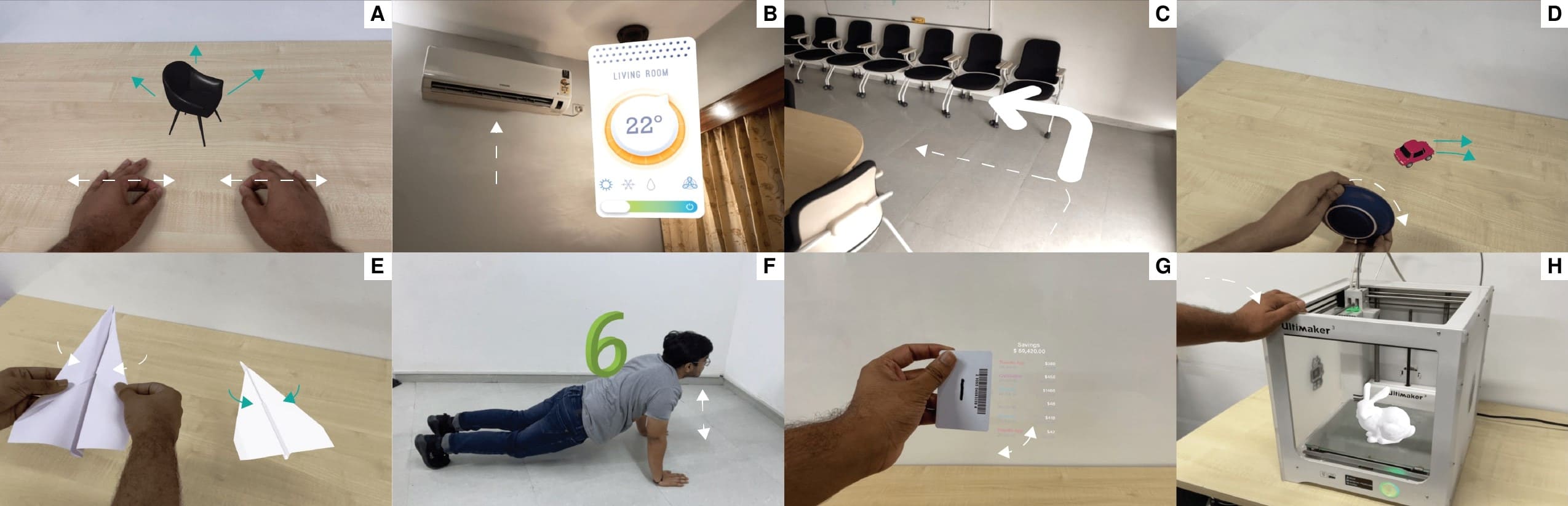}
\caption{Teachable Reality is an augmented reality prototyping tool to create interactive tangible AR applications that can use arbitrary everyday objects as user inputs. Some prototypes that can be created using Teachable Reality include: (A) An in-situ tangible UI that shows that a pinching gesture can control the scale of the virtual content. (B) A smart home AR application that displays a control panel when you look at the device. (C) A navigation system displays the next arrow showing the direction based on the user's current view. (D) An opportunistic AR controller - using a plate to steer and drive a virtual car. (E) An AR interface that displays 3D origami instructions as a step is completed. (F) An AR Assistant interface that counts the number of push-ups. (G) An intelligent Tangible AR interface that allows the rotation of a card to trigger a different layout. (H) An AR 3D printing interface that enables previewing the print when the user places their hand on the 3D printer.}
\Description{Figure 1 is a collection of 8 images showing instances of prototypes created by Teachable Reality. Teachable Reality is an augmented reality prototyping tool to create interactive tangible AR applications that can use arbitrary everyday objects as user inputs. Each image is labeled from A to H. 
Figure 1 A) shows a user's hands demonstrating the pinch gesture, which controls the size of a virtual chair in front of the user. 
Figure 1 B) shows a  smart home AR application where a virtual control panel appears when a user looks at the air conditioner. 
Figure 1 C) is a navigation system that displays the next arrow showing the direction based on the user's current view. 
Figure 1 D) demonstrates an opportunistic AR controller - a plate to steer and drive a virtual car on a tabletop. 
Figure 1 E) shows an AR interface that displays the next 3D origami instructions as every step is completed. 
Figure 1 F) shows an AR Assistant interface that counts the number of push-ups a user in view completes. 
Figure 1 G) is an intelligent Tangible AR interface that allows the rotation of a card to trigger a different layout for the information anchored to the card. 
Figure 1 H) shows a AR 3D printing interface that enables previewing the print when the user places their hand on the 3D printer.}
\label{fig:teaser}
\end{teaserfigure}

\maketitle

\section{Introduction}
Today, prototyping AR applications has become easier than ever before, thanks to various commercial prototyping tools (e.g., \textit{A-Frame}~\cite{noauthor_aframe_nodate}, \textit{RealityComposer}~\cite{inc_realitycomposer_nodate}, \textit{Adobe Aero}~\cite{noauthor_aero_nodate}) and research projects (e.g., \textit{Pronto}~\cite{leiva2020pronto}, \textit{ProtoAR}~\cite{nebeling2018protoar}, \textit{360Proto}~\cite{nebeling2019360proto}).
However, creating \textit{``functional tangible AR''} applications remains difficult as they need to capture and integrate with \textit{\textbf{real-world tangible interactions}}.
Currently, the common practice for such real-world integration is mainly based on either 1) marker-based tracking~\cite{kato1999marker, garrido2014automatic} or 2) a custom machine-learning pipeline (e.g., \textit{OpenCV}~\cite{noauthor_opencv_nodate}, \textit{MediaPipe}~\cite{noauthor_mediapipe_nodate}, etc.).
However, marker-based tracking has limited flexibility due to the nature of printed markers (e.g., cannot be used to detect object deformation or body motion~\cite{zheng2020tangible}, the marker always needs to be visible~\cite{dogan2022infraredtags}, etc.).
On the other hand, the custom machine learning (ML) approach allows great flexibility and customizability without the limitation of marker-based tracking (e.g., can incorporate physical motion~\cite{suzuki2020realitysketch}, gesture~\cite{wang2021gesturar}, and body-based interaction~\cite{wang2020capturar}), but it requires a significant amount of time and expertise to program such AR experiences.

This paper introduces \system{}, an alternative approach to prototyping tangible AR applications by leveraging \textit{\textbf{interactive machine teaching}}.
Interactive machine teaching~\cite{ramos2020interactive} is an emerging machine-learning approach that uses user-guided data for a custom classification pipeline (e.g., \textit{Teachable Machine}~\cite{carney2020teachable}).
By leveraging this, users can easily define their own in-situ tangible and gestural interactions in real-time, which allows the user to prototype functional AR applications without programming. 
Therefore, our approach enables \textit{quick} and \textit{easy} prototyping similar to marker-based approaches, while allowing \textit{flexible}, \textit{customizable}, and \textit{general-purpose} interactions, similar to the machine learning approach. While interactive machine teaching itself is not new, this paper contributes to the first integration of interactive machine teaching into AR authoring.
Based on formative interviews, we design our tool as an end-to-end system that allows the user to detect, train, bind, and author physical-virtual interactions entirely within a mobile AR interface without the need of going back and forth between programming on a desktop screen and testing in the real world.
In addition, we explore the design space of our proposed approach. We show the potential of our tool by demonstrating various application scenarios, including tangible and deformable interfaces (Figure \ref{fig:teaser}-D, E), context-aware assistant (Figure \ref{fig:teaser}-B, C), augmented and situated display (Figure \ref{fig:teaser}-G, H), and body-driven AR experiences (Figure \ref{fig:teaser}-A, F).

We evaluate our approach through two user studies: 1) a usability study with 13 participants and 2) expert reviews
with six tangible AR experts. The study results confirm that our approach can lower the barrier to creating functional AR
prototypes while allowing flexible and general-purpose prototyping experiences. We also found that our approach
can complement existing practices, such as marker-based or machine-learning approaches, by allowing rapid iteration
toward a high-fidelity prototype.
We discuss both benefits and limitations of our approach, pointing out the future opportunity for tangible AR prototyping tools. 

Finally, this paper contributes:
\begin{enumerate}
\item A new approach to authoring tangible AR prototypes by combining interactive machine teaching and in-situ AR scene authoring. 
\item A design space of our approach, which covers both input and output of a wide range of real-world tangible and gestural interactions for AR prototyping.
\item The insights from the two user studies which highlight the benefits and limitations of our proposed approach.
\end{enumerate}

\section{Related Work}

\subsection{AR Prototyping Tools}
To better contextualize \system{} within the landscape of the existing AR prototyping tools, we situate ourselves with the following dimensions (\underline{underlined category} is our focus). 

\subsubsection*{\textbf{1) Fidelity of Prototypes:} Low-fi vs. \underline{Medium-fi} vs. High-fi} 
Existing AR prototyping tools can be situated in the spectrum between low-fidelity and high-fidelity prototypes.
Low-fi prototyping tools like \textit{InVision}~\cite{noauthor_insvision_nodate}, \textit{Sketch}~\cite{noauthor_sketch_nodate}, and \textit{Adobe XD}~\cite{noauthor_XD_nodate} allow for quick initial exploration, whereas tools like \textit{A-Frame}~\cite{noauthor_aframe_nodate}, \textit{Unity}~\cite{heun2013reality}, and \textit{Unreal}~\cite{noauthor_unreal_nodate} are complex but enable high-fidelity interactive AR experiences by providing full-fledged AR development features.
Nebeling et al.~\cite{nebeling2018trouble} argue that there is a significant gap between low-fi and high-fi prototyping tools to create interactive AR applications. We aim to fill this gap by providing a medium-fi prototyping tool. This allows the users to create more realistic AR experiences than low-fi prototyping tools, while does not require complex programming like high-fi prototyping tools.

\subsubsection*{\textbf{2) Goals of Prototyping:}} \underline{Interaction Design} (Interactive) vs. Content Creation (Static)
AR prototyping workflow often employs two steps: 1) creating and placing virtual 3D content and 2) defining the interactions between users and virtual content. 
Many existing tools focus on the first category (e.g., \textit{HoloBuilder}~\cite{noauthor_holobuilder_nodate}, \textit{GravitySketch}~\cite{noauthor_gravity_nodate}, \textit{SketchUp}~\cite{noauthor_sketchup_nodate}, \textit{Lift-Off}~\cite{jackson2016lift}, \textit{RealityComposer}~\cite{inc_realitycomposer_nodate}, \textit{Adobe Aero}~\cite{noauthor_aero_nodate}, \textit{SceneCtrl}~\cite{yue2017scenectrl}, \textit{Window-Shaping}~\cite{huo2016window}, \textit{DistanciAR}~\cite{wang2021distanciar}).
On the other hand, \system{} focuses on \textit{\textbf{interaction design}}, similar to systems like \textit{DART}~\cite{macintyre2004dart} and \textit{ProtoAR}~\cite{nebeling2018protoar}, by assuming the user can reuse existing 3D models.

\subsubsection*{\textbf{3) Deployment of Prototype:}} \underline{Functional} vs. Mock-up
When prototyping \textit{interactive} AR experiences, the system needs to detect, track, and understand real-world interactions.
Many tools avoid this problem through mock-up prototyping (e.g., video-prototyping like \textit{Pronto}~\cite{leiva2020pronto}, \textit{Montage}~\cite{leiva2018montage} or Wizard-of-Oz prototyping like \textit{ProtoAR}~\cite{nebeling2018protoar}, \textit{360proto}~\cite{nebeling2019360proto}, \textit{360theater}~\cite{speicher2021designers}, \textit{WozARd}~\cite{alce2015wozard}).
In contrast, \system{}, like \textit{Rapido}~\cite{leiva2021rapido}, aims to create a \textit{\textbf{functional}} AR prototype, allowing real-world deployment and live user testing in an everyday environment, which is an important need for current AR designers and prototypers~\cite{ashtari2020creating}.

\subsubsection*{\textbf{4) Programming Approaches:} \underline{Programming by Demonstration} vs. Programming by Specification}
Existing approaches to creating functional AR prototypes often rely on simple textual or visual programming. 
For example, many marker-based AR prototyping tools use block-based or node-based programming, such as \textit{ARCadia}~\cite{kelly2018arcadia}, \textit{iaTAR}~\cite{lee2005immersive}, \textit{ComposAR}~\cite{seichter2008composar}, \textit{RealityEditor}~\cite{heun2013reality}, and
\textit{StoryMakeAR}~\cite{glenn2020storymakar}.
Alternatively, trigger-action authoring, which is often used with simplified visual programming, allows users to create interactive behaviors by binding a trigger event with a corresponding action, as seen in \textit{ProGesAR}~\cite{ye2022progesar}, \textit{Situated Game-Level Editing}~\cite{ng2018situated}, 
\textit{MRCAT}~\cite{whitlock2020mrcat}, and \textit{Aero}~\cite{noauthor_aero_nodate}.
In either case, most of these tools require the user to \textit{explicitly specify} the desired trigger and action, which can be difficult to work with real-world interactions due to complexity and ambiguity.
In contrast, our tool, while leveraging trigger-action authoring, allows the creation of interactive behaviors through \textit{\textbf{physical user demonstration}}, similar to \textit{Rapido}~\cite{leiva2021rapido}, \textit{CAPturAR}~\cite{wang2020capturar}, and \textit{GesturAR}~\cite{wang2021gesturar}.
Compared to these tools, however, our tool can support more \textit{flexible} and \textit{open-ended} demonstrations by leveraging interactive machine teaching.
For example, while \textit{Rapido}~\cite{leiva2021rapido} focuses on screen-based interactions, our tool allows the user to demonstrate tangible and physical interactions.
This approach not only allows gesture~\cite{wang2021gesturar} or body-based interaction~\cite{wang2020capturar} but also supports a range of user-defined tangible, gestural, and context-driven interactions, such as object deformation, environment detection, and face recognition.
While there may be a trade-off in tracking accuracy, our approach can significantly reduce the need for multiple different tools~\cite{ashtari2020creating, krauss2022elements} and fill the gap in the fragmented AR prototyping landscape~\cite{nebeling2018trouble}.

\subsection{Everyday Objects as User Interfaces}
Since the birth of tangible and graspable user interfaces~\cite{ishii1997tangible, fitzmaurice1995bricks}, HCI researchers have explored ways to use everyday objects and environments as user interfaces.
For example, in the context of AR/VR interfaces, researchers use everyday objects as haptic proxies~\cite{englmeier2020tangible, daiber2021everyday}, such as \textit{Annexing Reality}~\cite{hettiarachchi2016annexing}, \textit{VirtualBricks}~\cite{arora2019virtualbricks}, and \textit{GripMarks}~\cite{zhou2020gripmarks} or blend virtual experiences into surrounding environments~\cite{lindlbauer2016combining, kaimoto2022sketched}, such as \textit{WorldKit}~\cite{xiao2013worldkit} and \textit{IllumiRoom}~\cite{jones2013illumiroom}.
These prior works augment the tangible paper with fiducial markers (e.g., \textit{Replicate and Reuse}~\cite{gupta2020replicate}, \textit{Paper Trail}~\cite{rajaram2022paper}, \textit{HoloDoc}~\cite{li2019holodoc}, \textit{Tangible VR Books}~\cite{cardoso2021tangible}, \textit{Printed Paper Markers}~\cite{zheng2020tangible}), or augment surrounding objects and environments with smartphone cameras (e.g., \textit{LightAnchors}~\cite{ahuja2019lightanchors}), depth-cameras (e.g., \textit{RealFusion}~\cite{cecil2016realfusion}, \textit{3D Puppetry}~\cite{held20123d}), or embedded invisible tags (e.g., \textit{InfraTag}~\cite{dogan2022infraredtags}).
Similar to our work, some works also explore in-situ creation of tangible interfaces (e.g., \textit{iCon}~\cite{cheng2010icon}, \textit{Instant User Interfaces}~\cite{corsten2013instant}, \textit{Ephemeral Interaction}~\cite{walsh2014ephemeral}, \textit{Fillables}~\cite{corsten2013fillables}, \textit{Tangible Agile Mapping}~\cite{walsh2013tangible}, \textit{Opportunistic Interfaces for AR}~\cite{du2022opportunistic}).
However, one of the key limitations of these tools is the need for more flexibility and generalizability due to the pre-defined tangible inputs. In contrast, our tool leverages interactive machine teaching, allowing more flexible and customizable user inputs than existing tools.

\subsection{Interactive Machine Teaching}
Interactive machine teaching is an approach to creating an on-demand machine learning model based on user-guided data~\cite{ramos2020interactive}.
In recent years, systems like \textit{Teachable Machine}~\cite{carney2020teachable} have demonstrated the potential by allowing the user to quickly create a classification model through user demonstration.
\textit{LookHere}~\cite{zhou2022gesture} further expands this approach by exploiting users' deictic gestures to create a more accurate model.
Since interactive machine teaching is easily accessible for non-technical users, the existing research shows the potential of this approach for tangible storytelling~\cite{tseng2021plushpal}, human-robot interaction~\cite{williams2021teacher}, educational toolkits~\cite{cho2021deepblock}, and programming environments for children~\cite{jordan2021poseblocks, park2021tooee,sabuncuoglu2022prototyping}. 
However, to the best of our knowledge, there is no existing work that integrates interactive machine teaching into \textit{AR authoring} or even AR interfaces in general.
Since interactive machine teaching itself only supports the creation of the ML model, there is still a significant barrier to incorporating the ML model into AR applications.
In contrast, \system{} allows for a no-coding prototyping experience entirely within AR. Thus the user does not need to go back and forth between programming on a desktop and testing in the real world, enabling faster iteration and design exploration, all of which are informed by our formative study.

\section{Formative Study and Design Goals}
To better understand the need for such a system, we conducted formative interviews with six participants (P1-P6) who have experience in prototyping AR applications, tangible UIs, and interactive applications with Teachable Machine.
All interviews were recorded and later transcribed with the consent of the participants. During the 30-60 min formative study, we asked about the current practices and challenges of AR prototyping, especially when designing tangible interactions or integrating machine learning for input detection. Two authors conducted a thematic analysis of the transcriptions and identified emerging themes. Another author resolved and compiled them into five themes we describe below.

\subsubsection*{\textbf{1) Strong Need of Integrating Real-World Interactions for AR Prototypes}}
Overall, there is a strong need to integrate tangible objects and interactions for AR applications.
Participants shared their previous experiences with tangible AR prototype examples, such as tangible tabletop UI with projection mapping (P4), AR prototypes for sports or exercises (P6), and an AR collaboration tool using physical objects (P3).
All participants agreed that blended tangible interaction makes AR applications more unique and interesting.

\subsubsection*{\textbf{2) Lack of Flexibility in Marker-based Tracking Techniques}}
When creating such tangible AR applications, participants often used marker-based tracking (P1, P2, P3, P5).
However, they also complained about the limitations of marker-based tracking. 
For example, the hand occlusion problem diminishes the intended natural interaction (P1, P2). 
Moreover, participants also point out the lack of flexibility by saying that they need to think of the applications based on what fiducial markers can do rather than what they want (P3). 
Due to these limitations, the participants sometimes needed to rely on Arduino and electronic sensors to detect interactions (P1, P3, P6), which could introduce significant overhead (P1).
Overall, the participants think that integrating tangible interactions in AR prototypes is ``tricky'' (P1, P4).

\subsubsection*{\textbf{3) Integration of Computer Vision to AR is Not Well-Supported}}
Participants also used custom computer vision models for gesture detection (P2, P4, P6), but they pointed out that handling raw data to detect a custom gesture was really tedious (P4).
Some participants acknowledge that tools like Teachable Machine can lower the barrier, but they also mentioned that integration into AR applications is a challenge (P4, P6).
\textit{``P6: I used Teachable Machine, but it is still very time-consuming to integrate it as everything needs to be programmed from scratch''}.
In general, participants complained about the lack of available options to integrate real-world tracking into AR.
\textit{``P1: I don’t believe tools like RealityComposer has any ML support. So if I want to detect custom actions, that’s not an option.''}

\subsubsection*{\textbf{4) Need for Quick Prototyping Without Programming}}
When asked why they needed integrated tools, they answered that creating a functional prototype is a huge commitment as it can take days to even months (P2, P4, P5).
Because of this, most of the participants typically used low-fi prototyping methods such as stop motion (P2), Figma (P1), and  Wizard of Oz Powerpoint (P4).
\textit{``P1: Within the company, we often use Figma to convey concepts, but it's very hard to actually get a sense of what it's going to feel like.''}
They agreed that actual functional prototypes allow for a more creative ideation process and easier communication within the team.
Moreover, despite their extensive programming experience, they also want to avoid programming as much as possible for quick iteration.
Therefore, participants strongly agreed that there is a strong need for an integrated authoring tool without programming.

\subsubsection*{\textbf{5) Need for In-Situ Authoring and Live Testing}}
All participants agreed that the current prototyping workflow using platforms like Unity needs a lot of back and forth.
\textit{``P1: It just takes so long to build and push code on AR devices, then sometimes some of the features don't work as expected and so again''}
\textit{``P4: A lot of back and forth between the development devices on the Desktop and the testing devices like mobile phones and the Hololens.''}
Moreover, they need the virtual assets to be synchronized and directly manipulatable in the AR scene rather than on a separate computer screen (P4, P6). Therefore, it is essential to support in-situ authoring and a live testing environment that leverages both direct manipulation and real-time feedback.

\ \\
We identified five goals based on the themes that emerged from the formative study analysis, which inform our system design.

\begin{enumerate}
\item \textbf{\textit{Real-World Integration:}} should support rich real-world interaction for a blended AR experience.
\item \textbf{\textit{Flexible Tracking:}} should support flexible interaction and tracking for various application scenarios.
\item \textbf{\textit{Integrated AR Authoring:}} should integrate input detection and AR output authoring in the same environment.
\item \textbf{\textit{Direct Manipulation:}} should allow the user to prototype interactive experiences without programming.
\item \textbf{\textit{Live and Real-Time Testing:}} should support the immediate live testing in the real world for quick design iteration.
\end{enumerate}

\begin{figure*}[t!]
\centering
\includegraphics[width=\textwidth]{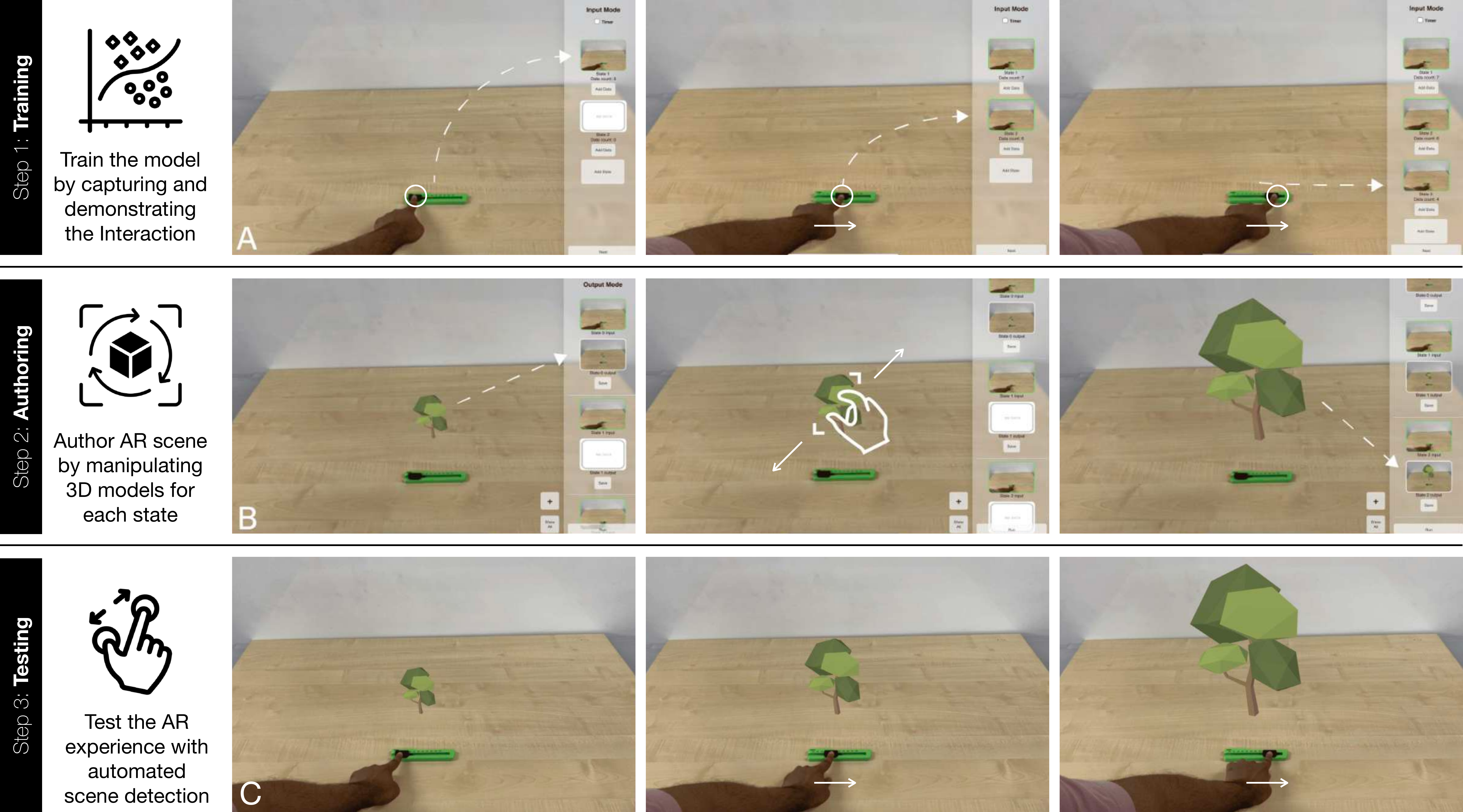}
\caption{Authoring Workflow of Teachable Reality: Using a paper cutter as a slider to control the scale of a virtual tree: A) User captures three states of the everyday object while demonstrating the tangible interaction. B) User saves the output of the virtual world corresponding to each input state. On importing the virtual asset, the user manipulates the virtual asset according to the desired output and saves it to every corresponding input state. C) User tests the created prototype, which animates between all the outputs that were saved.}
\label{fig:workflow}
\Description{Figure 2 is divided into 3 steps i.e., Training, Authoring, and Testing. In row A) the user demonstrates and captures the sliding interaction using a box cutter's slider. In row, B) the Authoring step is shown by the user modifying the size of a virtual tree and saving it corresponding to each state of the slider they had saved earlier. Finally, In row C) the testing step is demonstrated by how the size of the virtual tree changes as the box cutter's slider is manipulated in the real world.
}
\end{figure*}

\begin{figure*}[t!]
\centering
\includegraphics[width=\textwidth]{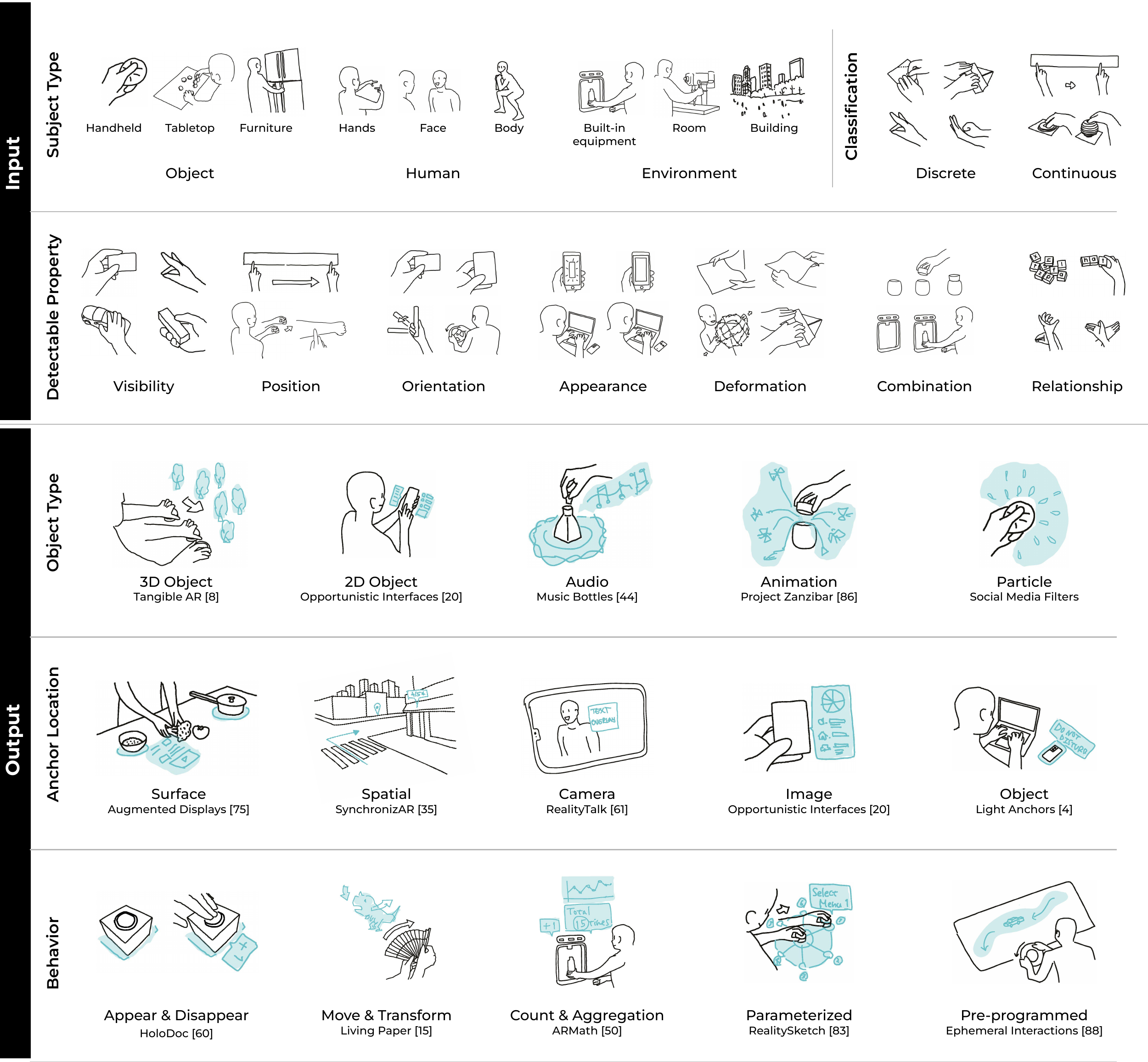}
\caption{Design space of supported modalities of input and output for tangible AR prototypes. }
\label{fig:design-space}
\Description{
Figure 3 shows the Design space and supported input and output modalities for tangible AR prototypes. The input design space explores types of subjects, classification, and detectable properties. The types of subjects can be an object, humans, or an environment. The types of classification can be discrete or continuous. In contrast, the types of detectable properties can be visibility, position, orientation, appearance, deformation, combination, and relationship.

The output design space is the types of virtual objects, anchored location, and behavior of virtual output. Types of virtual objects include 3D objects, 2D objects, audio, animation, and particle effects. Anchored locations for virtual content can be surface, spatial, camera-anchored, image-anchored, or object anchored. The behavior of virtual output includes appear and disappear animation, moving and transforming animation, counting and aggregation, parameterized output, and pre-programmed control.
}
\end{figure*}

\section{Teachable Reality}
\subsection{Overview}
\system{} is a mobile AR prototyping tool that combines  interactive machine teaching and in-situ AR authoring.
\system{} has three key features: 
1) \textit{interaction detection:} interaction detection based on an on-demand computer vision classification model, 
2) \textit{in-situ AR authoring:} AR authoring environment that lets the user quickly create desired interactive AR behaviors based on user-defined trigger action, and
3) \textit{live deployment and} testing: the user can quickly deploy the AR prototypes for iterative live testing in an everyday environment.
\system{} has a simple user interface. 
The main window shows the live current AR view that provides both authoring and live testing views. 
The right panel on each screen (Figure \ref{fig:workflow}) presents different options to the user during the two stages of authoring: 1) capture and store the different input states, and 2) save and display each state's corresponding output.

\subsection{Authoring Workflow}

\subsubsection*{\textbf{Step 1: Capturing and Demonstrating the Interaction}}
The first step is to capture a user's interaction with the tablet's camera.
When the user taps the \textit{Add Data} button, the camera starts capturing the scene from the main window so that the user can demonstrate the desired input interaction with an everyday object or environment. The user can add data for multiple states capturing stages of one interaction or multiple interactions according to their need.
For example, the user defines four different states based on the position of the black slider handle in a box cutter (Figure~\ref{fig:workflow}A).
Once the user finishes capturing and taps the \textit{Next} button, the system creates a computer vision classification model based on the provided data and starts automatically detecting each state based on the classifier, similar to \textit{Teachable Machine}~\cite{carney2020teachable}.

\subsubsection*{\textbf{Step 2: Authoring AR Scene for Each Interaction State}}
After registering each state, the user can author each AR scene with direct manipulation.
To do so, the user can tap the \textit{+} button. Then the user can choose a virtual asset from an asset library (prepared by the user or some default objects) to place into the scene.
When tapping the \textit{Save} button located below each state, the user can register the current AR scene as a corresponding AR scene. 
The placed virtual object can be manipulated with the touch gesture, such as drag-and-drop for position change, pinching gesture to change the scale, and twist gesture to change the orientation.
The user can quickly define the interactive behavior by moving the virtual object and saving the scene corresponding to each saved state. The interactive behavior consists of the trigger---the detection of each state and action---storing the corresponding AR scene.  
For example, Figure~\ref{fig:workflow}B illustrates the workflow where the user places a 3D model of a tree on a table, changes the size of the tree with pinching interaction, and then saves it to the corresponding state.
When placing the virtual object, the user can also choose different asset types (e.g., 3D object, 2D images, etc) and anchored locations (e.g., surface, object, image, camera, etc), as we discuss in the design space section (Figure~\ref{fig:design-space}).

\subsubsection*{\textbf{Step 3: Live Testing with Automated Scene Detection}}
Once the user finishes authoring the AR scene for each state, the prototype is deployed, and the user can start live-testing the prototype.
In the live preview mode, the system starts automatically detecting the different user-defined states.
When transitioning from one state to another, the system automatically animates the virtual object between the corresponding AR scenes, similar to the digital animation technique of auto-tweening.
For example, in Figure~\ref{fig:workflow}C, the size of the virtual tree changes based on the position of the blade slider of a box cutter, as if the user can use it as a tangible slider. The scale of the virtual tree smoothly due to the automated animation feature, as we described.
When transitioning between two stored positions of the slider, the scale of the tree interpolates between the two corresponding scenes the user had demonstrated.
For example, in Figure~\ref{fig:workflow}C, the size of the virtual tree changes based on the position of the blade slider of a box cutter, as if the user can use it as a tangible slider.

\section{Implementation}
To democratize tangible AR prototyping experiences, we release our prototype as an open source software~\footnote{\href{https://github.com/kyzylmonteiro/teachable-reality}{https://github.com/kyzylmonteiro/teachable-reality}}.
In this section, we describe the implementation detail for each core functionality.

\subsubsection*{AR Authoring Interface}
\system{} is a web-based mobile AR system that runs on any browser that supports the WebXR platform.
We tested the system with Google Chrome on Android (Google Pixel 6) and Safari on iOS (iPad Pro 12.9-inch).
It is developed using JavaScript, HTML, and CSS and runs entirely on the client side of the browser without needing a web server. 
The system uses 8th Wall~\cite{noauthor_8thwall_nodate}, A-Frame~\cite{noauthor_aframe_nodate}, and Three.js~\cite{noauthor_threejs_nodate} for the immersive AR authoring system. 
A-frame enables the placement, manipulation, and animation of virtual assets. While 8th Wall provided us access to spatial understanding, including surface detection and device position tracking. 

\subsubsection*{Image Capturing and Classification}
8th Wall uses the tablet's camera stream to detect the device's position and surface in a real-world environment based on a proprietary SLAM algorithm.
We also use the camera stream from 8th Wall for user-defined camera recording as well as the object and human pose detection.
For the recorded image classification, we leverage transfer learning using Tensorflow.js, which is the same backend as \textit{Teachable Machine}~\cite{carney2020teachable}.
The system runs Tensorflow.js on the client side for both the training and inference phase.
Since the system needs to train the ML model on-demand, the system trains the model with a separate thread using Web Worker in the background. 
To reduce the training time, the system leverages the MobileNet model~\cite{noauthor_mobilenet_nodate} as the base model pretrained on the ImageNet dataset~\cite{russakovsky2015imagenet}. 
Training time significantly varies depending on the scenario, thus it is difficult to generalize. However, the average training time with 5 states and 100 images for each state took approximately 16.5 sec with 30 trials of the different scenes (min: 8, max: 20.5 sec) with iPad Pro 12.9 inch (M1 chip, 8 Core CPU, 8 Core GPU, and 16GB RAM).

\subsubsection*{Object Tracking for Location Anchoring}
The system also detects and tracks the object for location anchors. 
For the surface and spatial anchoring, the user uses a detected surface and 3D coordination based on 8th Wall's built-in spatial anchoring features. 
For image anchoring, the system also uses the 8th Wall's image target feature.
For object tracking, the system tracks the object's position based on color tracking. 
When the user specifies the tracking color by tapping the object on a screen, the system obtains the RGB value of the 2D coordinate, then detects the largest contour of the object with OpenCV.js~\cite{noauthor_opencv_nodate}. Then, the system obtains the 3D position of the tracked color, by raycasting onto the virtual surface. 
Therefore, the system can only track the object's position on a surface.
For human-anchored positions, we use MediaPipe~\cite{noauthor_mediapipe_nodate} to obtain the human skeleton position data.
The system maintains the virtual object location based on the selected anchored origin. 
\section{Design Space}
As mentioned, \system{} adopts the \textit{trigger-action} authoring model, which consists of input (trigger) and output (action) for the interactive AR experiences. 
To better understand what kind of input and output our system can support, we present a design space exploration of \system{}'s supported modalities (Figure~\ref{fig:design-space}). To explore the design space, we investigated the existing literature on tangible interfaces to identify common elements of input and output for tangible AR applications. 
To create a generalizable and flexible design space, we first collected examples of tangible AR research, products, and art installations. Then, we abstract the common elements that can be seen in these examples through sketching and categorization. Figure~\ref{fig:design-space} illustrates these abstracted sketches along with a representative example for each category, by providing the name of the project and research paper.
While this design space may not be a systematic or exhaustive exploration of all possible tangible AR interfaces, we believe our design space, along with representative examples, provides an overview of what our approach enables and how each element could be used for various applications.

\subsection{Input: Types of Subject}
\begin{figure}[h!]
\centering
\includegraphics[width=\width]{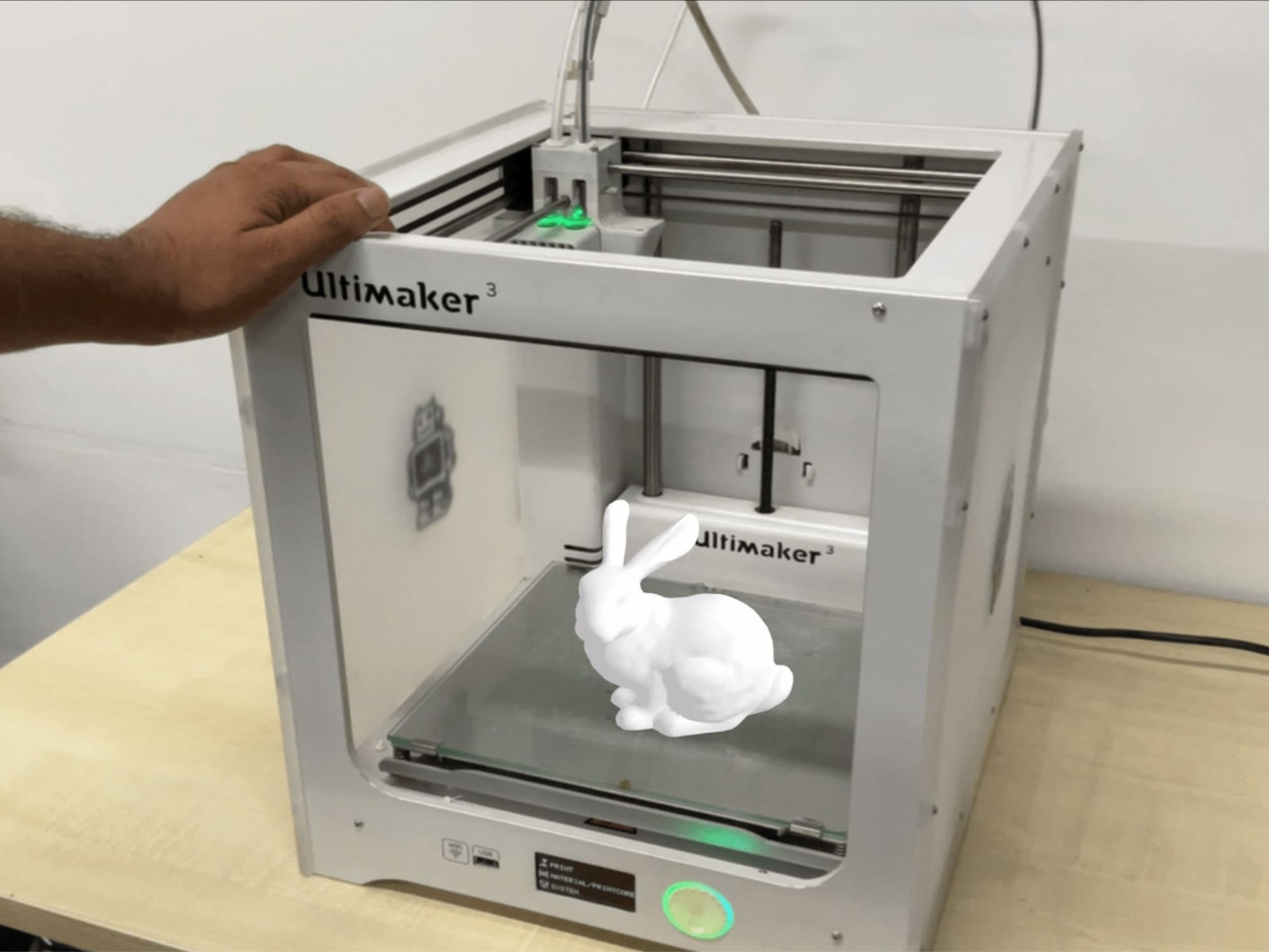}
\includegraphics[width=\width]{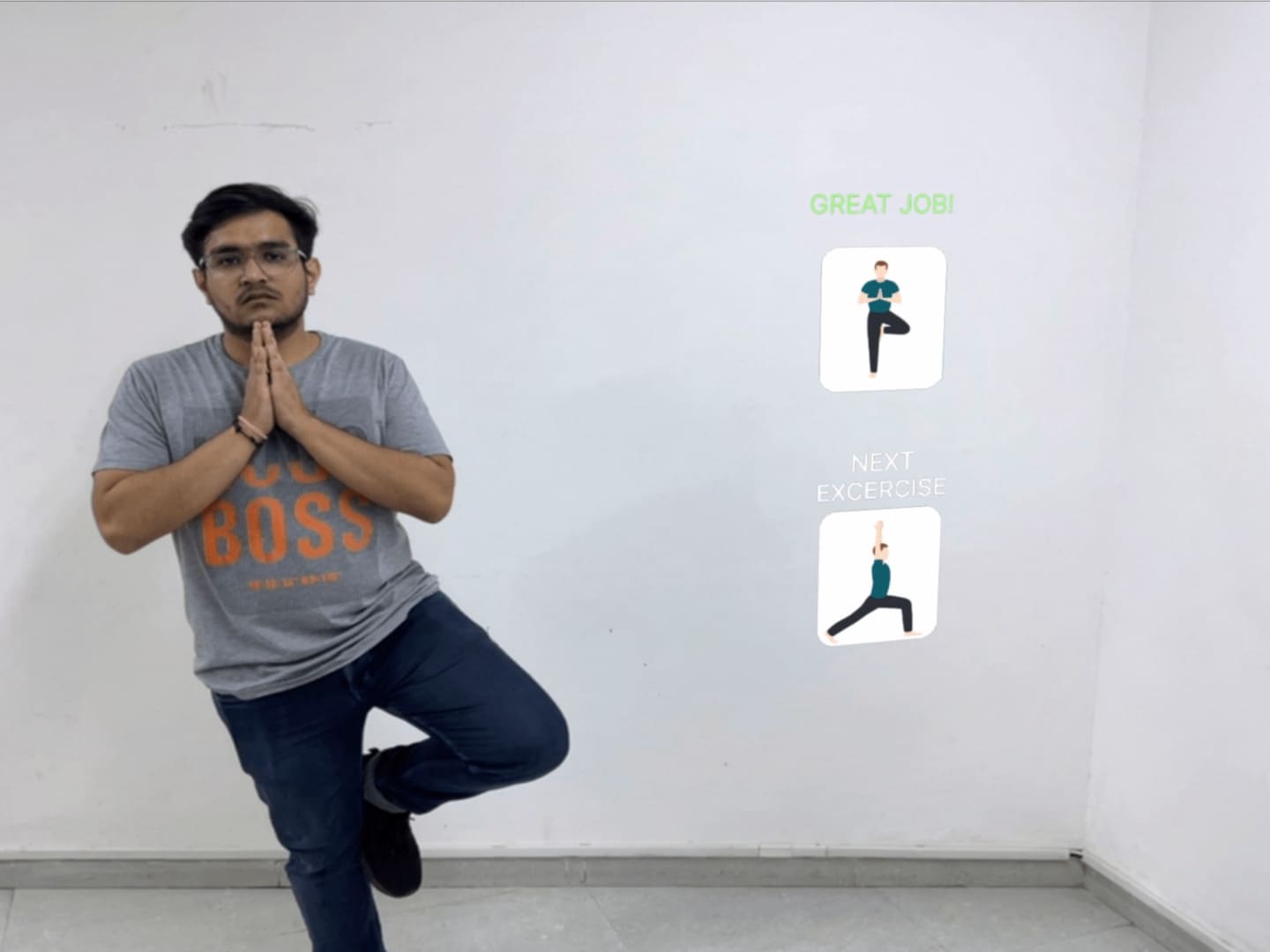}
\includegraphics[width=\width]{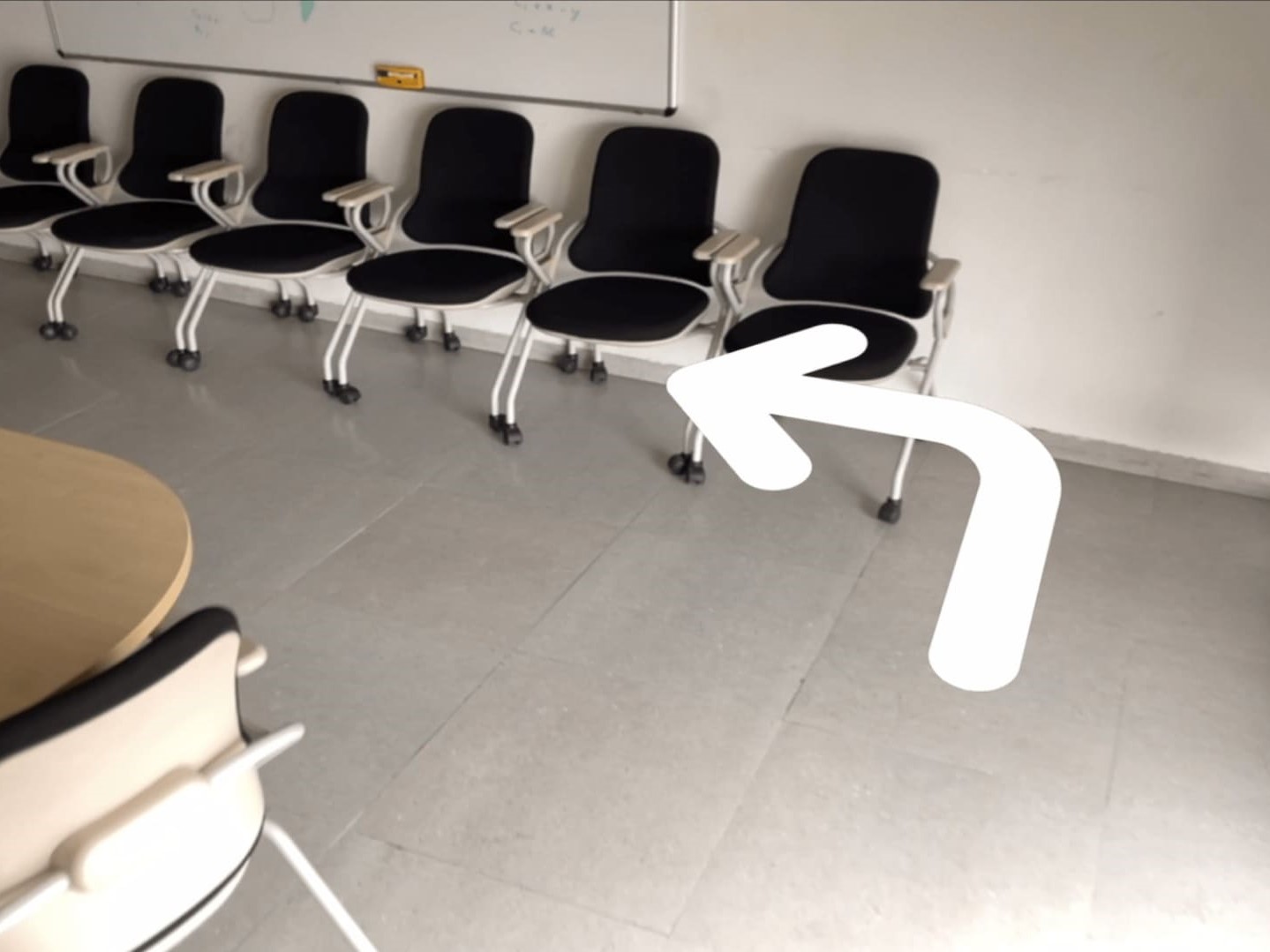}
\caption{Input - Types of the subject: The system supports various subjects as inputs, such as objects, humans, and environments.}
\label{fig:input-subject}
\Description{This figure has three images showing the different types of input supported. First image shows a 3D printer, an object as input. Second image shows a human as an input and the third image shows a room with chairs and table, i.e. an environment as input.}
\end{figure}

At a high level, the system can use any subject as an input as long as it is visible and detectable with the camera. 
While the design space of detectable subjects is vast, we show three main possible types of subjects (Figure~\ref{fig:input-subject}).

\subsubsection{Object} First, the system can detect a variety of physical objects from handheld- to room-scale objects, such as
toys, mugs, papers, books, and furniture, like \textit{HoloDoc}~\cite{li2019holodoc} and \textit{3D Puppetry}~\cite{held20123d}.
The system can detect various tangible interactions with these detected physical objects.

\subsubsection{Human} Also, the user can use a human as an input subject, including hand gestures, facial expressions, and body postures, similar to \textit{Interactive Body-Driven Graphics}~\cite{saquib2019interactive} or \textit{RealityTalk}~\cite{liao2022realitytalk}.
Our system itself does not have built-in gesture or posture recognition, but the user can train the model to recognize them in-situ.

\subsubsection{Environment} Moreover, while the system itself does not incorporate the device's position or location information, the user could also use a scene and environment as the user input. For example, the user could identify a location or room like a kitchen, bathroom, or living room based on a landmark that is visible with a camera. The user can also quickly create and test an AR navigation experience like \textit{Live View in Google Maps}~\cite{noauthor_mapsar_nodate}, as seen in Figure~\ref{fig:input-subject}C.

\subsection{Input: Types of Classification}
Depending on how the user trains the model, the user can also detect as two different types of inputs. 

\subsubsection{Discrete} 
Discrete input means that the detectable states are independent and there is no continuous relationship.
By default, \system{} treats all registered states as discrete and independent inputs.
For example, the binary state of visibility or different hand gestures are all discrete inputs.

\subsubsection{Continuous} 
\begin{figure}[h!]
\centering
\includegraphics[width=\width]{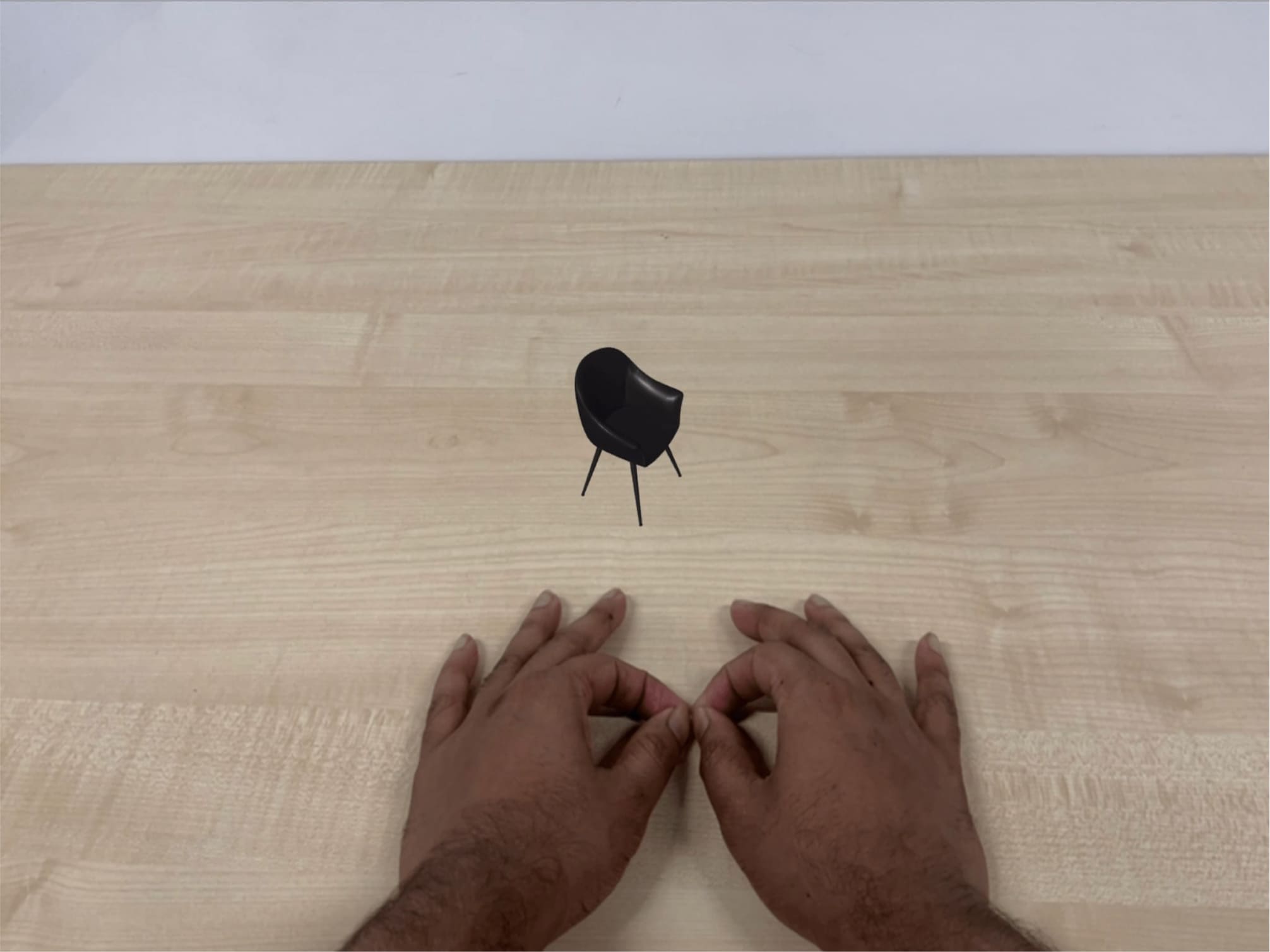}
\includegraphics[width=\width]{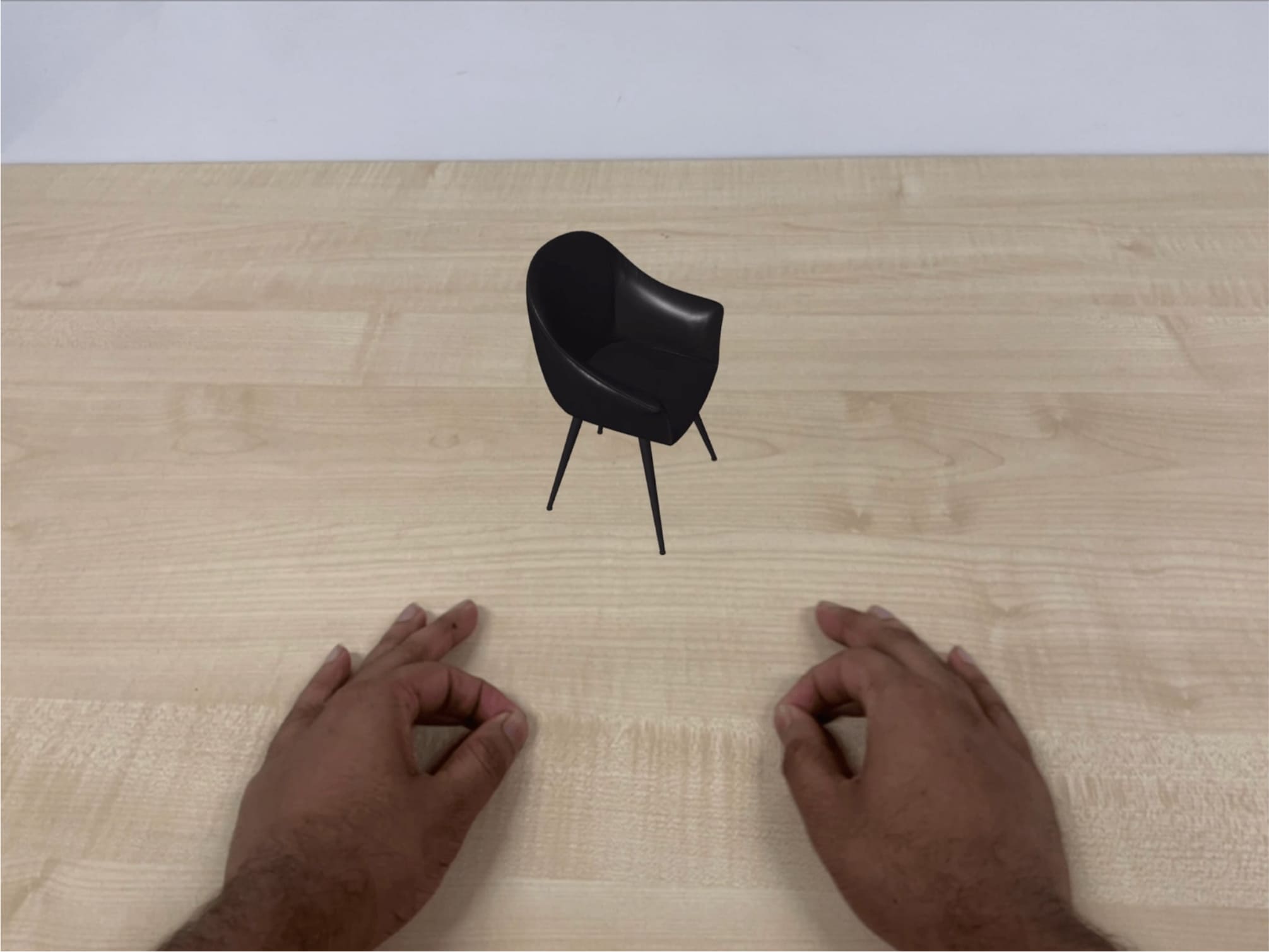}
\includegraphics[width=\width]{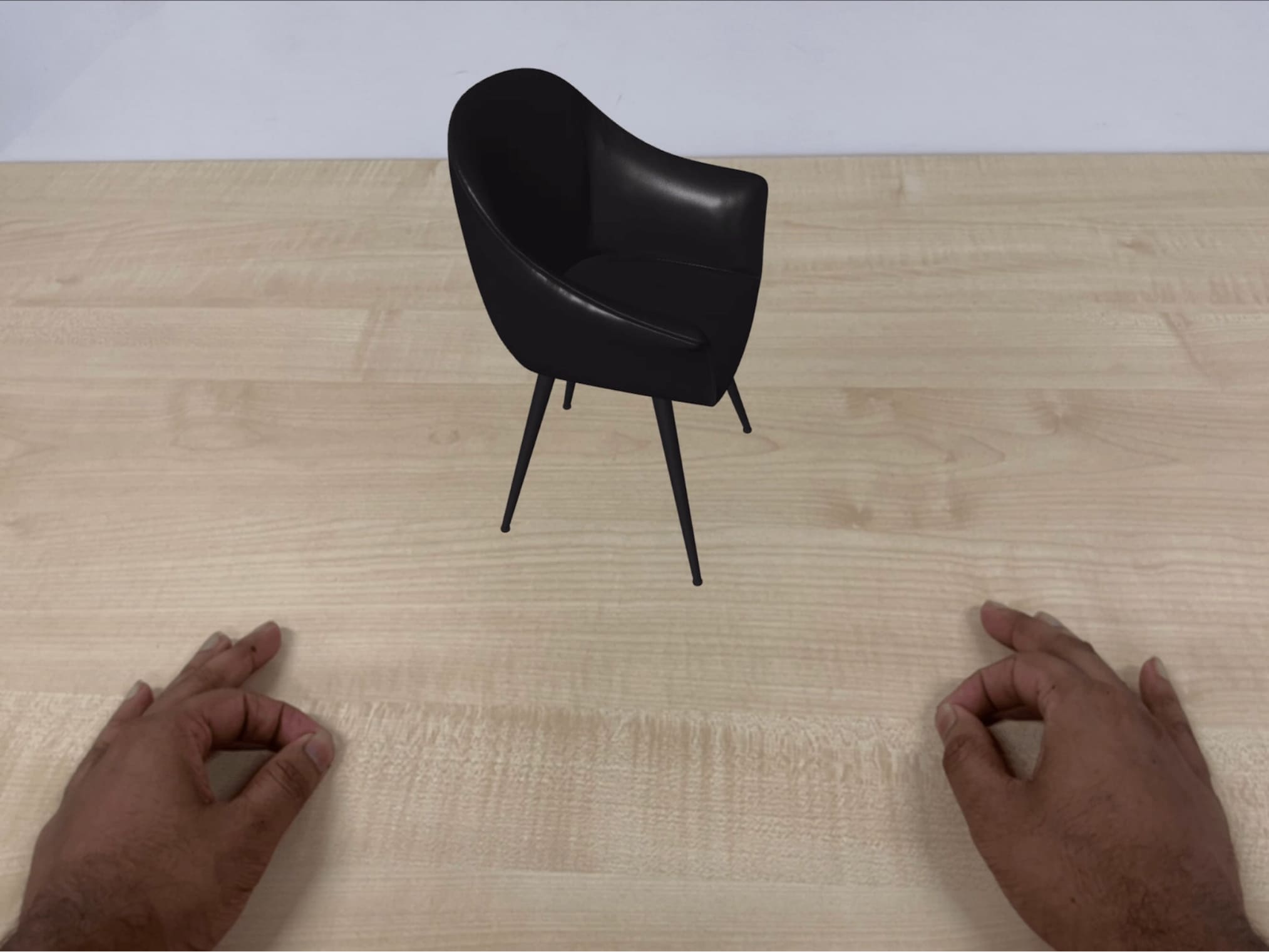}
\caption{Input - Continuous: Change the size of the virtual chair based on the distance between two hands.}
\label{fig:input-continuous}
\Description{Of the two types of input classification, Figure 5 demonstrates the continuous class by modifying the size of a virtual chair by pinching zoom-in zoom-out motion of hands.}
\end{figure}

But, by registering states in a sequential manner, the user can also define continuous input\removed{the continuous inputs}. For example, the user can use the continuous change of the position, orientation, or deformation of the object as a staggered input parameter.
By using this, the user could mimic and treat the input as a numerical and sequential value for a tangible controller (Figure~\ref{fig:input-continuous}). 

\subsection{Input: Types of Detectable Properties}
Depending on how the user trains the model, the system can also detect different states of the object for tangible interactions. 

\subsubsection{Visibility} 
First, the user can detect the presence or absence of the object based on visibility in the scene.
For example, the user can create a binary state to detect whether the user is holding an object in the field of view or not.

\subsubsection{Position} 
Alternatively, the user can detect the different positions of the object in the field of view. For example, the user can use the position of the handle to change the scale of a virtual object just like a tangible slider, similar to \textit{RealitySketch}~\cite{suzuki2020realitysketch}.

\subsubsection{Orientation} 
\begin{figure}[h!]
\centering
\includegraphics[width=\width]{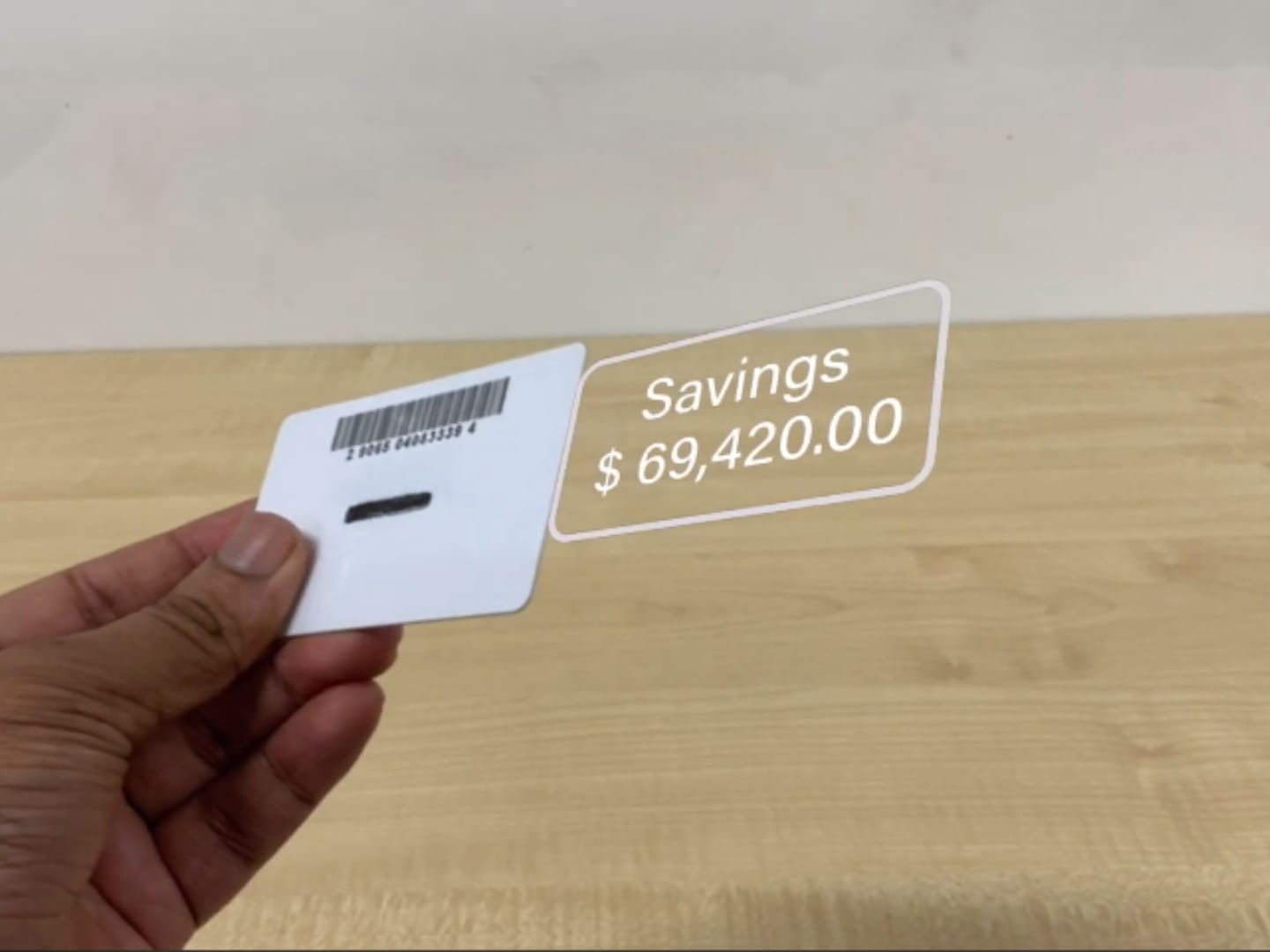}
\includegraphics[width=\width]{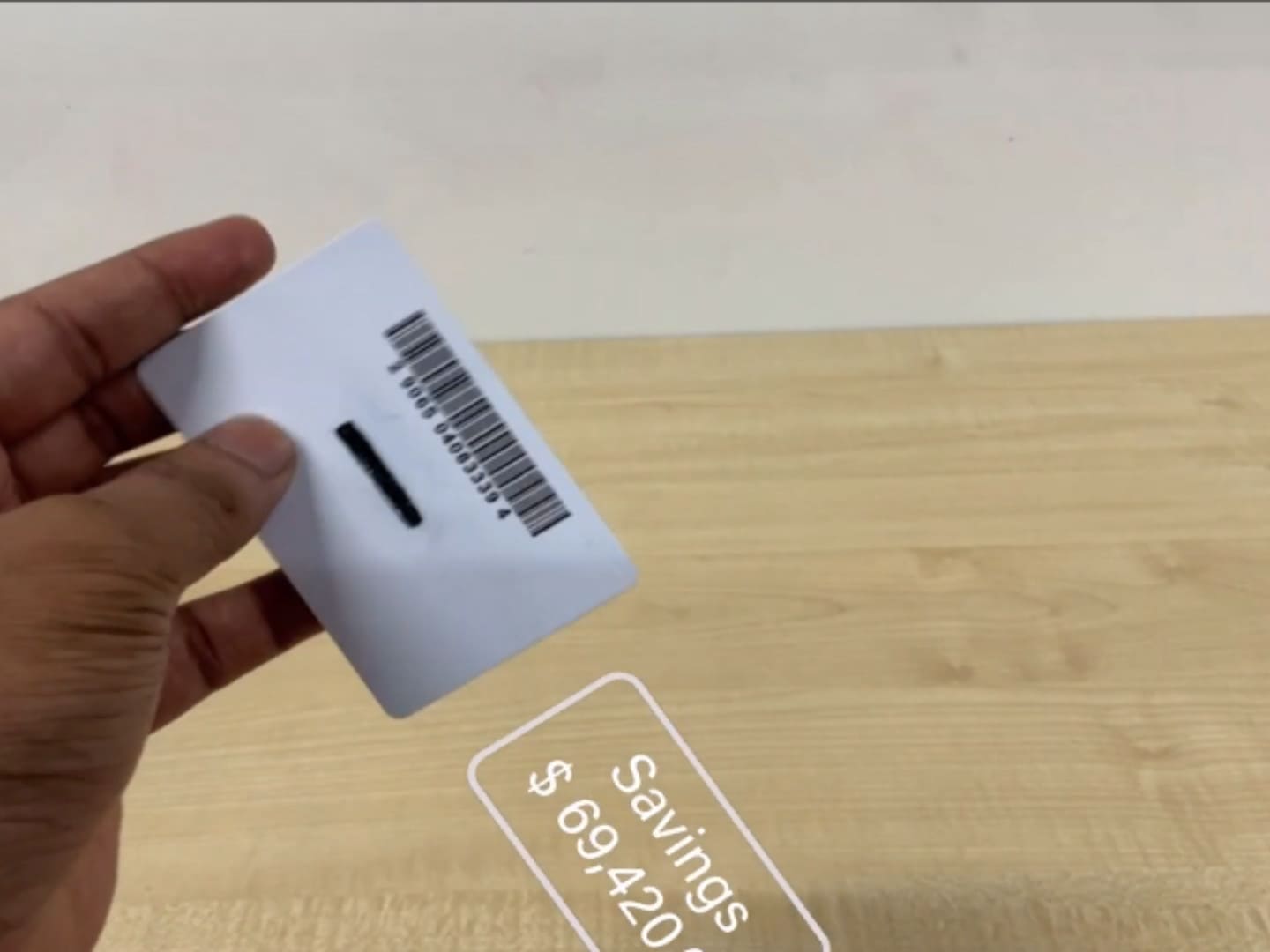}
\includegraphics[width=\width]{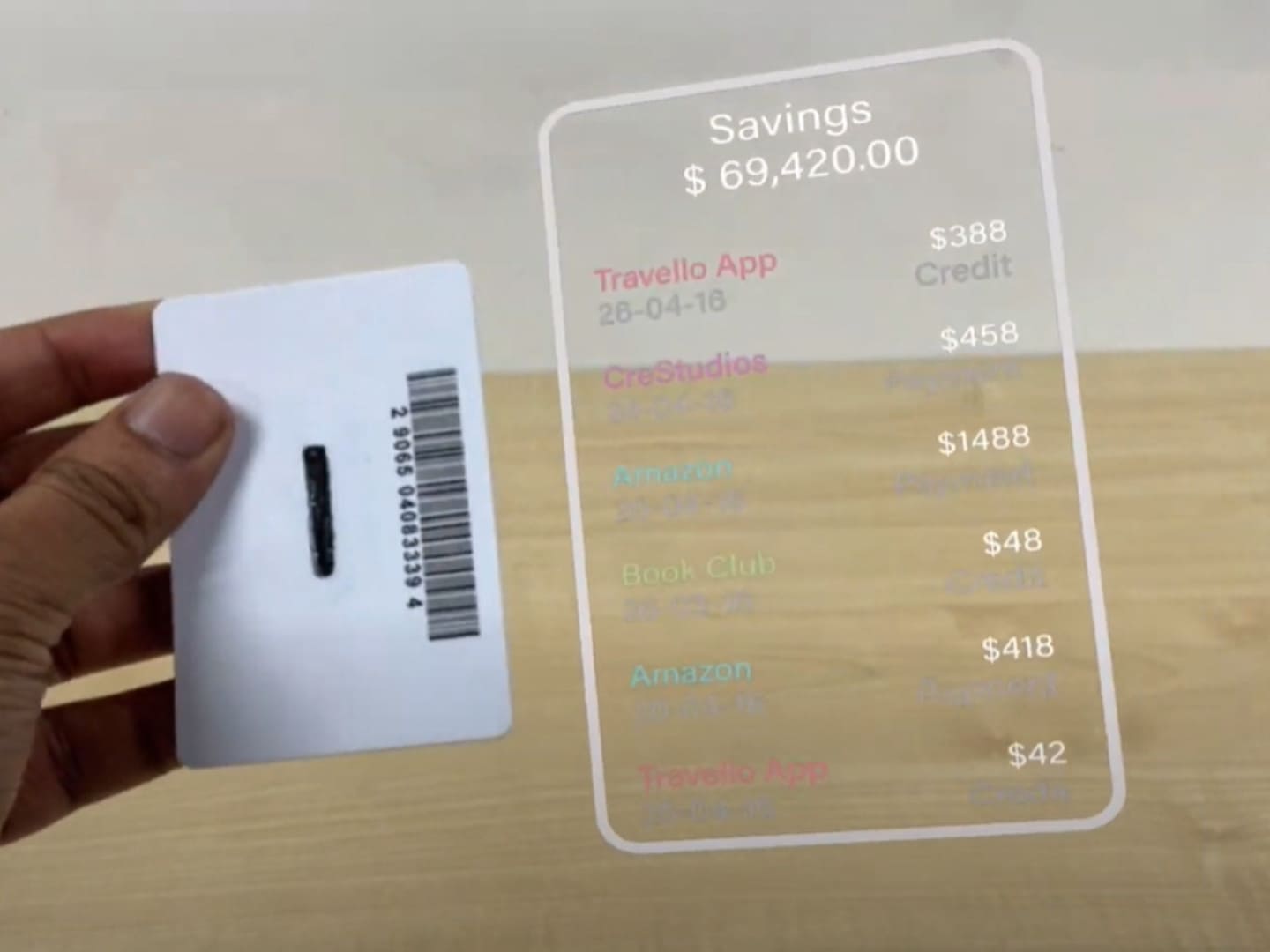}
\caption{Input - Orientation of the object: The user rotates the card to expand on their transaction.}
\label{fig:input-orientation}
\Description{This figure demonstrates orientation as a detectable property,it has three images one for each - vertical, diagonal and horizontal orientation of a debit card. These images shows that the layout of AR information anchored to the card changes as you rotate it.}
\end{figure}

The user can also use the orientation of the object as an input.
For example, the user can quickly create an AR application to show different information about a credit card, such as total balance or detailed monthly expenses, based on the orientation of the card (Figure~\ref{fig:input-orientation}).

\subsubsection{Appearance} 
The user can also detect the different appearances of the object. 
Based on the appearance, the user can show different virtual content based on the color of the block, the cover of the book, the application screen of the phone, and the content of a paper. 

\subsubsection{Deformation} 
\begin{figure}[h!]
\centering
\includegraphics[width=\width]{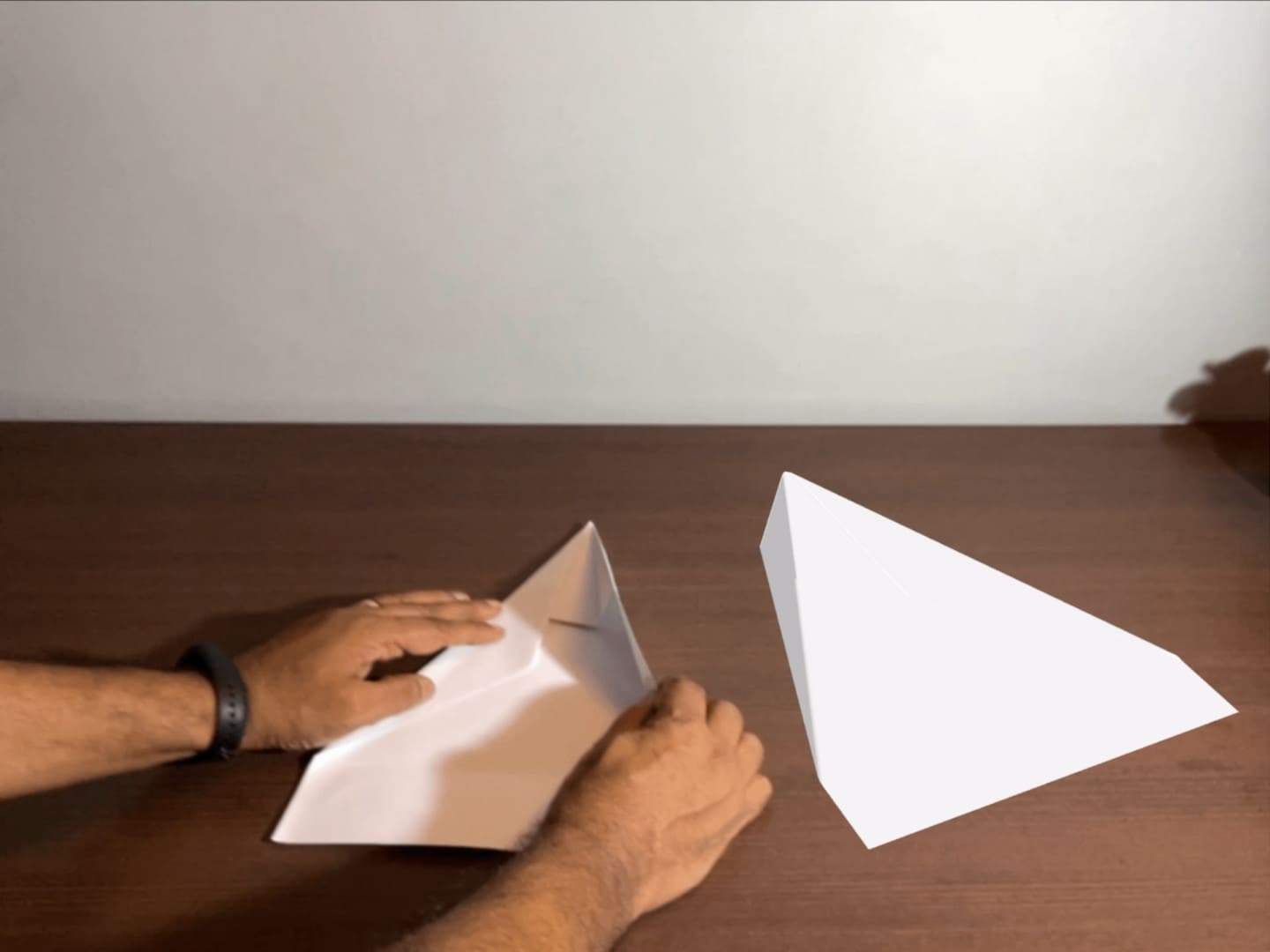}
\includegraphics[width=\width]{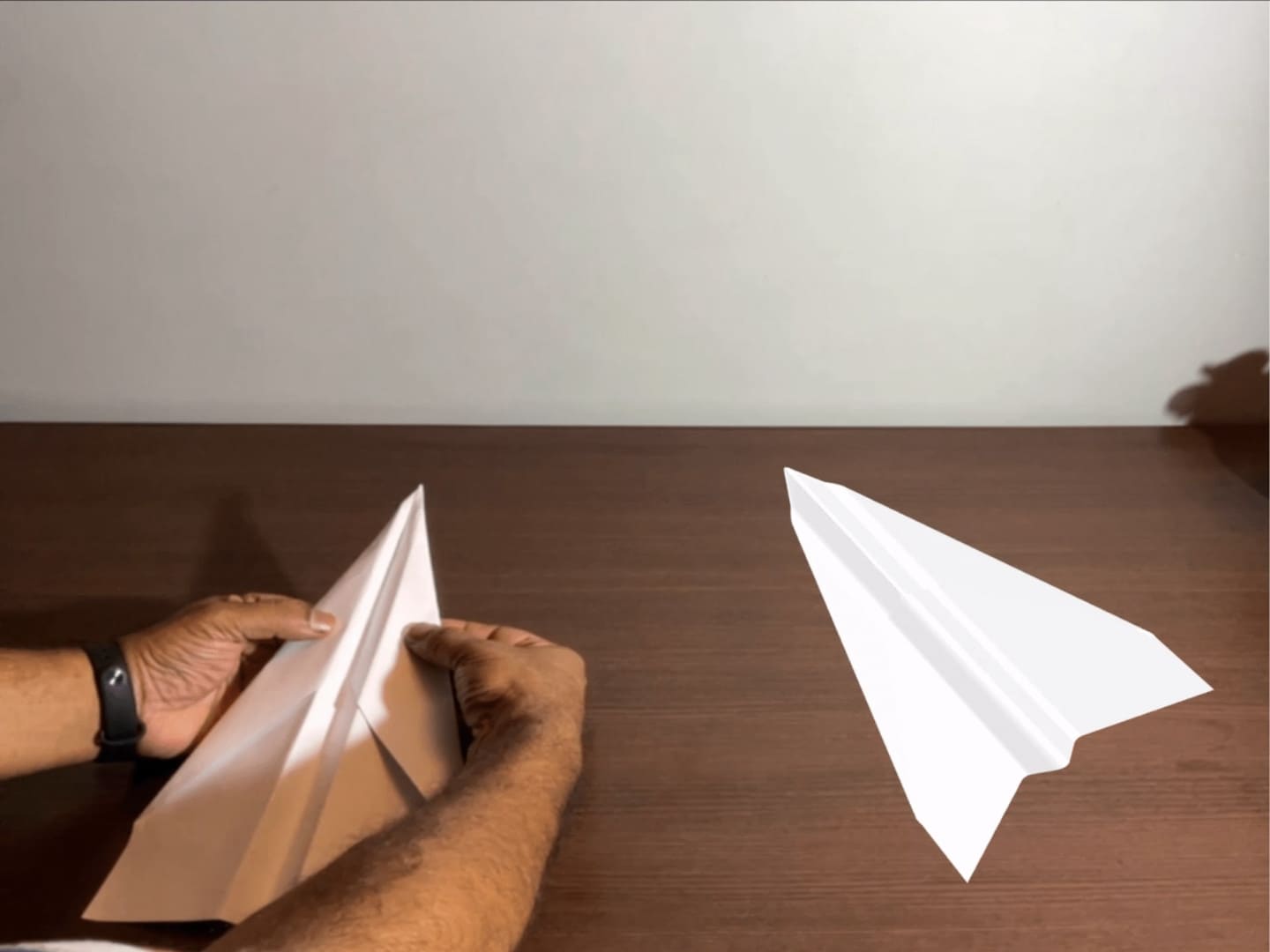}
\includegraphics[width=\width]{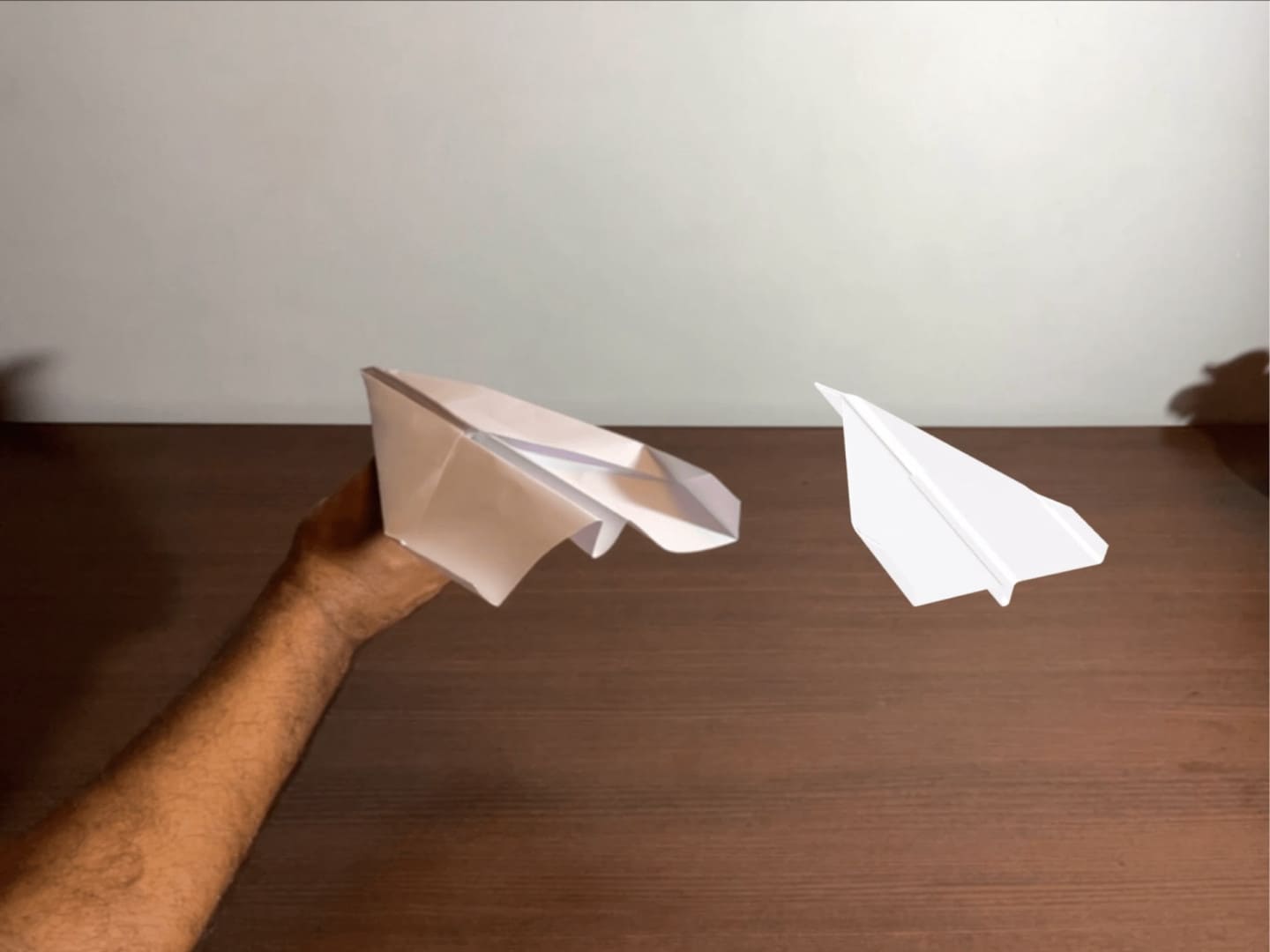}
\caption{Input - Deformation of the object: The user can use a deformable object like paper as input to an origami AR instruction animation to show the next step.}
\label{fig:input-deformation}
\Description{This figure demonstrates deformation as input,it has three images one for each step of folding a paper into a plane while a virtual paper is leading the user to the next step through animations.}
\end{figure}

Also, the user can detect deformation of the object, such as bendable paper, expandable Hoberman sphere, origami, and a slinky spring toy (Figure~\ref{fig:input-deformation}).
The user can use these deformable objects as input, or alternatively, the user can create an instruction based on the shape of each state (e.g., AR origami instruction).

\subsubsection{Combination} 
While the above categories are mostly focused on a single object's properties, the user can also detect a combination of objects.
For example, by identifying the combination of a hand and an object, the user can detect a simple touch interaction with a physical object. 

\subsubsection{Relationship} 
\begin{figure}[h!]
\centering
\includegraphics[width=\width]{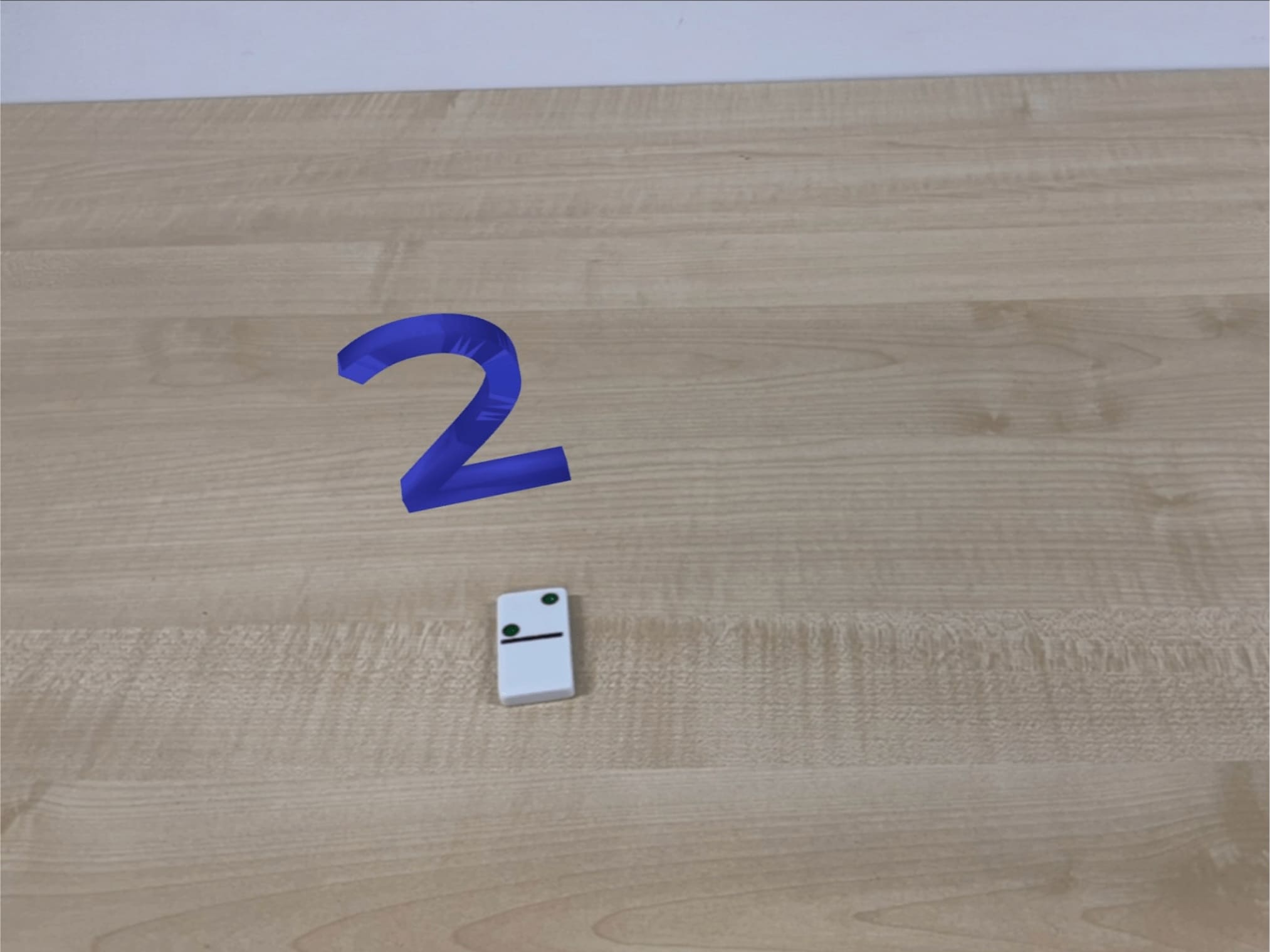}
\includegraphics[width=\width]{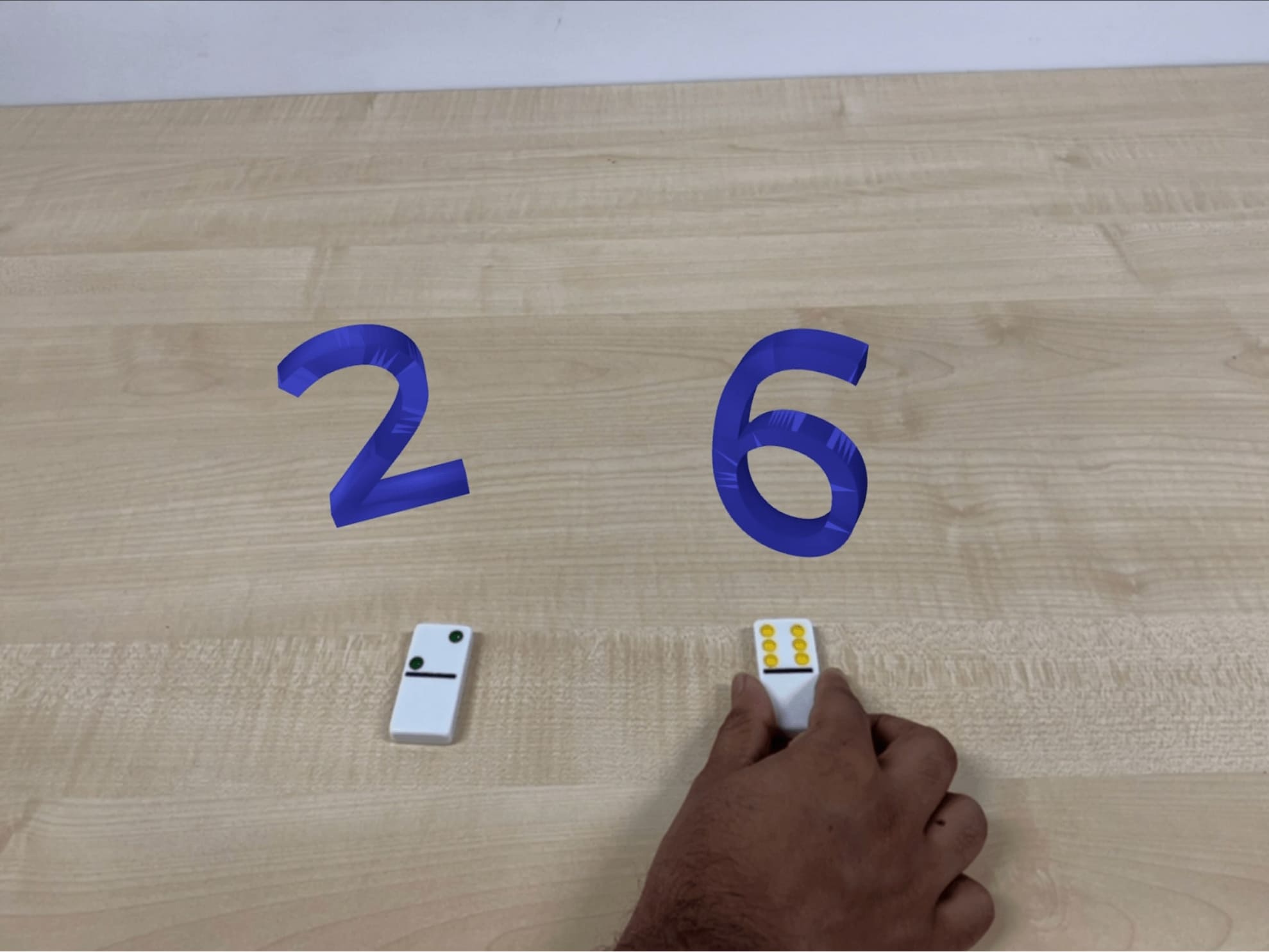}
\includegraphics[width=\width]{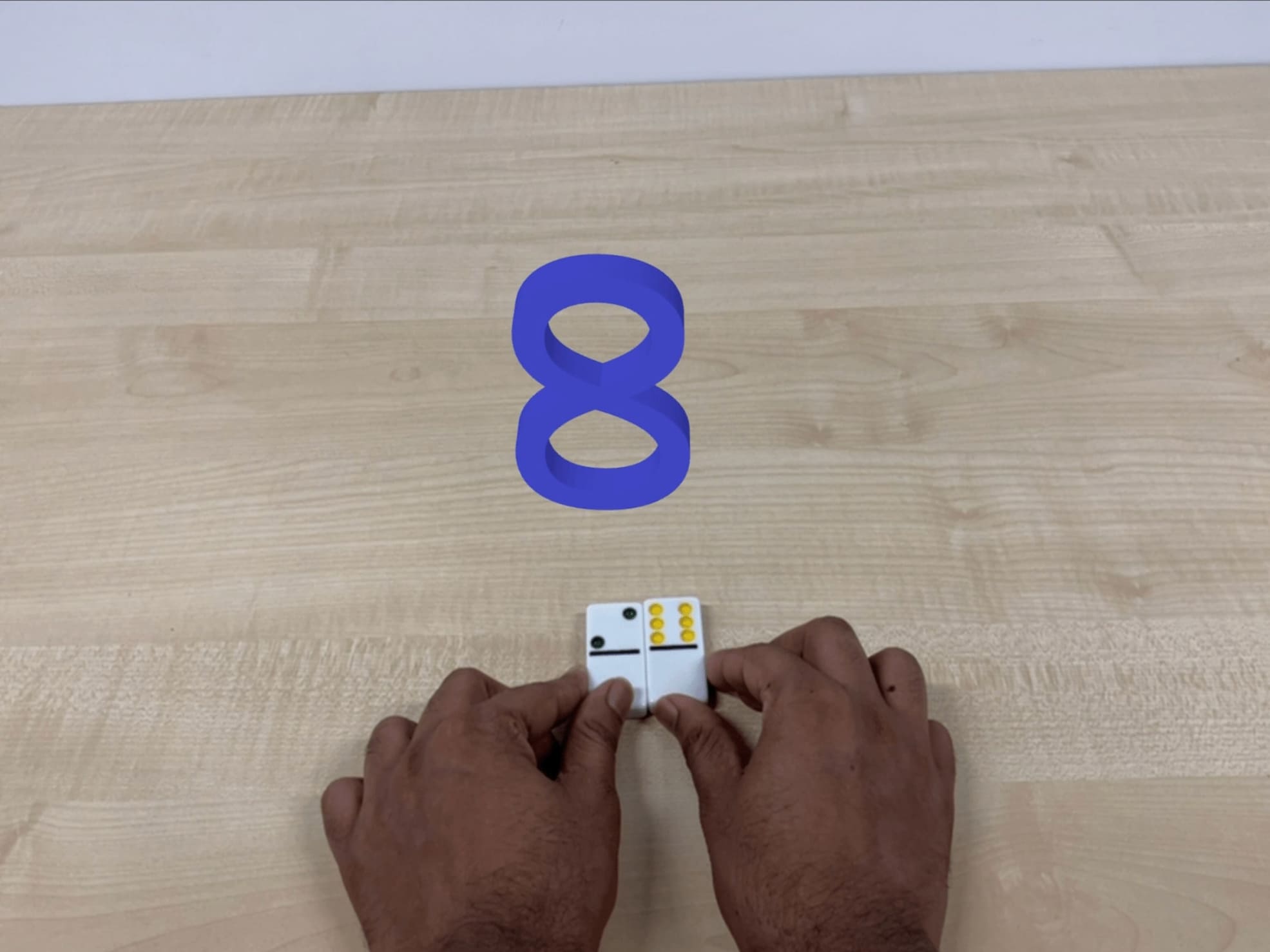}
\caption{Input - Relationship of the multiple objects: The system detects the different positional relationships between two blocks.}
\label{fig:input-relationship}
\Description{This figure demonstrates relationship between multiple objects as input,it has three images one with a numbered blocks of the value two and a virtual digit two appears. In the second image the user brings in another numbered block of value six and a virtual digit six appears. In the third figure the user pushes both these blocks together which adds the two virtual numbers he digit 8 is displayed.}
\end{figure}

While the combination focuses only on the presence or absence of multiple objects, the user can also use relationships between multiple objects.
For example, by detecting the finger's relative position to a paper, the user can mimic multiple touch point detection for the physical paper (Figure~\ref{fig:output-image-anchored}).
Alternatively, the user can also detect the distance between two objects, different grasping gestures for the object, or different arrangements of the multiple objects (Figure~\ref{fig:input-relationship}).

\subsection{Output: Types of Virtual Objects}
For AR output, the user can place various virtual objects into the AR scene.
Here, we describe different types of virtual objects that are supported by our system.

\subsubsection{3D Objects} 
\begin{figure}[h!]
\centering
\includegraphics[width=\width]{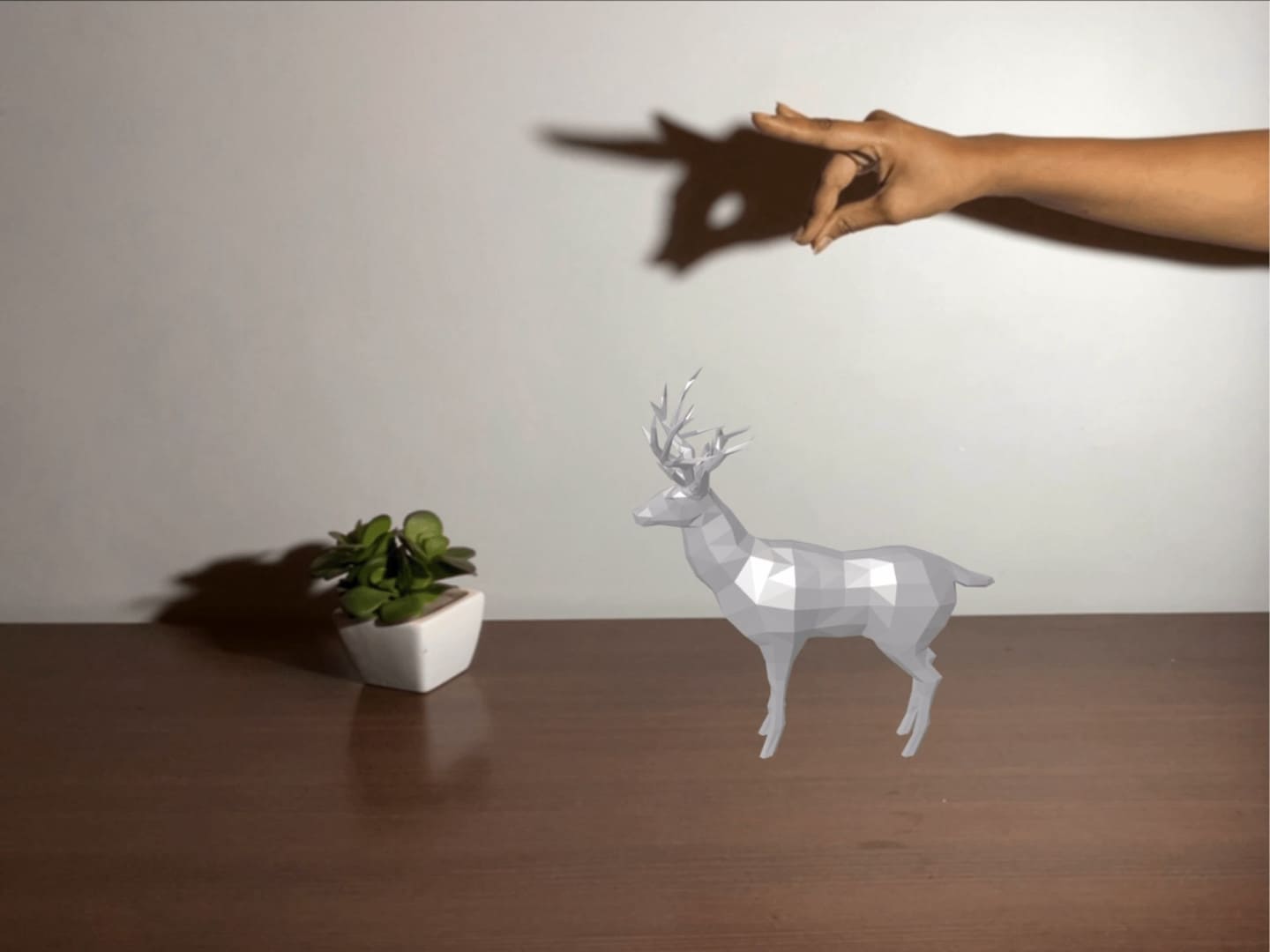}
\includegraphics[width=\width]{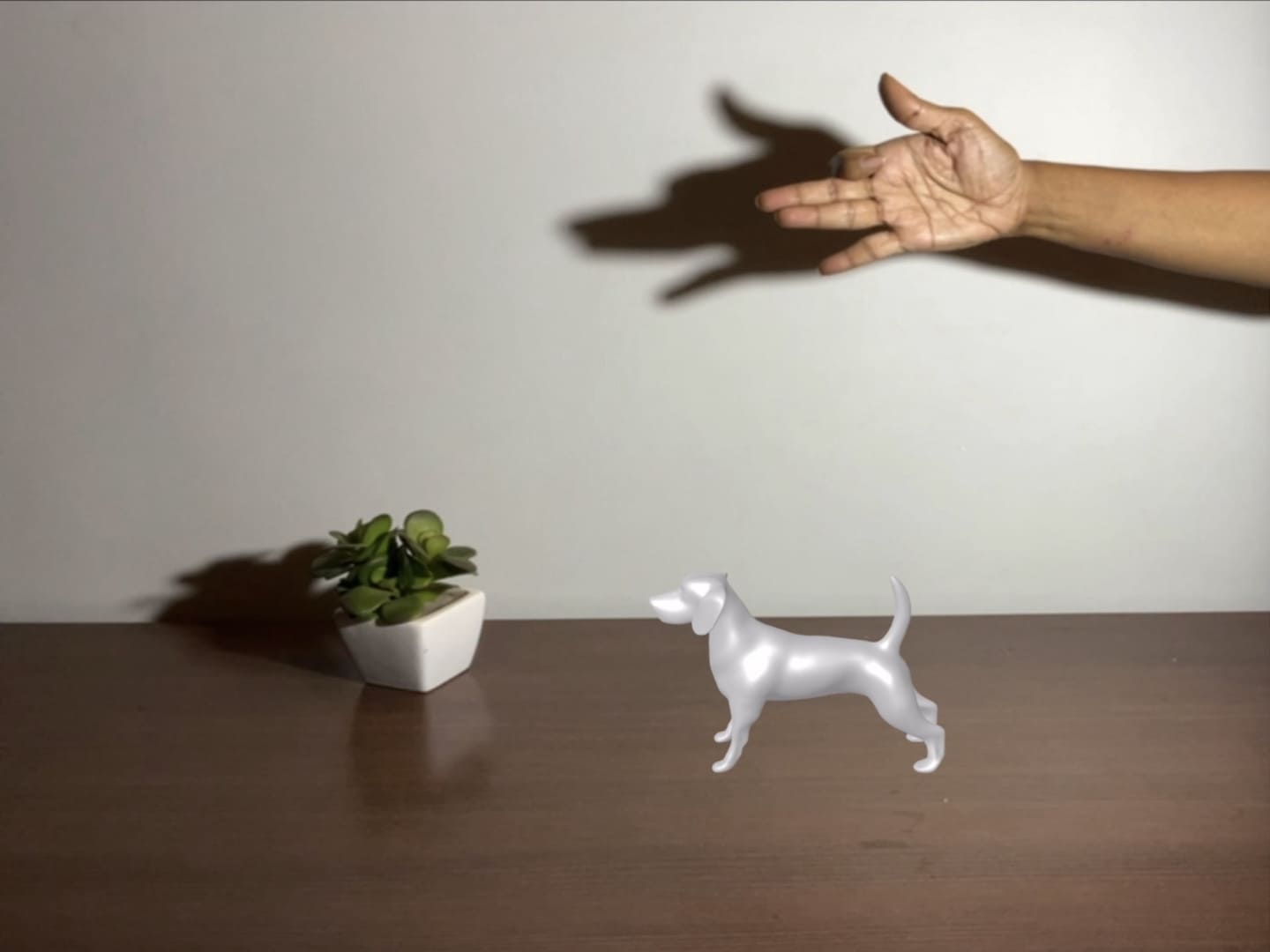}
\includegraphics[width=\width]{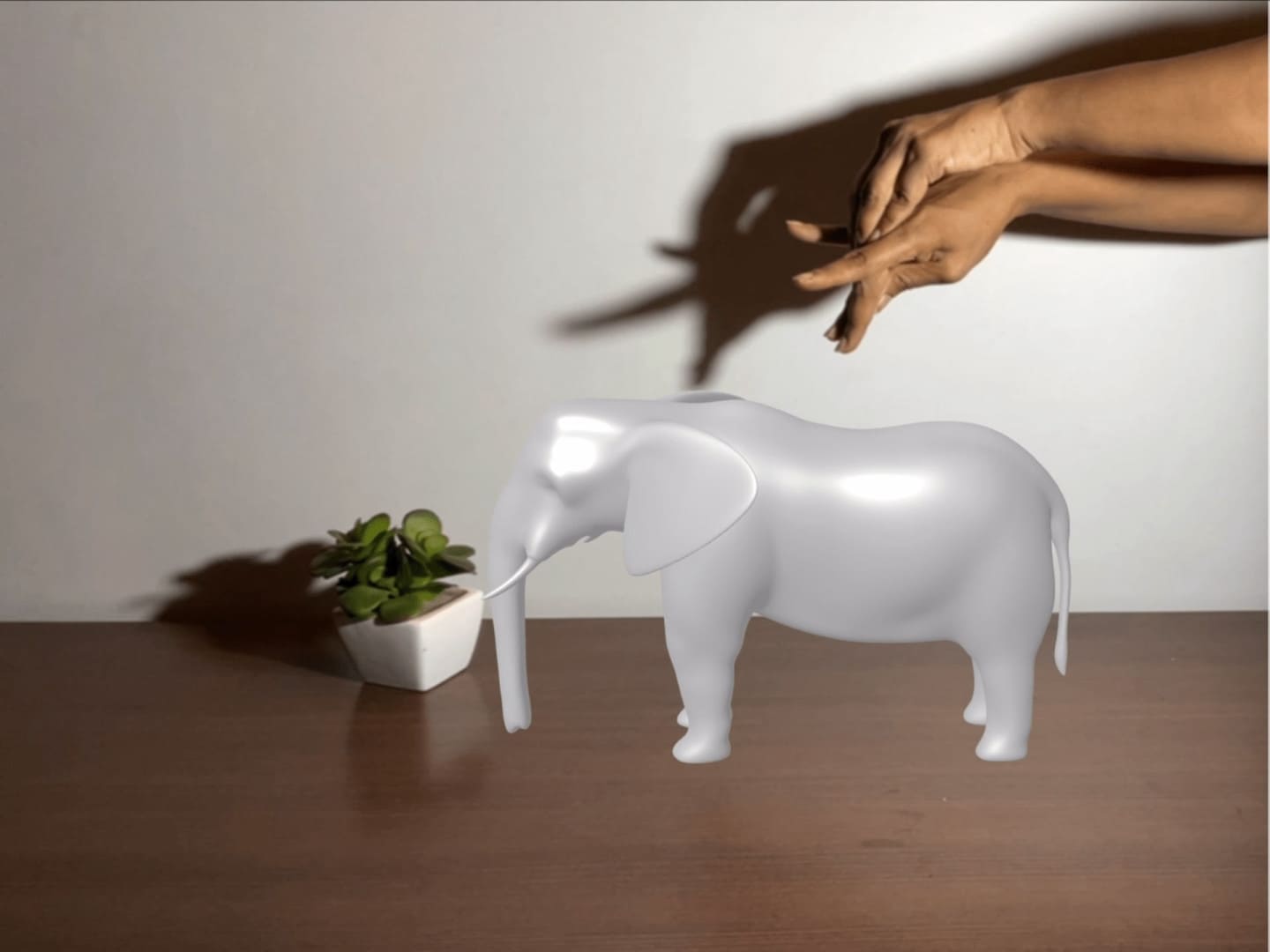}
\caption{Output - 3D Object: Showing different 3D objects like deer, dog, and elephant, based on the different gestures.}
\label{fig:output-3d-object}
\Description{This figure demonstrates a 3D object as an output. It has three images of a user making three hand shadow puppets - a deer, a dog, and an elephant and the corresponding animal's 3D virtual object appears in each image.}
\end{figure}

First, the user can place a virtual 3D object in the scene by importing from existing assets (Figure~\ref{fig:output-3d-object}). 
To do so, the user can simply press the \textit{3D Object} button, then the system lets the user select from the available 3D objects.
The user can also prepare their own assets to place in the scene.
Once placed, the user can change the position, orientation, and scale of the object through touch interaction.

\subsubsection{2D Images} The user can also place a 2D image in a scene and create prototypes similar to \textit{Opportunistic Interfaces}~\cite{du2022opportunistic}.
The user first taps the \textit{2D Image} button, then selects the image.
2D image is shown as a texture of the virtual plane in the 3D scene.
Therefore, similar to 3D objects, the user can also interactively change the position, orientation, and scale of the 2D objects. 

\subsubsection{Audio} 
\begin{figure}[h!]
\centering
\includegraphics[width=\width]{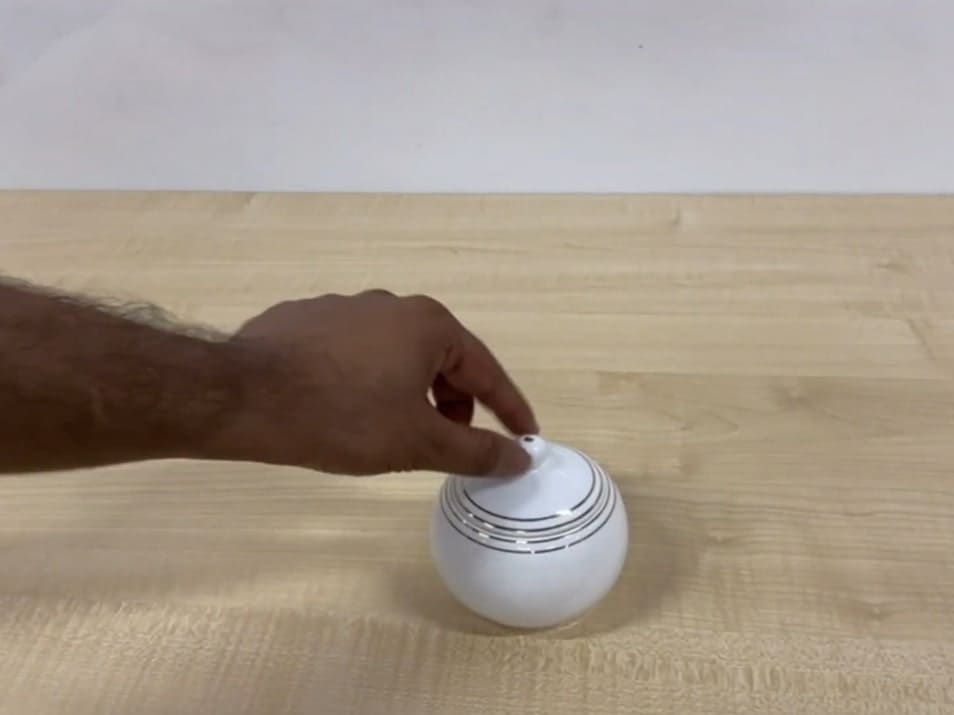}
\includegraphics[width=\width]{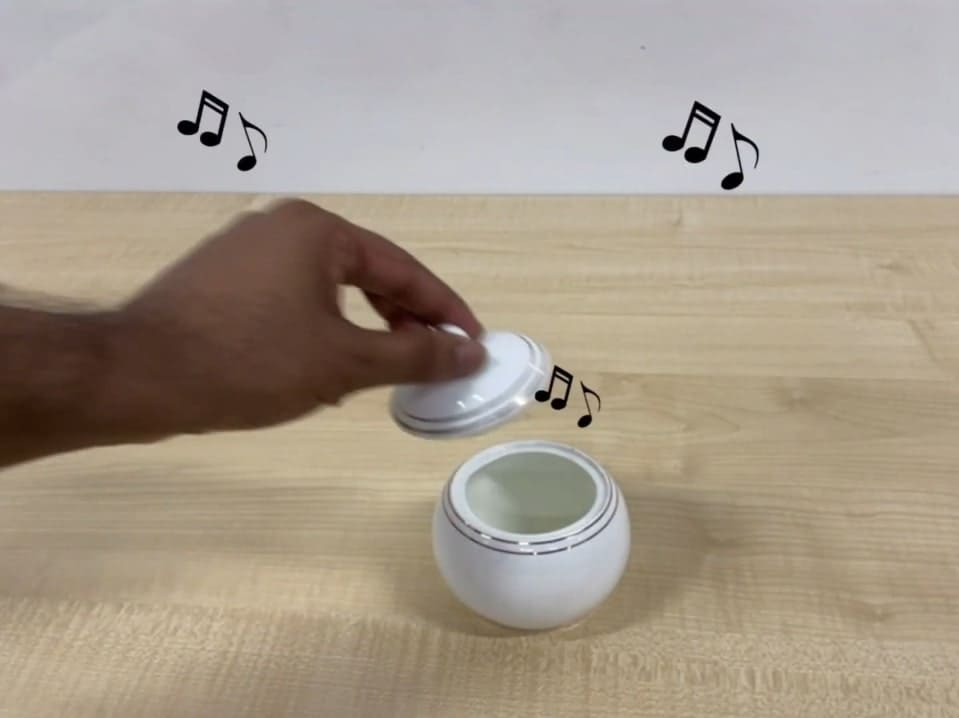}
\includegraphics[width=\width]{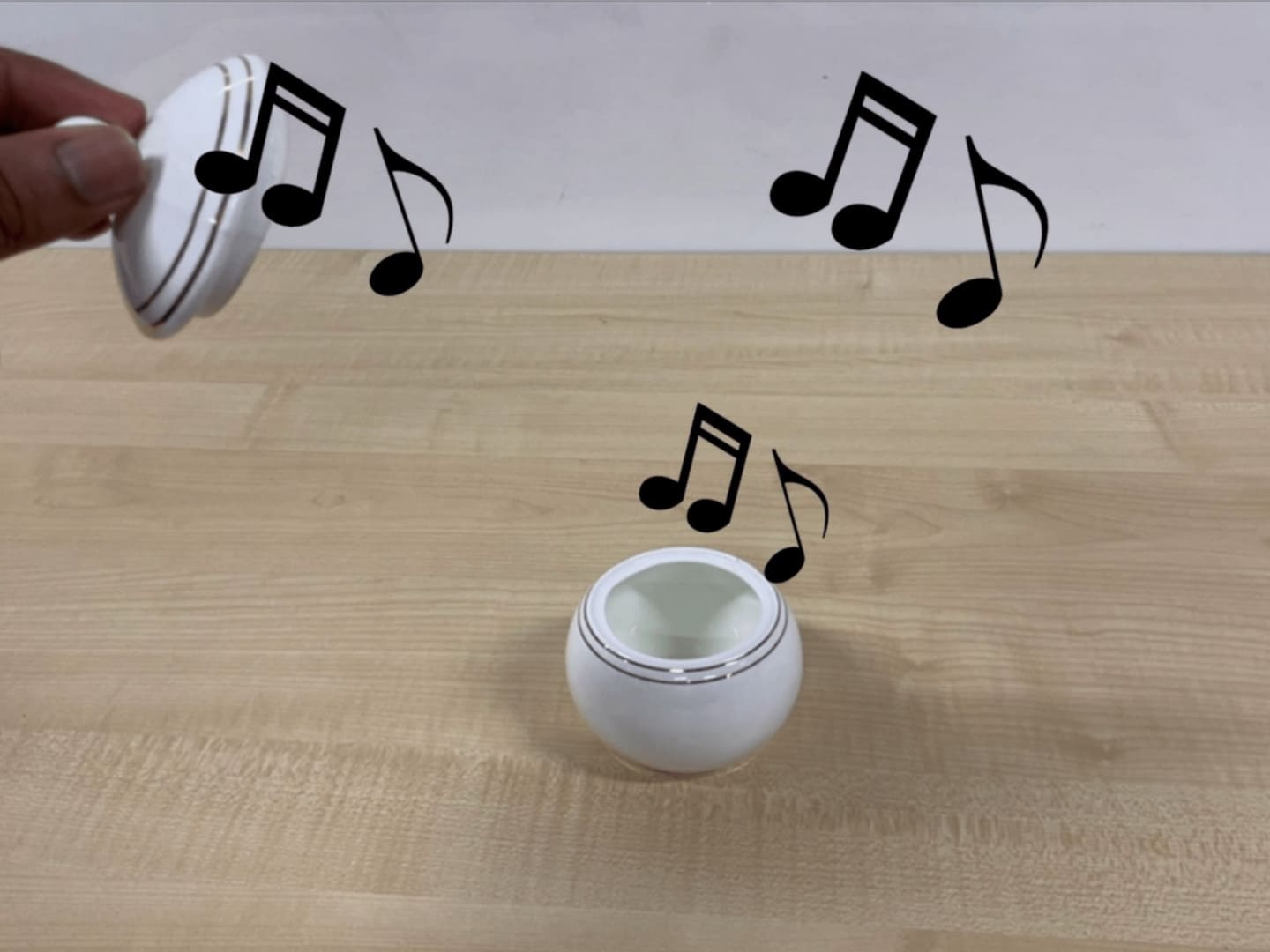}
\caption{Output - Audio: Playing audio when the lid is lifted.}
\label{fig:output-audio}
\Description{This figure demonstrates audio as an output. It has three images at different stages of a user lifting a lid, and the audio levels increase correspondingly.}
\end{figure}

The system also supports audio output to help create experiences.
To do so, the user can simply select an mp3 file, then the system plays the sound when the action is triggered (Figure~\ref{fig:output-audio}).
This way, the user can create a multi-modal output to enrich the AR experience.
For example, by detecting the different state of the physical bottle, the user can show a virtual animation and music output when opening the bottle cap, similar to \textit{Music Bottles}~\cite{ishii2004bottles}.

\subsubsection{Other Outputs} 
The user can also embed various types of pre-programmed assets, such as character animation like in \textit{Project Zanzibar}~\cite{villar2018project}, particle effects, or embedded screens. This enables users to create prototypes of experiences similar to social media filters.
Again, the system can load these various types of outputs based on the file import or iFrame.
By leveraging the embedded screens, the user can also show other useful outputs like interactive charts or data visualizations.

\subsection{Output: Anchored Location}
The system also supports different types of anchored locations where the imported virtual object should be placed.
When placing a virtual object, the interface lets the user select the anchored location type.
By moving the virtual object, the system maintains the position relative to the anchored location.

\subsubsection{Surface Anchored} 
\begin{figure}[h!]
\centering
\includegraphics[width=\width]{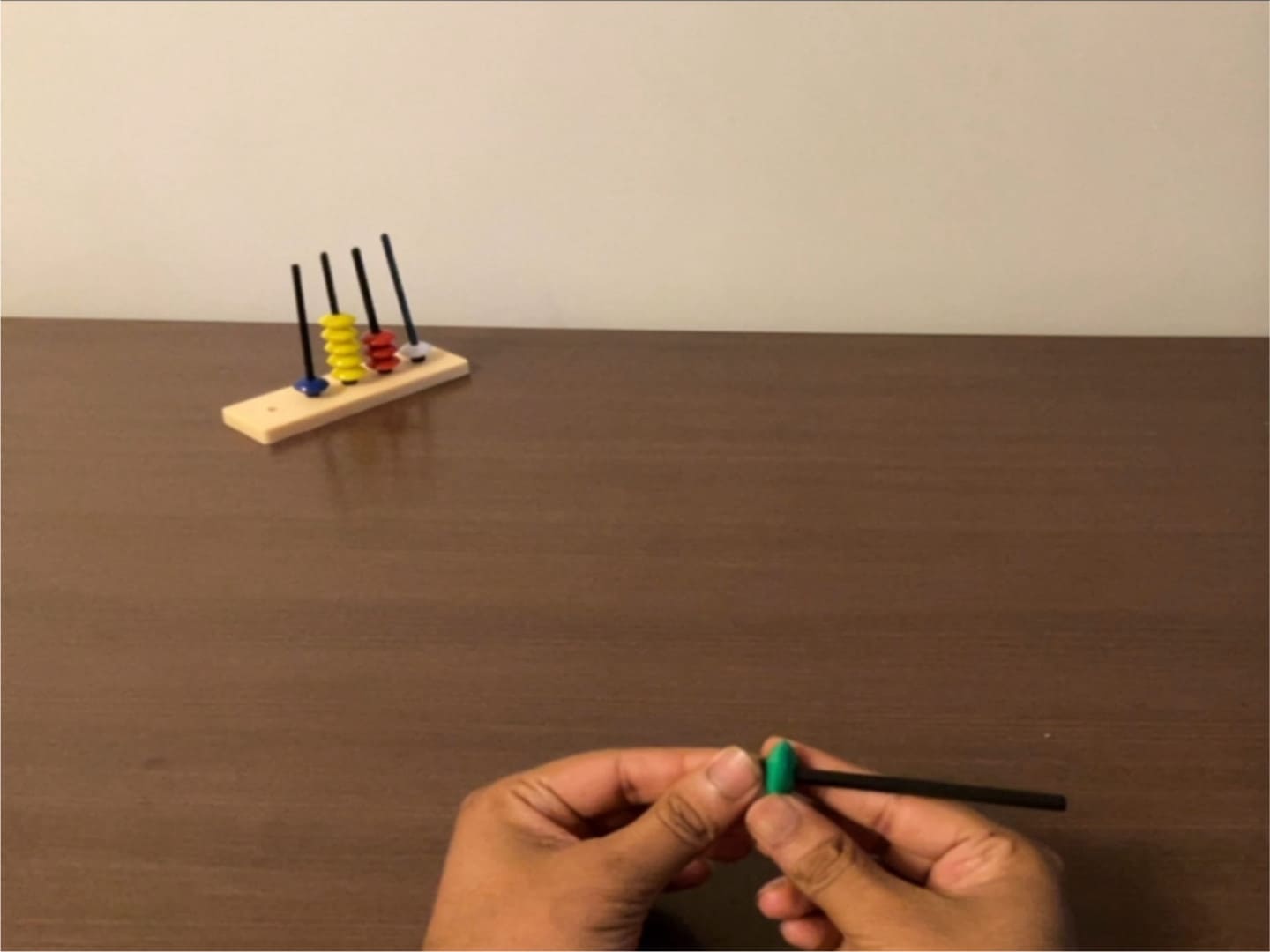}
\includegraphics[width=\width]{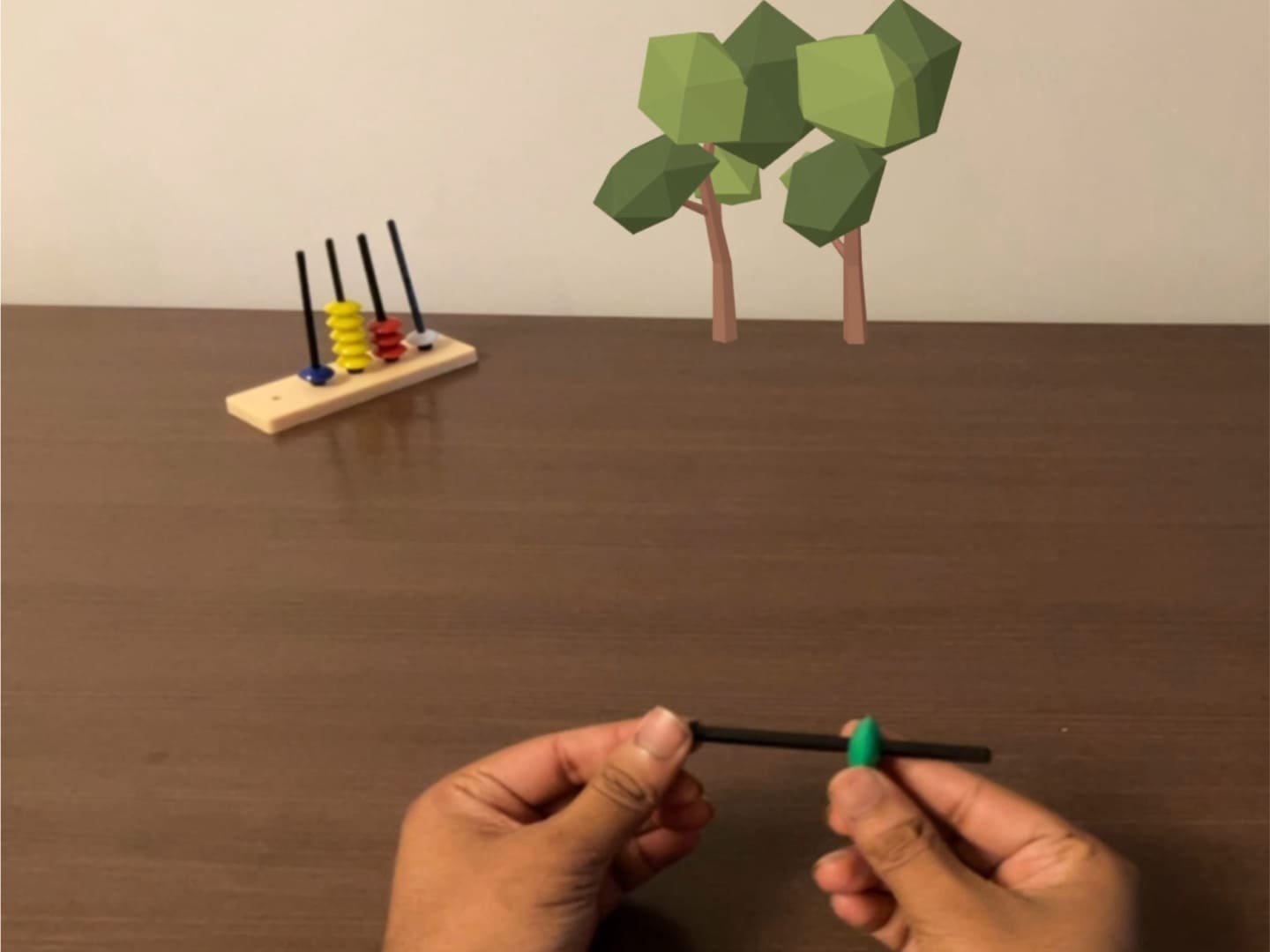}
\includegraphics[width=\width]{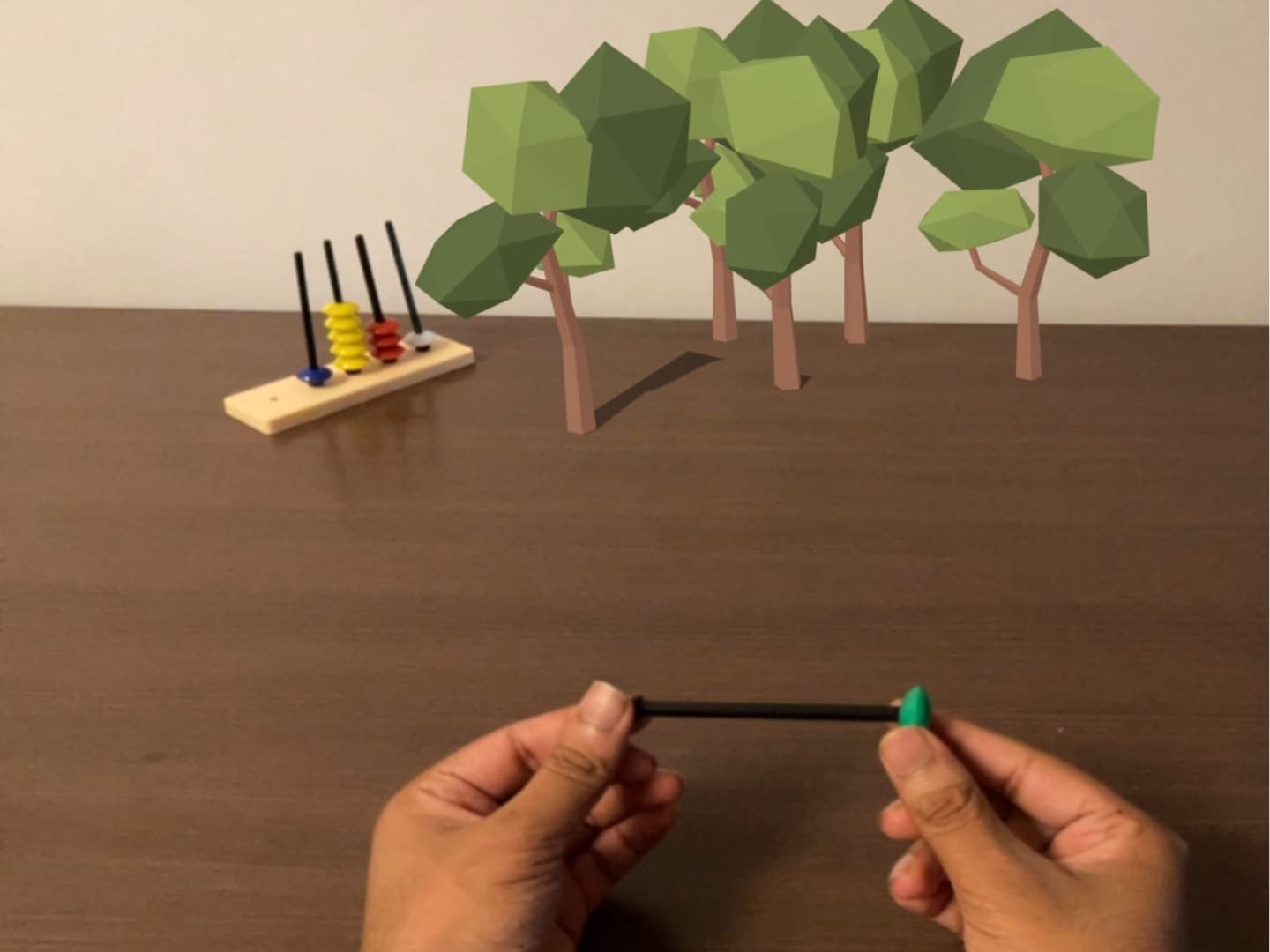}
\caption{Output - Surface Anchored: The user can spawn the trees as the position of the token moves.}
\label{fig:output-surface-anchored}
\Description{This figure demonstrates surface-anchored output. It has three images at different stages of a user sliding a token, and the number of virtual trees increases correspondingly.}
\end{figure}

By default, the user can place an object onto a detected surface (Figure~\ref{fig:output-surface-anchored}). 
Based on the system's built-in surface detection, the user can place a virtual object anchored on a horizontal or vertical surface like a table, floor, or wall like \textit{Augmented Displays}~\cite{reipschlager2020augmented}. 

\subsubsection{Spatial Anchored} Similarly, the user can also place a floating virtual object, which stays in a certain spatial position in mid-air. 
To do so, the user can simply tap the \textit{spatial} option, then the user can start manipulating the object without the bound of the detected surface.
Since the mobile AR system can track the spatial position, the spatially anchored object stays in the same position, regardless of the movement of the mobile phone. This allows the users to create prototypes for a system like spatial collaborations like \textit{SynchronizAR}~\cite{huo2018synchronizar}.

\subsubsection{Camera Anchored} Instead of placing on a spatially-anchored location, the user can also make information always visible by overlaying it in the user's field of view similar to the technique used by \textit{RealityTalk}~\cite{liao2022realitytalk}.
When the user taps the \textit{overlay} option, the virtual object is anchored on a screen, so that the user can move the position of the virtual object within the 2D screen.

\subsubsection{Image Anchored} 
\begin{figure}[h!]
\centering
\includegraphics[width=\width]{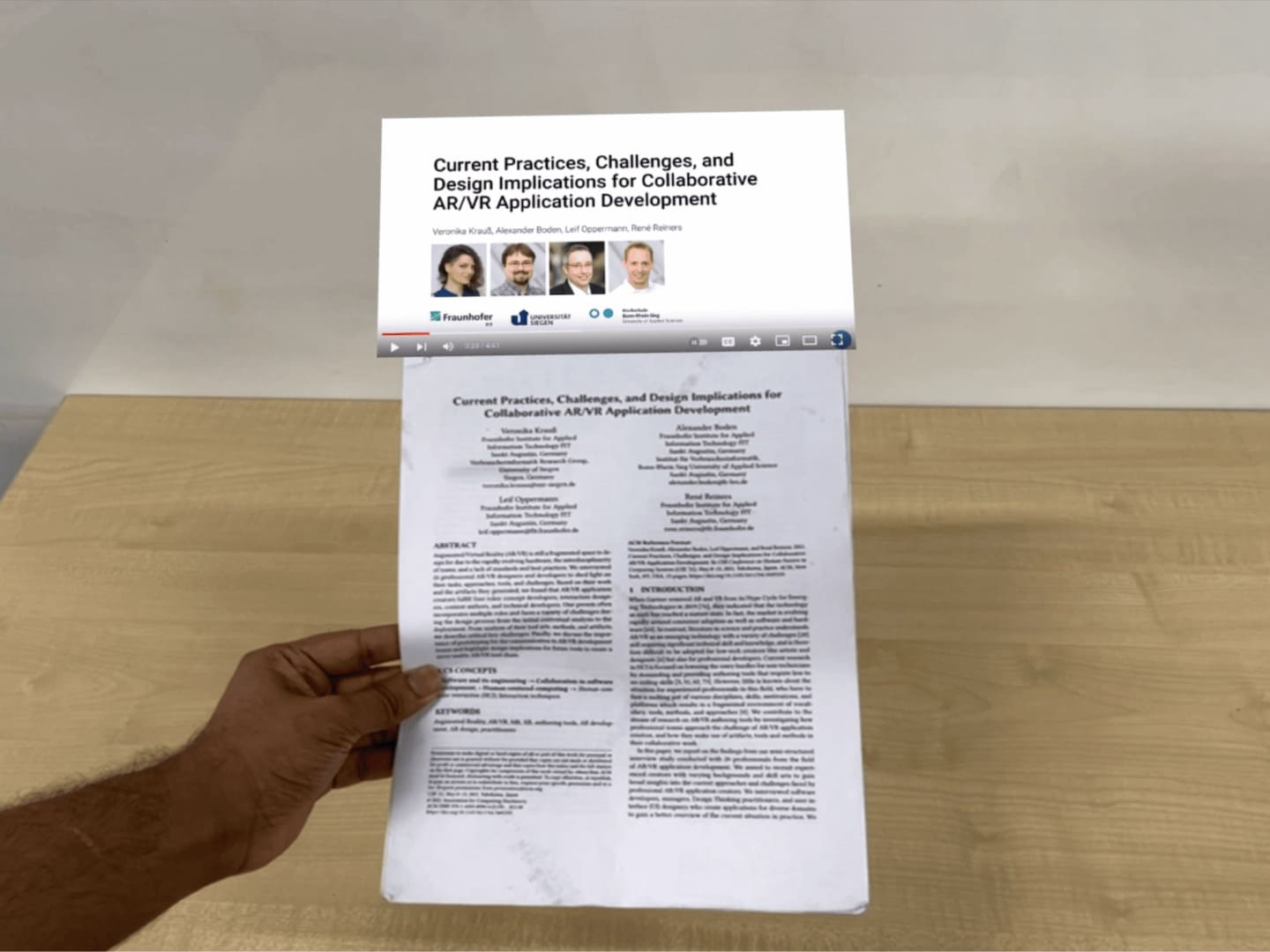}
\includegraphics[width=\width]{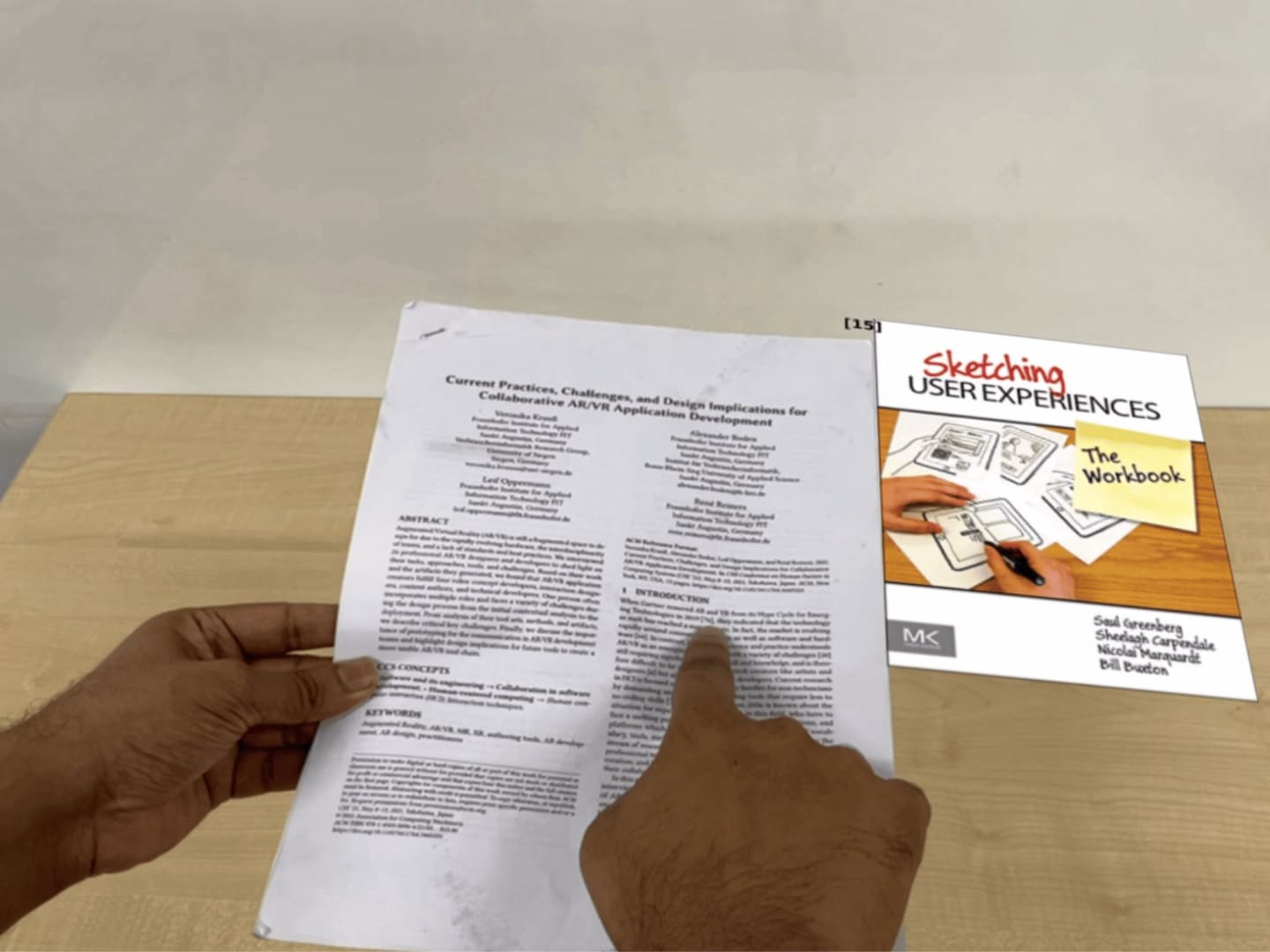}
\includegraphics[width=\width]{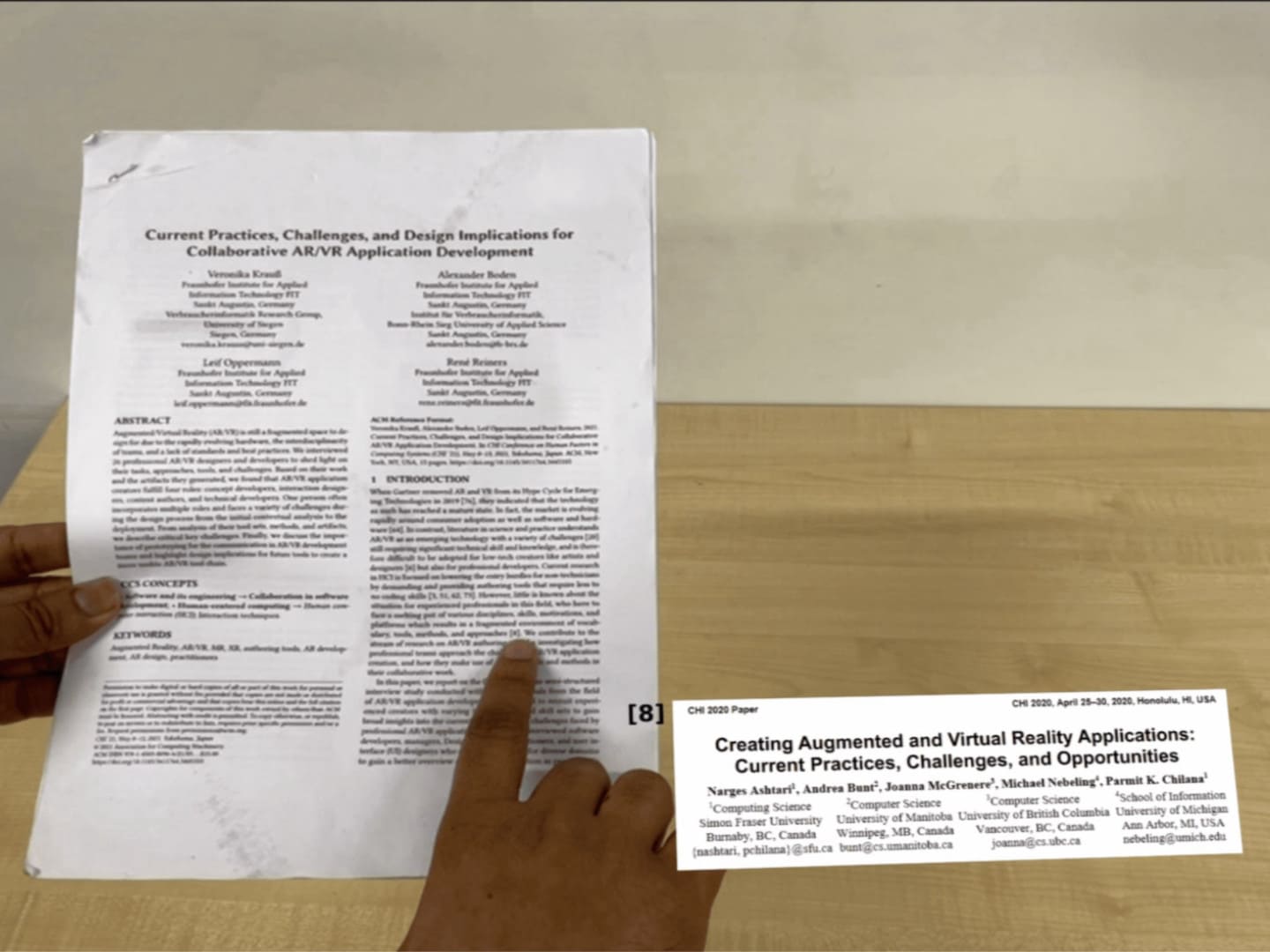}
\caption{Output - Image Anchored: The user can prototype an augmented display to show related information when reading papers.}
\label{fig:output-image-anchored}
\Description{This figure demonstrates image-anchored output. It has three images at different stages of a user reading a research paper. Initially the video presentation plays anchored to the paper while image 2 and 3 show related work as the user points to the citation.}
\end{figure}

The user can also place a virtual object anchored around the image based on the provided target image to create prototypes similar to \textit{Opportunistic Interfaces}~\cite{du2022opportunistic}.
In this case, the virtual object moves along with a paper (Figure~\ref{fig:output-image-anchored}).
To do so, the system leverages a common image target tracking based on the provided image.
When the user taps the \textit{image} option, the system lets the user specify the image target based on the selected or uploaded image.

\subsubsection{Object Anchored} 
\begin{figure}[h!]
\centering
\includegraphics[width=\width]{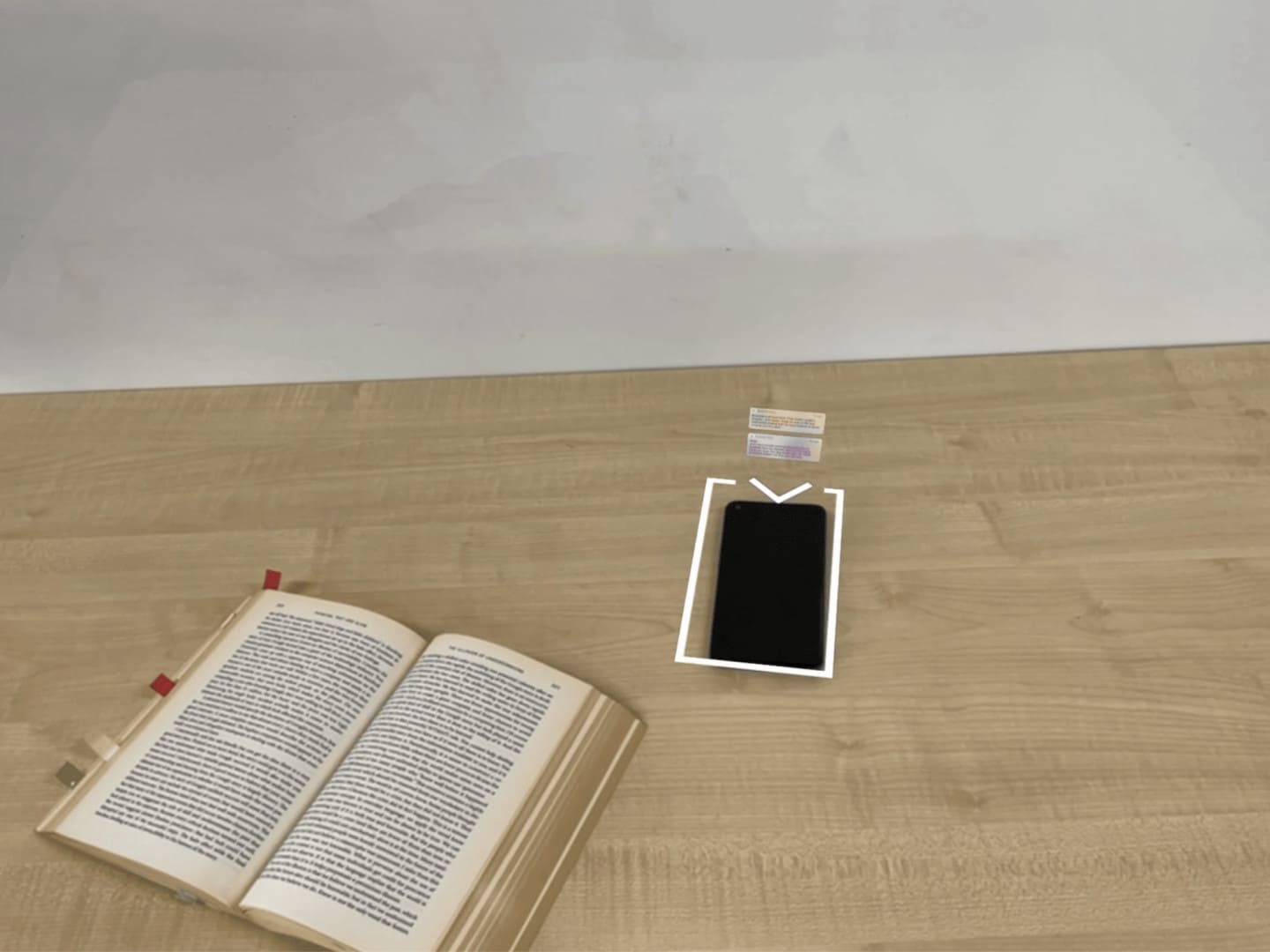}
\includegraphics[width=\width]{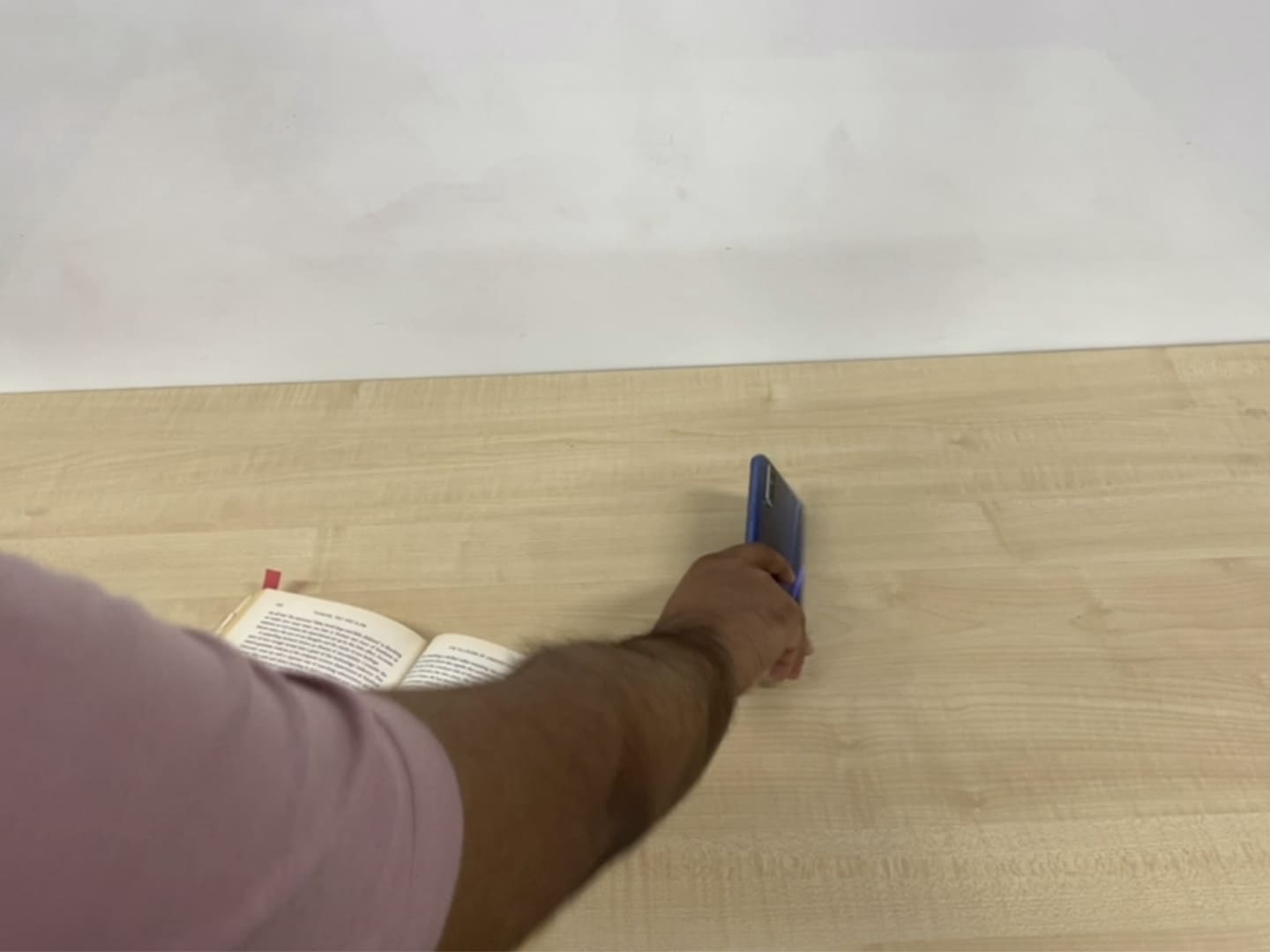}
\includegraphics[width=\width]{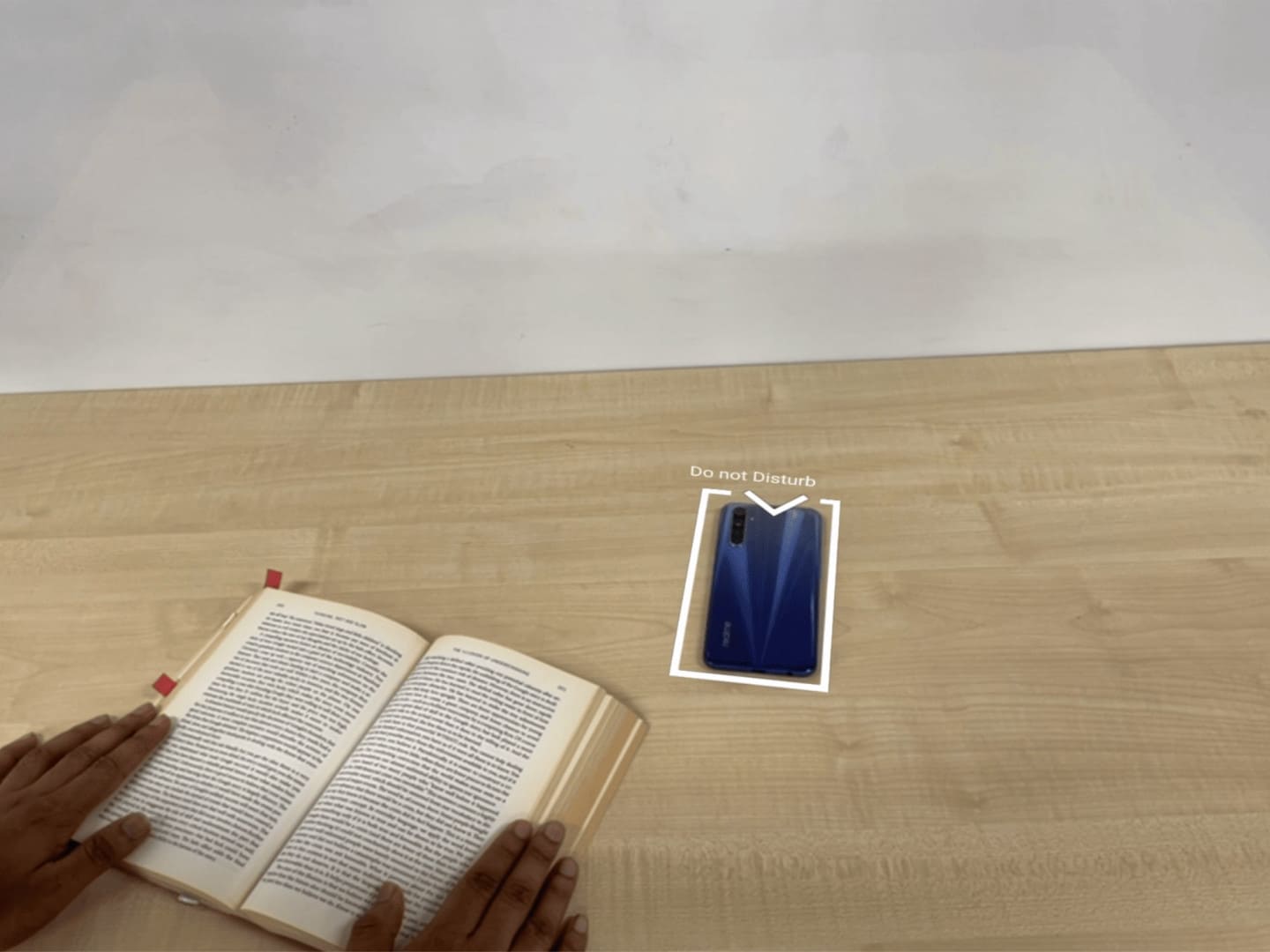}
\caption{Output - Object Anchored: The user can turn off silent mode by flipping the smartphone and hiding notifications.}
\label{fig:output-object-anchored}
\Description{This figure demonstrates Object Anchored output. The user turns off silent mode by flipping the smartphone and hiding notifications.}
\end{figure}

The user can also anchor a virtual object to a physical object  or a human (Figure~\ref{fig:output-object-anchored}).
In contrast to image anchored, object anchored can be any physical object or human, which can be useful for object-related information like annotation as seen in \textit{Light Anchors}~\cite{ahuja2019lightanchors}.
When the user taps the \textit{object} option, the user can then tap an object to specify the tracked object.
To track an object's position, the system uses simple 2D color tracking and raycasting to obtain the 3D position on a surface, similar to \textit{RealitySketch}~\cite{suzuki2020realitysketch}.
Therefore, the tracking works best with a solid colored object.

\subsection{Output: Behavior of Virtual Output}
As we mentioned, when transitioning from one state to another, the system automatically animates the object.
On top of that, the system also supports several additional output behaviors based on the trigger event.

\begin{figure}[h!]
\centering
\includegraphics[width=\width]{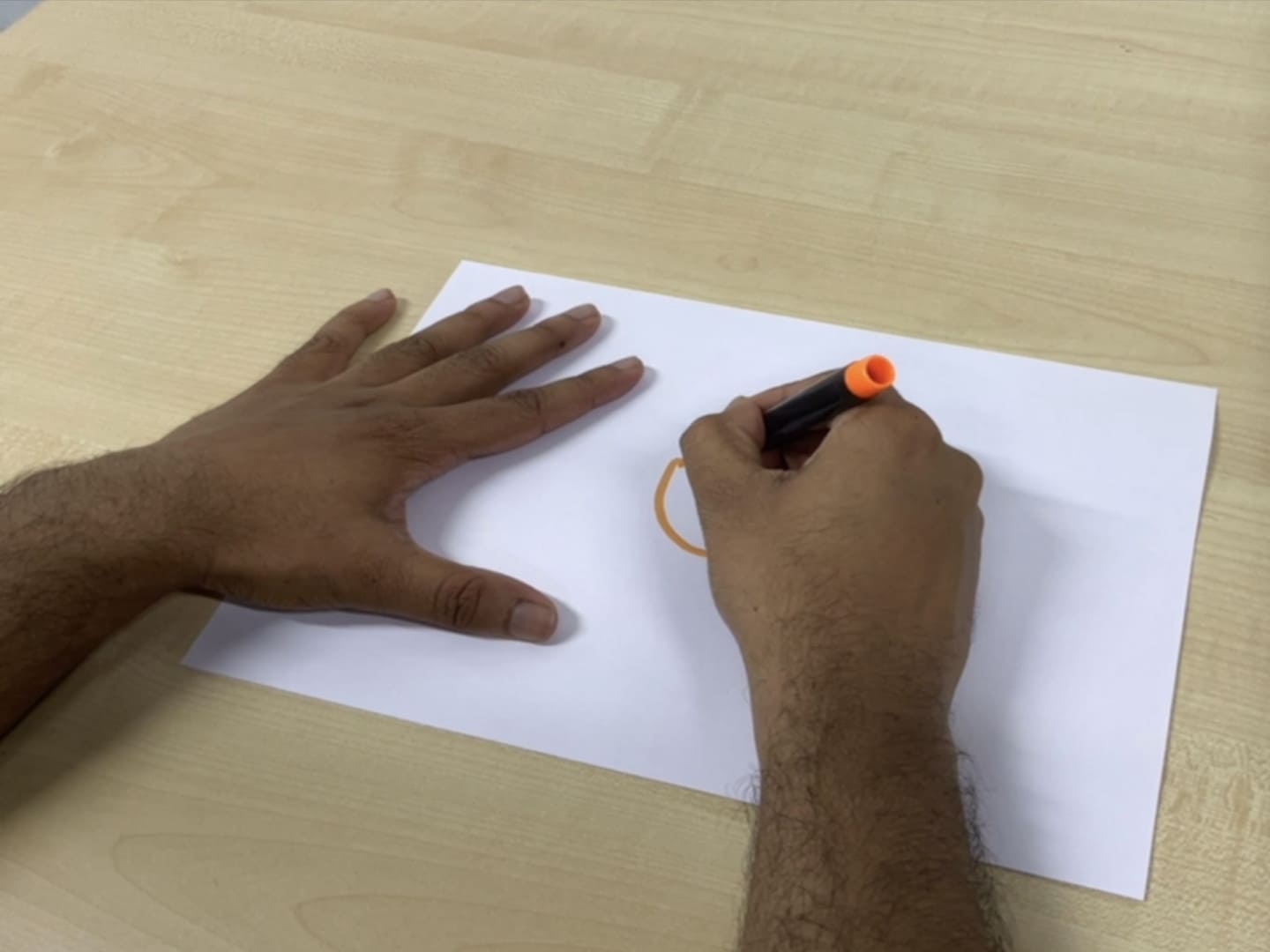}
\includegraphics[width=\width]{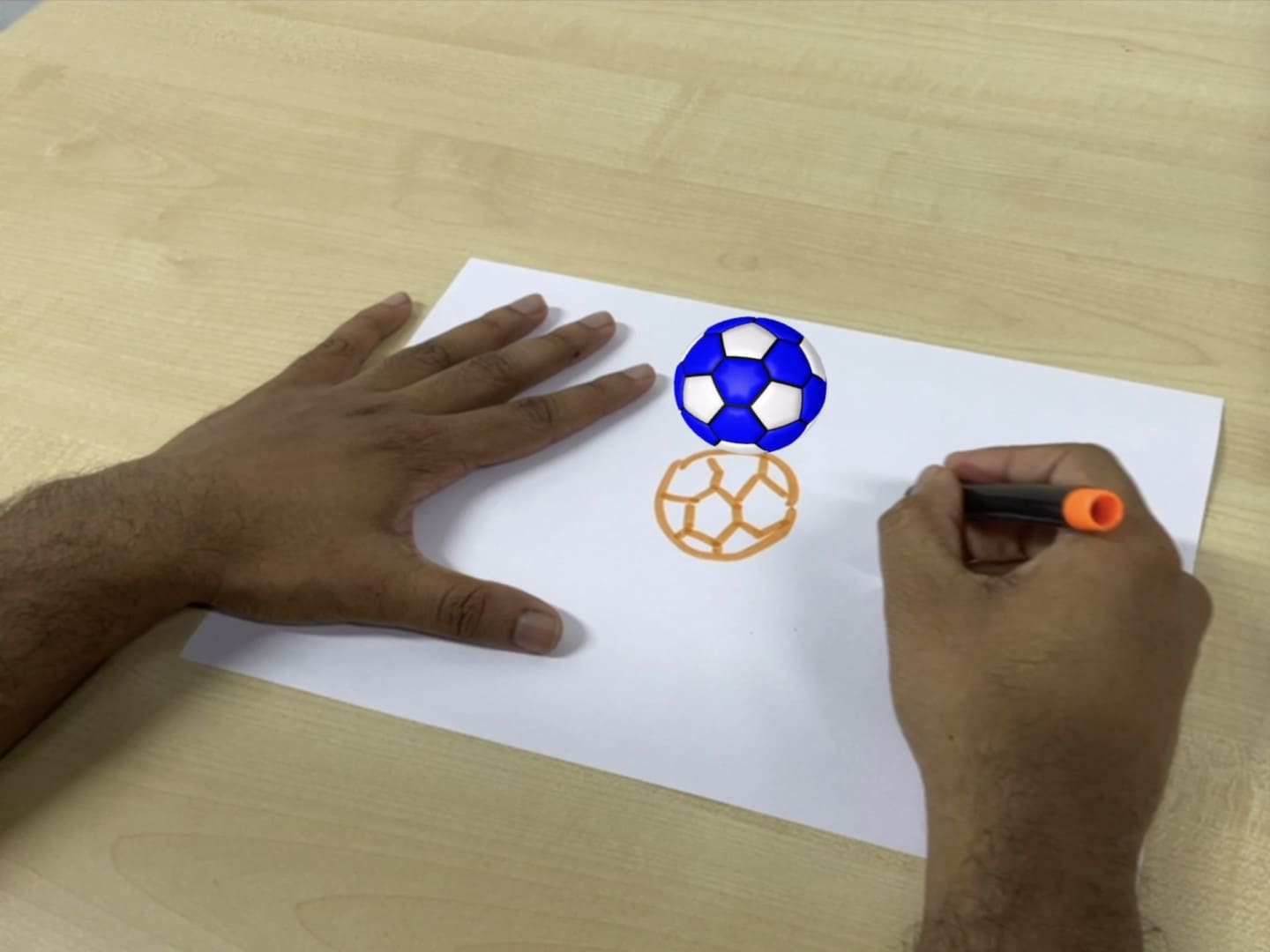}
\includegraphics[width=\width]{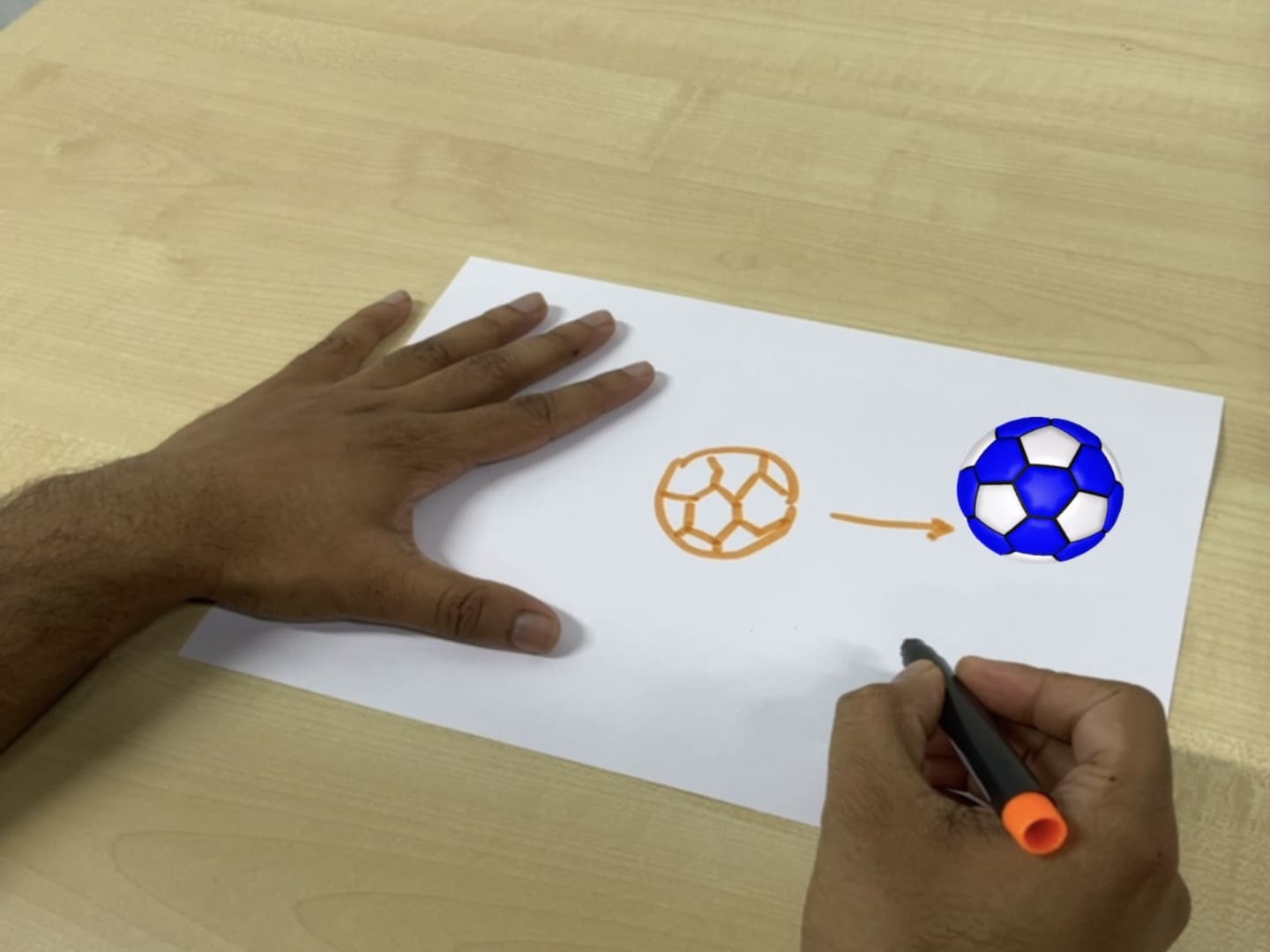}
\caption{Output - Appear-Disappear: A ball appears when a ball is sketched and Moving \& Transformation: The ball moves when the user sketches an arrow. }
\label{fig:output-move}
\Description{This figure demonstrates Appear-Disappear animation. A ball appears when a ball is sketched and to demonstrate Moving \& Transformation: The ball moves when the user sketches an arrow. }
\end{figure}

\subsubsection{Appear and Disappear Animation} 
The most basic output is to appear and disappear a virtual object given the state. 
By default, the system adds an animation when appearing and disappearing the virtual object by gradually changing its scale. This enables creation of prototypes similar to \textit{TangibleAR}~\cite{billinghurst2008tangible}, \textit{Holodoc}~\cite{li2019holodoc}, and \textit{Light Anchors}~\cite{ahuja2019lightanchors}.

\subsubsection{Moving and Transforming Animation}
Another basic output effect is the movement and transformation of the virtual object. 
By manipulating the virtual object's position, orientation, and scale for each state, the user can easily create a moving and transformation effect. This can enable the creation of controllers and a tangible user interface, similar to \textit{Living Paper}~\cite{claudino2020living}, \textit{Instant UI}~\cite{corsten2013instant}, \textit{Ephemeral Interactions}~\cite{walsh2014ephemeral}, and \textit{Bentroller}~\cite{shorey2017bendtroller}.
By default, the system animates the transition of the virtual object while moving or transforming.
For example, the system animates the ball's movement, when transitioning from one location to another (Figure~\ref{fig:output-move}).

\subsubsection{Counting and Aggregation}
\begin{figure}[h!]
\centering
\includegraphics[width=\width]{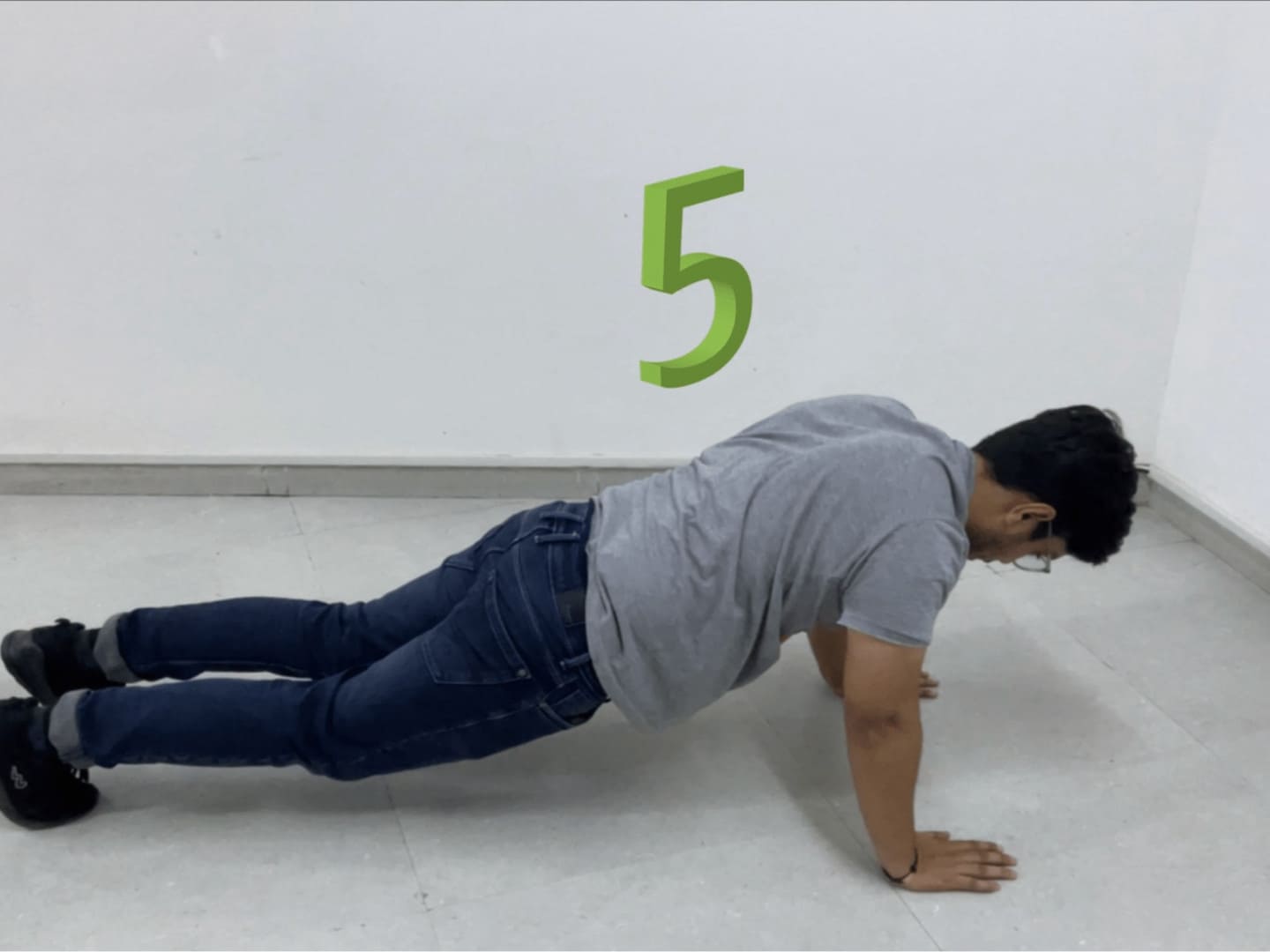}
\includegraphics[width=\width]{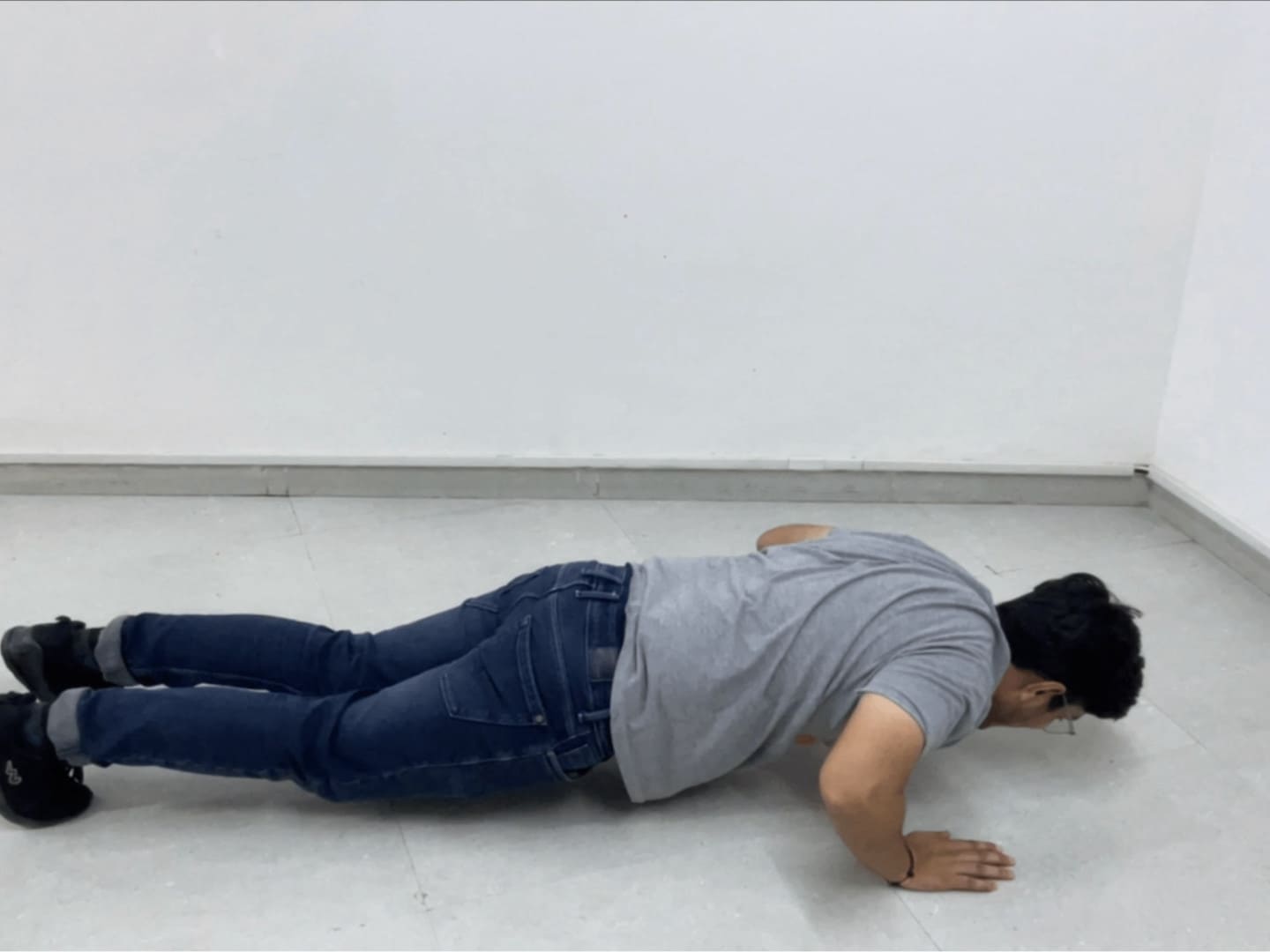}
\includegraphics[width=\width]{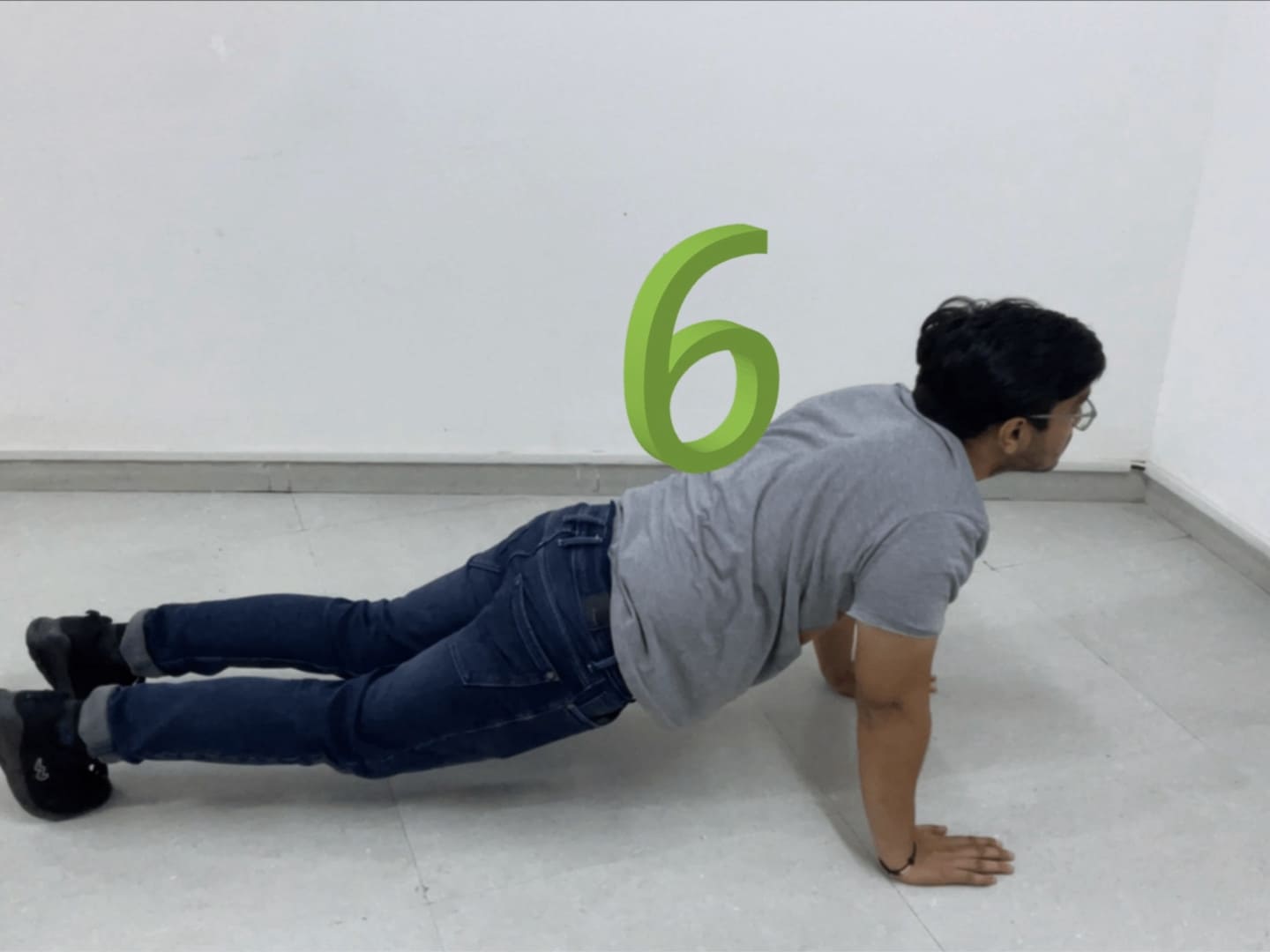}
\caption{Output - Counting: Counts the number of times a state is achieved}
\label{fig:output-count}
\Description{This figure demonstrates Counting as output. The AR system counts the number of times a state is achieved}
\end{figure}

The user can also use the detected count for each state. The system automatically counts how many times the specific state is triggered (transitioned from another state) so that the user can also use this value as an output parameter. 
For example, this can help the user to create a simple counter such as a count for push-ups or weight-lifting (Figure~\ref{fig:output-count}) or an AR prototype of \textit{ARMath}~\cite{kang2020armath}.
The user can also integrate this value into HTML to show some aggregated behavior like a graph.

\subsubsection{Parameterized Output}
Each state is basically discrete from the other, but the user can also define a continuous parameter, as we discussed.
For example, the user defines six states based on the position of the tangible object, then the user can use each state as a staggered parameter like [0.0, 0.2, 0.4, 0.6, 0.8, 1.0], given the start (0.0) and end value (1.0).
By binding this value to the virtual object's parameter, the user can also create a parameterized output similar to experiences created by \textit{RealitySketch}~\cite{suzuki2020realitysketch}.
For example, the user can associate the parameterized value to the orientation of the virtual object to create a circular slider.

\subsubsection{Pre-programmed Control}
Finally, the user can also integrate custom scripts for pre-programmed behaviors. 
For example, the user can change the orientation of the 3D car model based on the detected states of a tangible steering wheel (e.g., left, center, and right, based on the three states) similar to \textit{Ephemeral Interactions}~\cite{walsh2014ephemeral}, \textit{MarioKart Live}~\cite{noauthor_mariokartlive_nodate}, and \textit{Nintendo Labo}~\cite{noauthor_nintendolabo_nodate}.
The user can also add a simple script to move the car forward for each time interval as pre-programmed behavior.
Then, the user can create a simple virtual radio-controlled car with a tangible steering wheel.
The user can also associate custom script behavior to a certain detected state as well (e.g., hiding the steering wheel to stop the car). 

\begin{figure}[h!]
\centering
\includegraphics[width=\width]{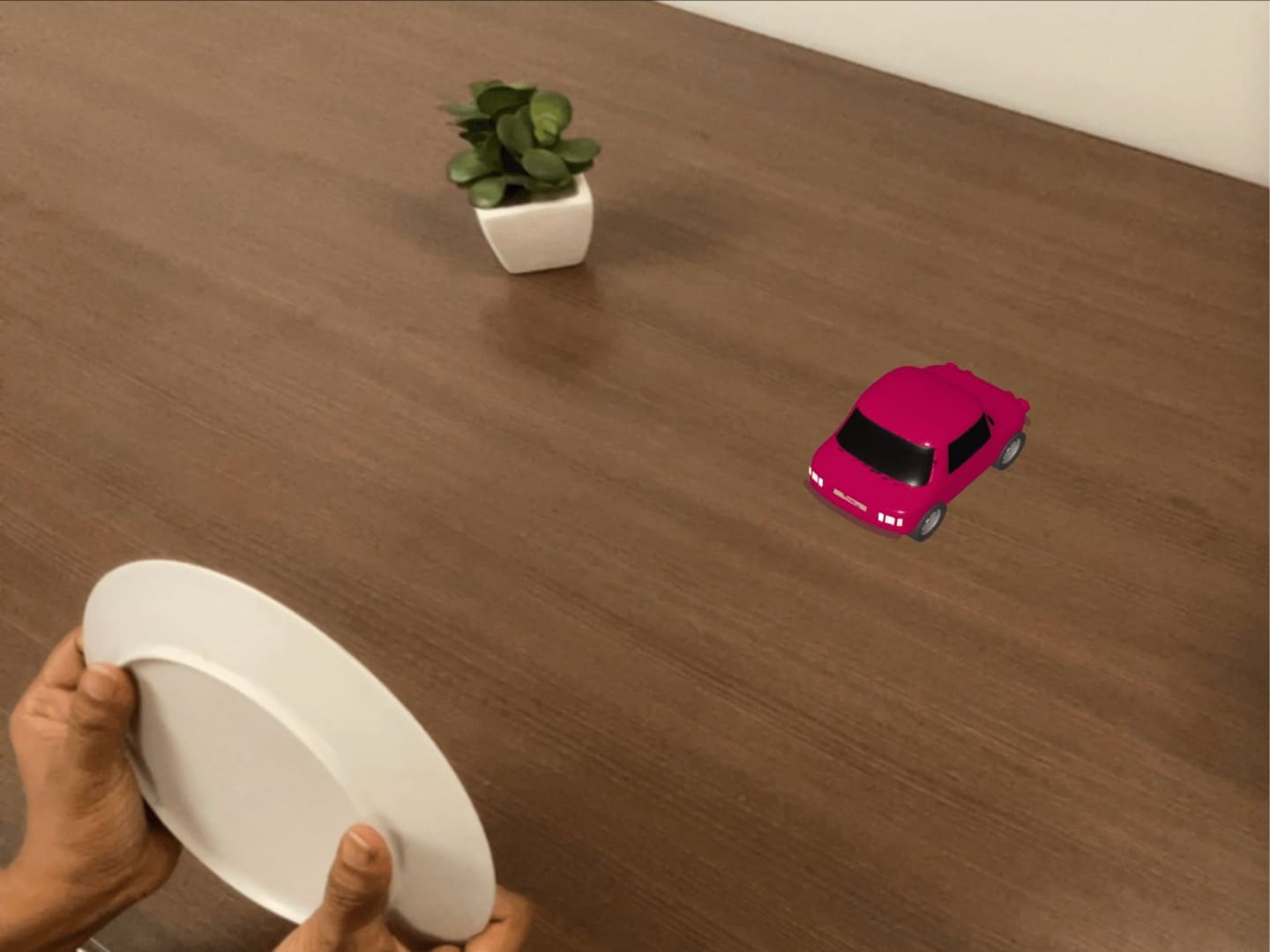}
\includegraphics[width=\width]{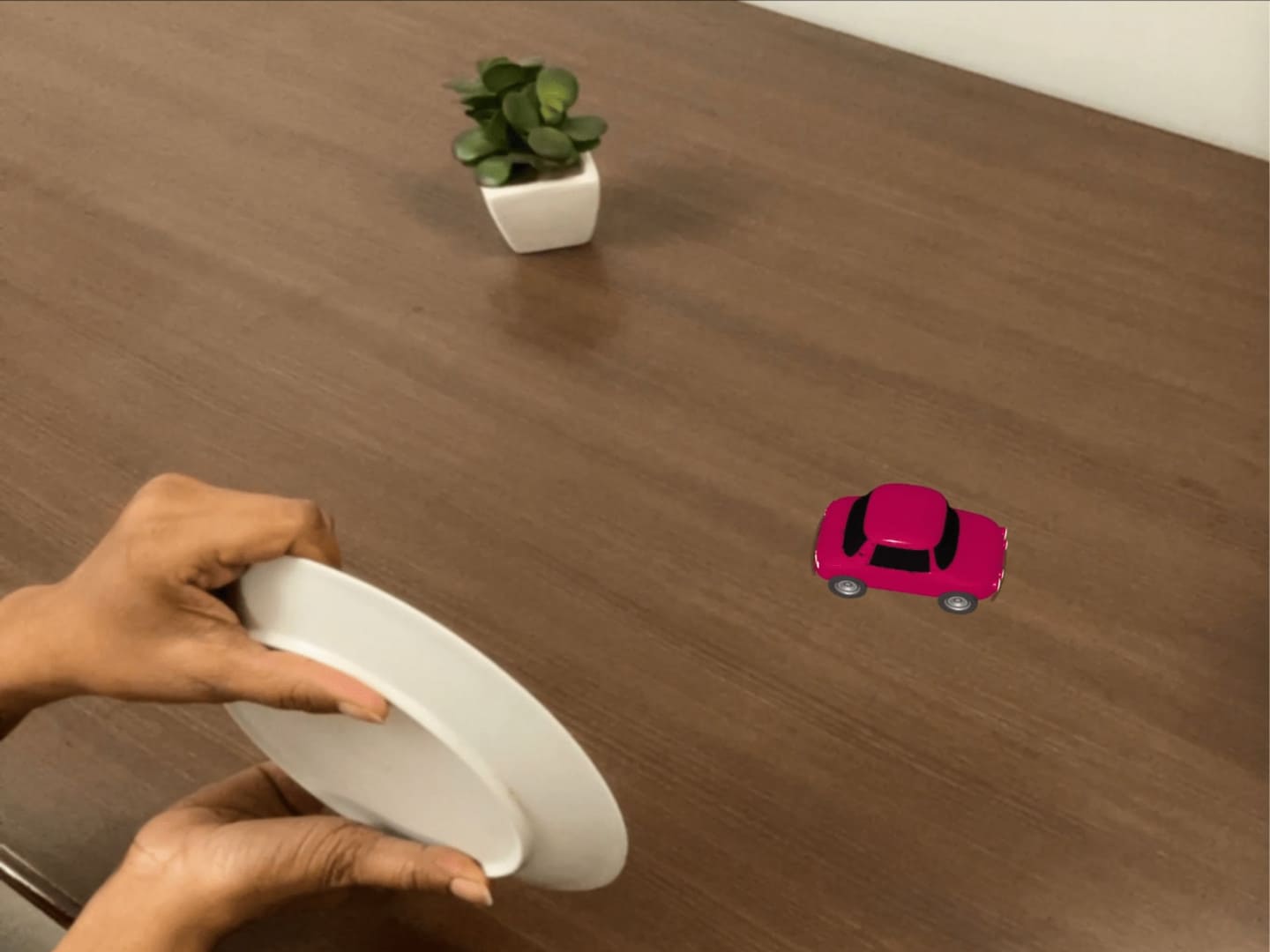}
\includegraphics[width=\width]{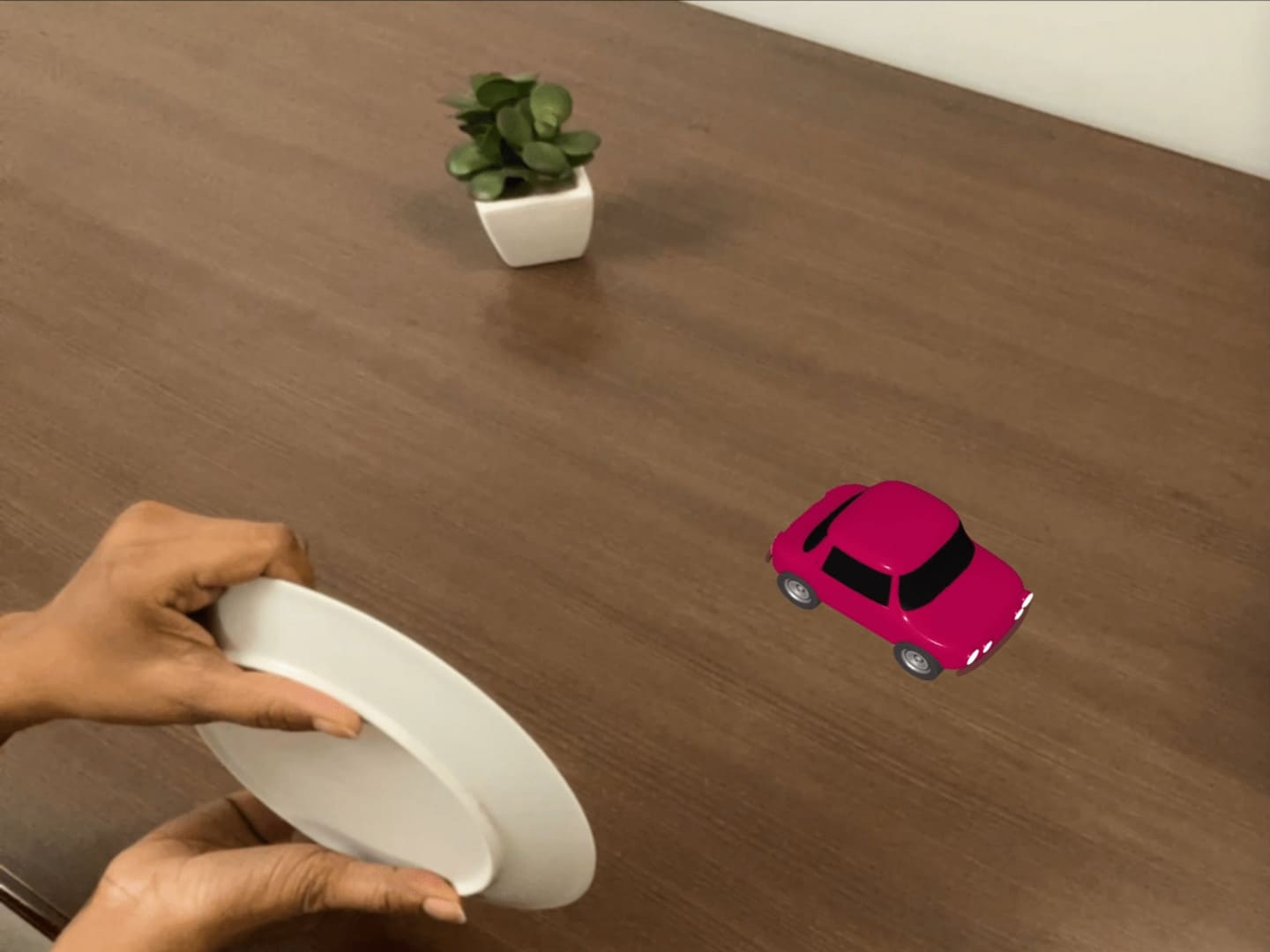}
\caption{Output - Pre-programmed Control: To drive a virtual RC car with a plate as a controller}
\label{fig:customScript}
\Description{This figure demonstrates Pre-programmed Control. A user drives a virtual RC car with a plate as a controller}
\end{figure}


\subsection{Applications}
Based on the combination of these numerous modalities, we identify some promising domains and application scenarios.

\noindent
\textbf{\textit{1) Tangible and Deformable Interfaces:}} The user can create an in-situ tangible controller with everyday objects similar to \textit{Instant UI}~\cite{corsten2013instant}, \textit{Ephemeral Interactions}~\cite{walsh2014ephemeral}, and \textit{RealitySketch}~\cite{suzuki2020realitysketch}, \textit{Music Bottles}~\cite{ishii2004bottles}. Also users can explore creating deformable user interfaces similar to \textit{FlexPad}~\cite{steimle2013flexpad}, \textit{Bendtroller}~\cite{shorey2017bendtroller}, and \textit{Non-Rigid HCI}~\cite{boem2019non}.

\noindent
\textbf{\textit{2) Context-Aware Assistant and Instruction:}} The user can also quickly prototype context-aware assistants, AR tutorials or instructions similar to \textit{Smart Makerspace}~\cite{knibbe2015smart}

\noindent
\textbf{\textit{3) Augmented and Situated Displays:}} The user can replicate situated displays like \textit{HoloDoc}~\cite{li2019holodoc},  \textit{BISHARE}~\cite{zhu2020bishare} where the AR scene shows the information related to the object the user is interacting with. 

\noindent
\textbf{\textit{4) Body-Driven AR Experiences:}} The system can support the quick prototype of body-driven applications, such as exergaming, entertainment, and exercise support. For example, the user can prototype an exercise assistant which identifies correct and incorrect postures of exercises.

\section{Evaluation}
We evaluated \system{} in two parts: (1) a usability study and (2) expert interviews.
Our study design was based on the ``usage evaluation'' strategy from Ledo et al.'s HCI toolkit evaluation \cite{ledo2018evaluation}.
Usability studies with end users aim to help verify whether the system is conceptually clear, easy to use, and useful. However, since \system{} is the first to explore rapid prototyping for tangible augmented reality, there is no clear baseline for us to make comparisons with. To address this, we conducted expert interviews to gain insights into current common practices and existing tools. We expect both usability studies and expert interviews will help us identify the benefits and limitations of the current system and gain insights for future iterations.

\begin{figure*}[h!]
\centering
\includegraphics[width=0.8\textwidth]{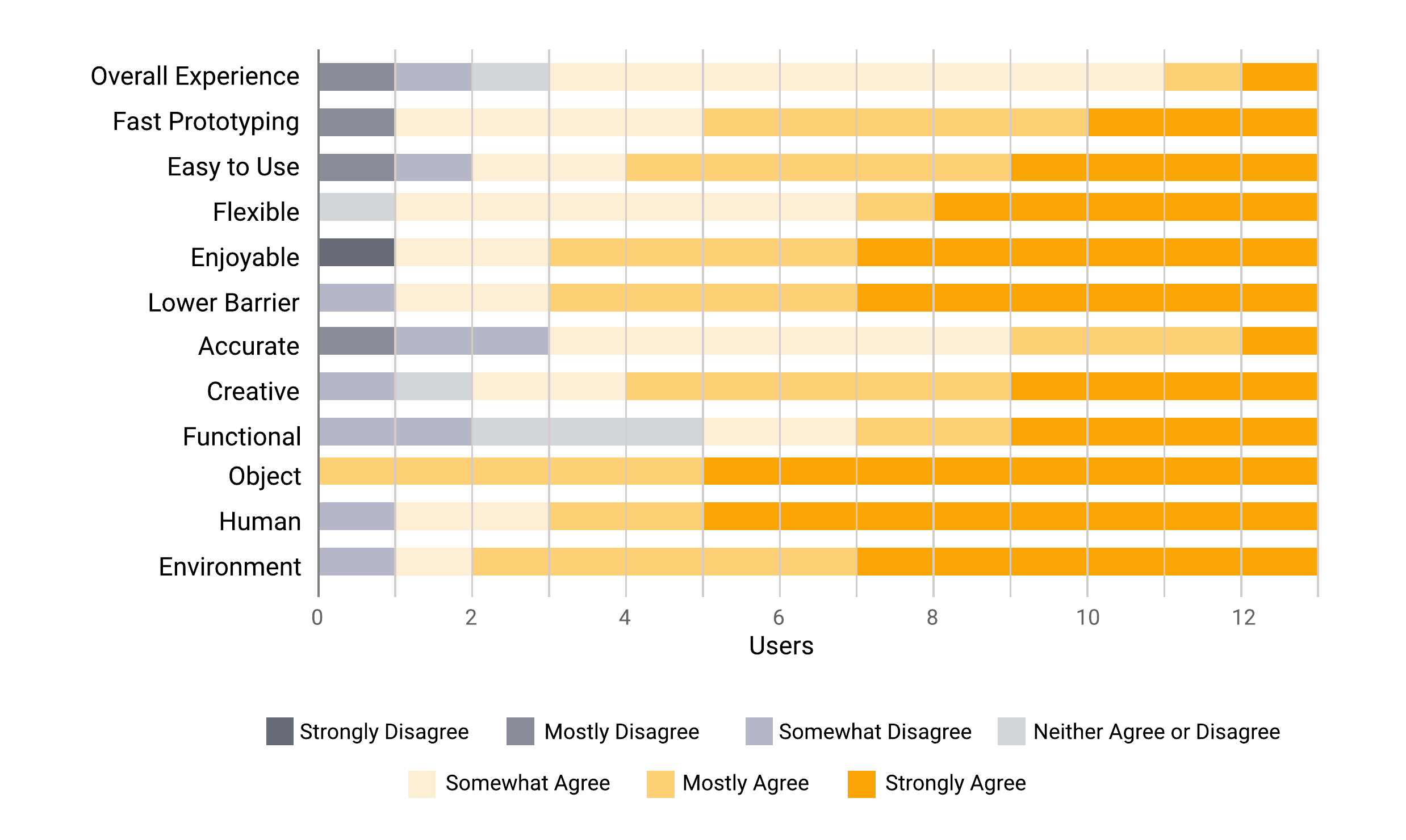}
\caption{Usability Questionnaire Responses of 13 participants}
\label{fig:user-study}
\Description{Results of the Usability Questionnaire Responses of 13 participants}
\end{figure*}

\subsection{Usability Study}
\subsubsection{Method}
We evaluate the usability of our system by asking the user to perform four prototyping tasks. We recruited 13 participants (7 male, 6 female, ages 20-37) from our local community. They all had varying experiences with AR applications on mobile phones, tablets, and head-mounted devices. 

\subsubsection{Introduction and Set-up}
 A step-by-step walk-through of the system was given to each participant, using a simple example of spawning virtual objects based on the appearance of a physical object. The participants were given an iPad with a keyboard and a stand to use for the study.

\subsubsection{Pre-Defined Tasks}
The participants were then asked to perform three pre-defined tasks. For each of the pre-defined tasks, the participants were shown a video demonstrating what had to be created for the task. We chose three simple tasks that enabled the users to explore and use most features of the system. The tasks were as follows: 

\begin{enumerate}
    \item \textit{Object:} Control a virtual object scale with a physical object.
    \item \textit{Environment:} Decorate the environment with a virtual object. 
    \item \textit{Body:} Control the position of a virtual object with body pose. 
    
\end{enumerate}

\subsubsection{Open-Ended Task}
For the open-ended final task, we asked participants to create prototypes they would like to create using the system. We gave participants inspiration by showing them example prototypes that were created using \system{}. We also provided them with 60 3D objects and 15 physical objects for further inspiration for their creation. We gave minimal assistance to the participants for this task. All tasks were screen recorded to obtain objective measurements (recognition errors, task completion time, etc). After the four tasks, the participants were asked to fill out a questionnaire evaluating their experience. The study lasted approximately 30-45 minutes per participant, and the participants were compensated with \$10 CAD Amazon Gift card.

\subsubsection{Results}
\ \\
\subsub{1) Overall Experience} Overall, participants responded positively about their prototyping experience. Participants understood how \system{} worked at a high level, found it easy to use, and felt that it was useful. We asked the participants to rate various aspects of \system{} on a 7-point Likert scale followed by some subjective questions. Figure \ref{fig:user-study} summarizes the 7-point Likert questionnaire response of the usability study.

\subsub{2) Strengths}
Participants found that the workflow was clear (P6, P8), fun (P7), easy to understand (P1, P7, P8), and intuitive (P3, P10). Participants also found the system very enjoyable and rated the system 5.92 on a scale of 1 to 7 for the enjoyability (SD = 1.65). \textit{P1: ``I wanted to play with it all day!''}. Participants also stated that they were able to materialize their idea very quickly (P3, P8, M = 5.61, SD = 1.32) for fast prototyping. Participants took 163 seconds (SD = 83s) on average to complete a task (Object-based - 141s, Environment 157s, Human 114s, Open-ended - 238s). Participants agreed the system lowers the barrier  (M = 6.07, SD = 1.18) for users from a non-development background. All participants had varying experiences with AR but most participants appreciated \system{}'s no coding interface. \textit{P10: ``The system is straightforward [and] doesn’t use any highly technical language, so it would be a great tool for someone unfamiliar with programming.''}. The flexibility of the tool was also recognized by the participants. Especially the capability of creating prototypes with physical objects was seen to be useful (M = 6.61, SD = 0.50). P5: ``I definitely think there are lots of opportunities in being able to pick up anything off your desk and being to instrument it.'' Participants appreciated the authoring of input and output. P7: \textit{``Overall the idea of taking photos of what I want it to respond to was intuitive.''} The participants found our approach of integrating interactive machine teaching and AR to enable creativity and that the features add more opportunities for expression (P3). P10: \textit{``It provides lots of opportunities for creativity with minimal effort which I really enjoy.''}

\subsub{3) Areas for Improvements} 
The participants rated the system's accuracy 4.84 on an average (SD = 1.40), which was a concern among the participants. Recognition errors (like glitches in plane tracking,  or delay in state recognition) were seen often (M = 2 times per task), however, these were reported to be minor and did not impair the experience. While most agreed that the interaction of the workflow and use of assets were intuitive, participants gave us suggestions to improve the user experience, which include but are not limited to - the option to delete a virtual asset, an undo-redo button, increasing the size of the buttons, add instructions in the system. Novice participants also pointed out that the tool required them to be creative and that providing some preset animations for the output could be helpful.

\subsection{Expert Interview}
\subsubsection{Method}
We also conducted an expert review with six experts to gain in-depth feedback. 
They all have AR prototyping experiences (Min: 2, Max: 17 years). The experts are faculty in computer science and related areas (E2, E3), PhD or Post-Doc researchers in AR fields (E5, E6), and full-time professional AR prototypers and researchers in large tech companies (E1, E4).
We first demonstrated the system with a simple example and showed applications to give a better understanding of the capabilities of the tool. Then, we conducted an in-depth and open-ended discussion about our approach and use scenarios. The interview lasted approximately one hour for each expert and we provided \$20 CAD for their participation.

\subsubsection{Insights and Feedback}
\ \\
\subsub{1) Overall Impression}
All participants were excited by the potential of our tool.
Participants found the tool to be quite intuitive (E1, E4), useful (E4, E5), playful (E1, E6), compelling (E2), general-purpose (E1), and versatile (E3). 
They could see that tool could be easily integrated into their current workflow (E1, E2, E4, E6). 
All participants agreed that the live testing feature of \system{} could significantly reduce the number of iterations needed during development (E1-E6), by making the prototyping process a lot easier and faster (E1, E2, E4, E6).
E2: \textit{``Having this tool available would make my prototyping much faster.''}
The output authoring interactions like drag-and-drop or pinching were also considered intuitive (E4) and even fun (E1, E6). Besides the general impressions, the participants also commented on specific strengths (listed below) and a few limitations (noted in the next section) of our technical approach.

\subsub{2) Authoring Workflow and Trigger Action Approach}
In terms of the workflow, all participants found the authoring workflow easy to understand.
For example, one expert pointed out that the mental model of our workflow is close to the existing creative thinking process (E4), which allows for easy adoption even for non-technical people (E3).
E4: \textit{``It's taking a principle of animation and storyboarding. It's basically just turning a storyboard into real life. So it's really easy to understand.''} 

\subsub{3) Comparison to Marker-Based Tracking}
When compared to marker-based approaches, the participants appreciate the no-setup nature of our approach. Without the need for programming or printing markers, our system significantly reduces the hurdle of prototyping.
The participants also appreciated the general-purpose capability, compared to the existing tools (E1).
In fact, the participants also found the examples created by the system quite varied (E6), versatile (E3), and impressive (E4), which can broaden the prototyping opportunities that are currently not possible (E2).
A participant (E3), who has an extensive experience with the marker-based approach, mentioned that \system{} could be a viable alternative to the current practice.

\subsub{4) Comparison to Teachable Machine}
Since one expert (E3) had used Teachable Machine before, we asked about the difference between the Teachable Machine and our system. 
E3 pointed out that the no-coding authoring and live testing make for a significant advantage.
E3: \textit{``One thing that stands out is the fact  that it is deployable immediately. If you use Teachable Machine, it's collecting the data, and then training, training, training, and then checking, which is quite ineffective. One benefit of using your tool is that I can collect the data, manipulate, save virtual content, and deploy the output with a single interface.''}
In fact, all participants also appreciate \system{}'s no-coding approach.
E1: \textit{``Even though I know how to program, I like to avoid coding. I always lean towards not coding.}'' 
The participants agreed that our tool can lower the barrier for AR allowing designers with visual design backgrounds to create interactive prototypes (E3, E4).

\subsub{5) Communication Tools as a Potential Use Case}
When asked about potential use cases, the experts suggested the strong potential for a \textit{communication tool} (E1, E2, E4).
E4: \textit{``I would use it, especially if I'm working in a really collaborative environment if I had an idea, and I wanted to demonstrate it really quickly and have somebody try it out, before I spend a lot of time actually implementing the thing in code.''} 
Currently, they share an idea through a video storyboard, but these storyboards are still just an \textit{approximation} and there is still a communication gap between designers and programmers (E1, E4).
\system{} could solve this problem by allowing designers and non-technical people to demonstrate and share their idea quickly between teams. 
E5 and E6 also mentioned that they would definitely use \system{} for their own current projects if possible.

\section{Limitations and Future Work}
\subsection*{Addressing Limitations of Computer Vision Based Approaches}
While our choice of computer-vision techniques sufficed for novice study participant's prototyping needs and determined suitable by experts for demonstrating a concept (E4), we acknowledge concerns raised generally about its reliability when tested in the wild in uncontrolled environments (E3, E5). In that sense, the participants thought that our tool was not a replacement of existing methods, but rather a complement for early stage yet functional prototypes.

\subsubsection*{Increase Accuracy of Input Detection}

Experts raised concerns regarding the reliability of the accuracy of the system (E3, E5).  As they had been using computer vision techniques very frequently they were concerned about how accurate the system was in the wild. On the other hand, they also acknowledged that the accuracy may not be a priority concern as the purpose of this tool is just to demonstrate a concept for oneself or other peers (E4). 

We identify the detection accuracy mostly relies on how the user trains the model, and it is difficult to get stable performance. In particular, the inaccurate detection performance for environments or cluttered scenes makes it difficult to test for context-aware applications.
To address these problems, future work should leverage user-supported training. For example, \textit{LookHere}~\cite{zhou2022gesture} uses gestural interactions to specify which region or object the model should focus on. In this way, the model accuracy can greatly improve.
Alternatively, instead of an RGB camera, we could also enhance the model training based on the depth camera information like LiDAR input.
This allows more accurate tracking and detection.

\subsubsection*{More Robust Object Tracking}
In our current implementation, we use simple object tracking based on selected color matching. 
We adapt this simple tracking method based on \textit{RealitySketch}~\cite{suzuki2020realitysketch}, which reports that color tracking provides the fastest tracking method for the real-time sketched animation, compared to other sophisticated algorithms \textit{YOLO}~\cite{redmon2016you}, \textit{Faster R-CNN}~\cite{girshick2015fast}, \textit{Mask R-CNN}~\cite{he2017mask}).
However, this tracking method is not robust and generalizable enough for many situations. 
For example, if the scene has multiple similar colored objects, the tracking may not work well.
In the future, we expect the recent advances in computer vision will provide more general, robust, yet fast object tracking methods for our purpose, similar to robust and fast body tracking algorithms (MediaPipe) which we used in our prototype.

\subsection*{Enhancing Usefulness and Usability}
\subsubsection*{Enabling Complex Prototypes}
In addition, while the system could create a large number of simple prototypes, experts discussed how future work could look at complex narratives. These would include non-linear narratives, which could have multiple branching of states and more complex logic. E1 suggested a state machine like approach which could look at activating and deactivating states according to the progression of the user's interactions. E5 suggested a block programming approach which would give access to more control over the prototype. We leave these to future work as, though these approaches will be useful, they will be expensive as it would include training multiple models on demand synchronously and activating and deactivating them based on a user's interactions.

\subsubsection*{Supporting Multi-Modal Input \& Output}
In this paper, we mostly focus on camera-based input using a computer vision model, but our approach can be generalized to many different inputs. 
The system can combine multiple inputs for more accurate and expressive interaction. For example, a system that can enable creation of a navigation application that uses the combination of camera and location-based inputs. While our scope for output of prototypes was limited to AR, a potential direction for future work can be to explore other forms of output in addition to AR like haptics. For example, A system that enables creation of a haptic experience which compliments the AR content. This could help create more holistic tangible AR prototypes. With this in mind, future work can showcase a more exhaustive design space and compare the interaction space to present each input and output type’s comparison and, as a result, the pros and cons of each interaction type and modality. 

\subsubsection*{Immersive AR Authoring with HMDs}
Our current implementation uses screen-based mobile AR, but the integration with head-mounted displays (HMDs) will allow more blended and immersive prototyping experiences like \textit{GesturAR}~\cite{wang2021gesturar}.
In this case, the possible limitation is the lack of computational power, which make the training slow.
To avoid this problem, we can leverage cloud-based training for HMDs.

\subsubsection*{Towards Explainable and Understandable Input Detection}

The experts also mentioned that there is a hidden learning curve for how to better train the computer vision model, especially for those without such knowledge. 
\textit{``E1: You might have to remind that designers don't really have a sense of how these image recognition works. They might be surprised when it doesn't detect a state at a particular angle''}.
Experts also pointed out that using ML techniques introduces the ``black-box'' nature into the input tracking of the system (E2, E3, E4), by saying that if the system trains with some unrelated objects in the background, then it gives incorrect results. 
They mention that an Explainable AI interface over the tool would be highly beneficial, which we leave to future work.

\section{Conclusion}
This paper presents \system{}, an augmented reality (AR) prototyping tool to create interactive tangible AR applications that can use arbitrary everyday objects as user inputs.
When creating functional tangible AR prototypes, current tools often need to rely on markers-based inputs or pre-defined interactions (gesture, location, body posture, electronic devices, etc), which limits the flexibility, customizability, and generalizability of possible interactions.
In contrast, \system{} incorporates \textit{\textbf{interactive machine teaching}} to immersive AR authoring, which captures the user's demonstrated action as in-situ tangible input, by leveraging on-demand and user-defined compute vision classification.
The user can use these classified inputs to create interactive AR applications based on an action-trigger programming model.
We showcase various applications examples, which include tangible and deformable interfaces, context-aware assistants, augmented and situated displays and body-driven experiences.
The results of our user study confirm the flexibility of our approach to quickly and easily create various tangible AR prototypes.
\begin{acks}
This research was funded in part by the Natural Sciences and Engineering Research Council of Canada (NSERC RGPIN-2021-02857) and Mitacs Globalink Research Internship. We also thank all of the experts and participants for our user studies.
\end{acks}

\ifdouble
  \balance
\fi
\bibliographystyle{ACM-Reference-Format}
\bibliography{references}


\begin{thebibliography}{100}


\ifx \showCODEN    \undefined \def \showCODEN     #1{\unskip}     \fi
\ifx \showDOI      \undefined \def \showDOI       #1{#1}\fi
\ifx \showISBNx    \undefined \def \showISBNx     #1{\unskip}     \fi
\ifx \showISBNxiii \undefined \def \showISBNxiii  #1{\unskip}     \fi
\ifx \showISSN     \undefined \def \showISSN      #1{\unskip}     \fi
\ifx \showLCCN     \undefined \def \showLCCN      #1{\unskip}     \fi
\ifx \shownote     \undefined \def \shownote      #1{#1}          \fi
\ifx \showarticletitle \undefined \def \showarticletitle #1{#1}   \fi
\ifx \showURL      \undefined \def \showURL       {\relax}        \fi
\providecommand\bibfield[2]{#2}
\providecommand\bibinfo[2]{#2}
\providecommand\natexlab[1]{#1}
\providecommand\showeprint[2][]{arXiv:#2}

\bibitem[\protect\citeauthoryear{??}{noa}{2022}]%
        {noauthor_holobuilder_nodate}
 \bibinfo{year}{2022}\natexlab{}.
\newblock \bibinfo{title}{HoloBuilder}.
\newblock
\newblock
\urldef\tempurl%
\url{https://www.holobuilder.com/}
\showURL{%
\tempurl}


\bibitem[\protect\citeauthoryear{8th Wall}{8th Wall}{2022}]%
        {noauthor_8thwall_nodate}
\bibfield{author}{\bibinfo{person}{8th Wall}.} \bibinfo{year}{2022}\natexlab{}.
\newblock \bibinfo{title}{Niantic Inc.}
\newblock
\newblock
\urldef\tempurl%
\url{https://www.8thwall.com/}
\showURL{%
\tempurl}


\bibitem[\protect\citeauthoryear{A-Frame}{A-Frame}{2022}]%
        {noauthor_aframe_nodate}
\bibfield{author}{\bibinfo{person}{A-Frame}.} \bibinfo{year}{2022}\natexlab{}.
\newblock \bibinfo{title}{A-Frame}.
\newblock
\newblock
\urldef\tempurl%
\url{https://aframe.io/}
\showURL{%
\tempurl}


\bibitem[\protect\citeauthoryear{Ahuja, Pareddy, Xiao, Goel, and
  Harrison}{Ahuja et~al\mbox{.}}{2019}]%
        {ahuja2019lightanchors}
\bibfield{author}{\bibinfo{person}{Karan Ahuja}, \bibinfo{person}{Sujeath
  Pareddy}, \bibinfo{person}{Robert Xiao}, \bibinfo{person}{Mayank Goel}, {and}
  \bibinfo{person}{Chris Harrison}.} \bibinfo{year}{2019}\natexlab{}.
\newblock \showarticletitle{Lightanchors: Appropriating point lights for
  spatially-anchored augmented reality interfaces}. In
  \bibinfo{booktitle}{\emph{Proceedings of the 32nd Annual ACM Symposium on
  User Interface Software and Technology}}. \bibinfo{pages}{189--196}.
\newblock


\bibitem[\protect\citeauthoryear{Alce, Wallerg{\aa}rd, and Hermodsson}{Alce
  et~al\mbox{.}}{2015}]%
        {alce2015wozard}
\bibfield{author}{\bibinfo{person}{G{\"u}nter Alce}, \bibinfo{person}{Mattias
  Wallerg{\aa}rd}, {and} \bibinfo{person}{Klas Hermodsson}.}
  \bibinfo{year}{2015}\natexlab{}.
\newblock \showarticletitle{WozARd: a wizard of Oz method for wearable
  augmented reality interaction—a pilot study}.
\newblock \bibinfo{journal}{\emph{Advances in human-computer interaction}}
  \bibinfo{volume}{2015} (\bibinfo{year}{2015}).
\newblock


\bibitem[\protect\citeauthoryear{Arora, Saini, Mehra, Jain, Shrey, and
  Parnami}{Arora et~al\mbox{.}}{2019}]%
        {arora2019virtualbricks}
\bibfield{author}{\bibinfo{person}{Jatin Arora}, \bibinfo{person}{Aryan Saini},
  \bibinfo{person}{Nirmita Mehra}, \bibinfo{person}{Varnit Jain},
  \bibinfo{person}{Shwetank Shrey}, {and} \bibinfo{person}{Aman Parnami}.}
  \bibinfo{year}{2019}\natexlab{}.
\newblock \showarticletitle{Virtualbricks: Exploring a scalable, modular
  toolkit for enabling physical manipulation in vr}. In
  \bibinfo{booktitle}{\emph{Proceedings of the 2019 CHI Conference on Human
  Factors in Computing Systems}}. \bibinfo{pages}{1--12}.
\newblock


\bibitem[\protect\citeauthoryear{Ashtari, Bunt, McGrenere, Nebeling, and
  Chilana}{Ashtari et~al\mbox{.}}{2020}]%
        {ashtari2020creating}
\bibfield{author}{\bibinfo{person}{Narges Ashtari}, \bibinfo{person}{Andrea
  Bunt}, \bibinfo{person}{Joanna McGrenere}, \bibinfo{person}{Michael
  Nebeling}, {and} \bibinfo{person}{Parmit~K Chilana}.}
  \bibinfo{year}{2020}\natexlab{}.
\newblock \showarticletitle{Creating augmented and virtual reality
  applications: Current practices, challenges, and opportunities}. In
  \bibinfo{booktitle}{\emph{Proceedings of the 2020 CHI conference on human
  factors in computing systems}}. \bibinfo{pages}{1--13}.
\newblock


\bibitem[\protect\citeauthoryear{Billinghurst, Kato, Poupyrev,
  et~al\mbox{.}}{Billinghurst et~al\mbox{.}}{2008}]%
        {billinghurst2008tangible}
\bibfield{author}{\bibinfo{person}{Mark Billinghurst},
  \bibinfo{person}{Hirokazu Kato}, \bibinfo{person}{Ivan Poupyrev},
  {et~al\mbox{.}}} \bibinfo{year}{2008}\natexlab{}.
\newblock \showarticletitle{Tangible augmented reality}.
\newblock \bibinfo{journal}{\emph{Acm siggraph asia}} \bibinfo{volume}{7},
  \bibinfo{number}{2} (\bibinfo{year}{2008}), \bibinfo{pages}{1--10}.
\newblock


\bibitem[\protect\citeauthoryear{Boem and Troiano}{Boem and Troiano}{2019}]%
        {boem2019non}
\bibfield{author}{\bibinfo{person}{Alberto Boem} {and}
  \bibinfo{person}{Giovanni~Maria Troiano}.} \bibinfo{year}{2019}\natexlab{}.
\newblock \showarticletitle{Non-rigid HCI: A review of deformable interfaces
  and input}. In \bibinfo{booktitle}{\emph{Proceedings of the 2019 on Designing
  Interactive Systems Conference}}. \bibinfo{pages}{885--906}.
\newblock


\bibitem[\protect\citeauthoryear{Cardoso and Ribeiro}{Cardoso and
  Ribeiro}{2021}]%
        {cardoso2021tangible}
\bibfield{author}{\bibinfo{person}{Jorge~CS Cardoso} {and}
  \bibinfo{person}{Jorge~M Ribeiro}.} \bibinfo{year}{2021}\natexlab{}.
\newblock \showarticletitle{Tangible VR book: exploring the design space of
  marker-based tangible interfaces for virtual reality}.
\newblock \bibinfo{journal}{\emph{Applied Sciences}} \bibinfo{volume}{11},
  \bibinfo{number}{4} (\bibinfo{year}{2021}), \bibinfo{pages}{1367}.
\newblock


\bibitem[\protect\citeauthoryear{Carney, Webster, Alvarado, Phillips, Howell,
  Griffith, Jongejan, Pitaru, and Chen}{Carney et~al\mbox{.}}{2020}]%
        {carney2020teachable}
\bibfield{author}{\bibinfo{person}{Michelle Carney}, \bibinfo{person}{Barron
  Webster}, \bibinfo{person}{Irene Alvarado}, \bibinfo{person}{Kyle Phillips},
  \bibinfo{person}{Noura Howell}, \bibinfo{person}{Jordan Griffith},
  \bibinfo{person}{Jonas Jongejan}, \bibinfo{person}{Amit Pitaru}, {and}
  \bibinfo{person}{Alexander Chen}.} \bibinfo{year}{2020}\natexlab{}.
\newblock \showarticletitle{Teachable machine: Approachable Web-based tool for
  exploring machine learning classification}. In
  \bibinfo{booktitle}{\emph{Extended abstracts of the 2020 CHI conference on
  human factors in computing systems}}. \bibinfo{pages}{1--8}.
\newblock


\bibitem[\protect\citeauthoryear{Cecil~Piya}{Cecil~Piya}{2016}]%
        {cecil2016realfusion}
\bibfield{author}{\bibinfo{person}{Vinayak Cecil~Piya}.}
  \bibinfo{year}{2016}\natexlab{}.
\newblock \showarticletitle{RealFusion: An interactive workflow for repurposing
  real-world objects towards early-stage creative ideation}. In
  \bibinfo{booktitle}{\emph{Graphics interface}}.
\newblock


\bibitem[\protect\citeauthoryear{Cheng, Liang, Chen, Laing, and Kuo}{Cheng
  et~al\mbox{.}}{2010}]%
        {cheng2010icon}
\bibfield{author}{\bibinfo{person}{Kai-Yin Cheng}, \bibinfo{person}{Rong-Hao
  Liang}, \bibinfo{person}{Bing-Yu Chen}, \bibinfo{person}{Rung-Huei Laing},
  {and} \bibinfo{person}{Sy-Yen Kuo}.} \bibinfo{year}{2010}\natexlab{}.
\newblock \showarticletitle{iCon: utilizing everyday objects as additional,
  auxiliary and instant tabletop controllers}. In
  \bibinfo{booktitle}{\emph{Proceedings of the SIGCHI conference on Human
  factors in computing systems}}. \bibinfo{pages}{1155--1164}.
\newblock


\bibitem[\protect\citeauthoryear{Cho, Kim, Go, Kim, Kim, and Kim}{Cho
  et~al\mbox{.}}{2021}]%
        {cho2021deepblock}
\bibfield{author}{\bibinfo{person}{Jinsung Cho}, \bibinfo{person}{Geunmo Kim},
  \bibinfo{person}{Hyunmin Go}, \bibinfo{person}{Sungmin Kim},
  \bibinfo{person}{Jisub Kim}, {and} \bibinfo{person}{Bongjae Kim}.}
  \bibinfo{year}{2021}\natexlab{}.
\newblock \showarticletitle{DeepBlock: Web-based Deep Learning Education
  Platform}.
\newblock \bibinfo{journal}{\emph{The Journal of the Institute of Internet,
  Broadcasting and Communication}} \bibinfo{volume}{21}, \bibinfo{number}{3}
  (\bibinfo{year}{2021}), \bibinfo{pages}{43--50}.
\newblock


\bibitem[\protect\citeauthoryear{Claudino~Daffara, Brewer, Thoravi~Kumaravel,
  and Hartmann}{Claudino~Daffara et~al\mbox{.}}{2020}]%
        {claudino2020living}
\bibfield{author}{\bibinfo{person}{Stephanie Claudino~Daffara},
  \bibinfo{person}{Anna Brewer}, \bibinfo{person}{Balasaravanan
  Thoravi~Kumaravel}, {and} \bibinfo{person}{Bjoern Hartmann}.}
  \bibinfo{year}{2020}\natexlab{}.
\newblock \showarticletitle{Living Paper: Authoring AR Narratives Across
  Digital and Tangible Media}. In \bibinfo{booktitle}{\emph{Extended Abstracts
  of the 2020 CHI Conference on Human Factors in Computing Systems}}.
  \bibinfo{pages}{1--10}.
\newblock


\bibitem[\protect\citeauthoryear{Corsten, Avellino, M{\"o}llers, and
  Borchers}{Corsten et~al\mbox{.}}{2013a}]%
        {corsten2013instant}
\bibfield{author}{\bibinfo{person}{Christian Corsten}, \bibinfo{person}{Ignacio
  Avellino}, \bibinfo{person}{Max M{\"o}llers}, {and} \bibinfo{person}{Jan
  Borchers}.} \bibinfo{year}{2013}\natexlab{a}.
\newblock \showarticletitle{Instant user interfaces: repurposing everyday
  objects as input devices}. In \bibinfo{booktitle}{\emph{Proceedings of the
  2013 ACM international conference on Interactive tabletops and surfaces}}.
  \bibinfo{pages}{71--80}.
\newblock


\bibitem[\protect\citeauthoryear{Corsten, Wacharamanotham, and
  Borchers}{Corsten et~al\mbox{.}}{2013b}]%
        {corsten2013fillables}
\bibfield{author}{\bibinfo{person}{Christian Corsten}, \bibinfo{person}{Chat
  Wacharamanotham}, {and} \bibinfo{person}{Jan Borchers}.}
  \bibinfo{year}{2013}\natexlab{b}.
\newblock \showarticletitle{Fillables: everyday vessels as tangible controllers
  with adjustable haptics}.
\newblock In \bibinfo{booktitle}{\emph{CHI'13 Extended Abstracts on Human
  Factors in Computing Systems}}. \bibinfo{pages}{2129--2138}.
\newblock


\bibitem[\protect\citeauthoryear{Daiber, Degraen, Zenner, D{\"o}ring,
  Steinicke, Ariza~Nunez, and Simeone}{Daiber et~al\mbox{.}}{2021}]%
        {daiber2021everyday}
\bibfield{author}{\bibinfo{person}{Florian Daiber}, \bibinfo{person}{Donald
  Degraen}, \bibinfo{person}{Andr{\'e} Zenner}, \bibinfo{person}{Tanja
  D{\"o}ring}, \bibinfo{person}{Frank Steinicke}, \bibinfo{person}{Oscar~Javier
  Ariza~Nunez}, {and} \bibinfo{person}{Adalberto~L Simeone}.}
  \bibinfo{year}{2021}\natexlab{}.
\newblock \showarticletitle{Everyday Proxy Objects for Virtual Reality}. In
  \bibinfo{booktitle}{\emph{Extended Abstracts of the 2021 CHI Conference on
  Human Factors in Computing Systems}}. \bibinfo{pages}{1--6}.
\newblock


\bibitem[\protect\citeauthoryear{Dogan, Taka, Lu, Zhu, Kumar, Gupta, and
  Mueller}{Dogan et~al\mbox{.}}{2022}]%
        {dogan2022infraredtags}
\bibfield{author}{\bibinfo{person}{Mustafa~Doga Dogan}, \bibinfo{person}{Ahmad
  Taka}, \bibinfo{person}{Michael Lu}, \bibinfo{person}{Yunyi Zhu},
  \bibinfo{person}{Akshat Kumar}, \bibinfo{person}{Aakar Gupta}, {and}
  \bibinfo{person}{Stefanie Mueller}.} \bibinfo{year}{2022}\natexlab{}.
\newblock \showarticletitle{InfraredTags: Embedding Invisible AR Markers and
  Barcodes Using Low-Cost, Infrared-Based 3D Printing and Imaging Tools}. In
  \bibinfo{booktitle}{\emph{CHI Conference on Human Factors in Computing
  Systems}}. \bibinfo{pages}{1--12}.
\newblock


\bibitem[\protect\citeauthoryear{Du, Olwal, Le~Goc, Wu, Tang, Zhang, Zhang,
  Tan, Tombari, and Kim}{Du et~al\mbox{.}}{2022}]%
        {du2022opportunistic}
\bibfield{author}{\bibinfo{person}{Ruofei Du}, \bibinfo{person}{Alex Olwal},
  \bibinfo{person}{Mathieu Le~Goc}, \bibinfo{person}{Shengzhi Wu},
  \bibinfo{person}{Danhang Tang}, \bibinfo{person}{Yinda Zhang},
  \bibinfo{person}{Jun Zhang}, \bibinfo{person}{David~Joseph Tan},
  \bibinfo{person}{Federico Tombari}, {and} \bibinfo{person}{David Kim}.}
  \bibinfo{year}{2022}\natexlab{}.
\newblock \showarticletitle{Opportunistic Interfaces for Augmented Reality:
  Transforming Everyday Objects into Tangible 6DoF Interfaces Using Ad hoc UI}.
  In \bibinfo{booktitle}{\emph{CHI Conference on Human Factors in Computing
  Systems Extended Abstracts}}. \bibinfo{pages}{1--4}.
\newblock


\bibitem[\protect\citeauthoryear{Englmeier, D{\"o}rner, Butz, and
  H{\"o}llerer}{Englmeier et~al\mbox{.}}{2020}]%
        {englmeier2020tangible}
\bibfield{author}{\bibinfo{person}{David Englmeier}, \bibinfo{person}{Julia
  D{\"o}rner}, \bibinfo{person}{Andreas Butz}, {and} \bibinfo{person}{Tobias
  H{\"o}llerer}.} \bibinfo{year}{2020}\natexlab{}.
\newblock \showarticletitle{A tangible spherical proxy for object manipulation
  in augmented reality}. In \bibinfo{booktitle}{\emph{2020 IEEE Conference on
  Virtual Reality and 3D User Interfaces (VR)}}. IEEE,
  \bibinfo{pages}{221--229}.
\newblock


\bibitem[\protect\citeauthoryear{Fitzmaurice, Ishii, and Buxton}{Fitzmaurice
  et~al\mbox{.}}{1995}]%
        {fitzmaurice1995bricks}
\bibfield{author}{\bibinfo{person}{George~W Fitzmaurice},
  \bibinfo{person}{Hiroshi Ishii}, {and} \bibinfo{person}{William~AS Buxton}.}
  \bibinfo{year}{1995}\natexlab{}.
\newblock \showarticletitle{Bricks: laying the foundations for graspable user
  interfaces}. In \bibinfo{booktitle}{\emph{Proceedings of the SIGCHI
  conference on Human factors in computing systems}}.
  \bibinfo{pages}{442--449}.
\newblock


\bibitem[\protect\citeauthoryear{Garrido-Jurado, Mu{\~n}oz-Salinas,
  Madrid-Cuevas, and Mar{\'\i}n-Jim{\'e}nez}{Garrido-Jurado
  et~al\mbox{.}}{2014}]%
        {garrido2014automatic}
\bibfield{author}{\bibinfo{person}{Sergio Garrido-Jurado},
  \bibinfo{person}{Rafael Mu{\~n}oz-Salinas},
  \bibinfo{person}{Francisco~Jos{\'e} Madrid-Cuevas}, {and}
  \bibinfo{person}{Manuel~Jes{\'u}s Mar{\'\i}n-Jim{\'e}nez}.}
  \bibinfo{year}{2014}\natexlab{}.
\newblock \showarticletitle{Automatic generation and detection of highly
  reliable fiducial markers under occlusion}.
\newblock \bibinfo{journal}{\emph{Pattern Recognition}} \bibinfo{volume}{47},
  \bibinfo{number}{6} (\bibinfo{year}{2014}), \bibinfo{pages}{2280--2292}.
\newblock


\bibitem[\protect\citeauthoryear{Girshick}{Girshick}{2015}]%
        {girshick2015fast}
\bibfield{author}{\bibinfo{person}{Ross Girshick}.}
  \bibinfo{year}{2015}\natexlab{}.
\newblock \showarticletitle{Fast r-cnn}. In
  \bibinfo{booktitle}{\emph{Proceedings of the IEEE international conference on
  computer vision}}. \bibinfo{pages}{1440--1448}.
\newblock


\bibitem[\protect\citeauthoryear{Glenn, Ipsita, Carithers, Peppler, and
  Ramani}{Glenn et~al\mbox{.}}{2020}]%
        {glenn2020storymakar}
\bibfield{author}{\bibinfo{person}{Terrell Glenn}, \bibinfo{person}{Ananya
  Ipsita}, \bibinfo{person}{Caleb Carithers}, \bibinfo{person}{Kylie Peppler},
  {and} \bibinfo{person}{Karthik Ramani}.} \bibinfo{year}{2020}\natexlab{}.
\newblock \showarticletitle{StoryMakAR: Bringing stories to life with an
  augmented reality \& physical prototyping toolkit for youth}. In
  \bibinfo{booktitle}{\emph{Proceedings of the 2020 CHI Conference on Human
  Factors in Computing Systems}}. \bibinfo{pages}{1--14}.
\newblock


\bibitem[\protect\citeauthoryear{Google}{Google}{2022a}]%
        {noauthor_mapsar_nodate}
\bibfield{author}{\bibinfo{person}{Google}.} \bibinfo{year}{2022}\natexlab{a}.
\newblock \bibinfo{title}{Live AR View, Google Maps}.
\newblock
\newblock
\urldef\tempurl%
\url{https://arvr.google.com/ar/}
\showURL{%
\tempurl}


\bibitem[\protect\citeauthoryear{Google}{Google}{2022b}]%
        {noauthor_mediapipe_nodate}
\bibfield{author}{\bibinfo{person}{Google}.} \bibinfo{year}{2022}\natexlab{b}.
\newblock \bibinfo{title}{MediaPipe}.
\newblock
\newblock
\urldef\tempurl%
\url{https://mediapipe.dev/}
\showURL{%
\tempurl}


\bibitem[\protect\citeauthoryear{Google}{Google}{2022c}]%
        {noauthor_mobilenet_nodate}
\bibfield{author}{\bibinfo{person}{Google}.} \bibinfo{year}{2022}\natexlab{c}.
\newblock \bibinfo{title}{Mobile Net v3}.
\newblock
\newblock
\urldef\tempurl%
\url{https://tfhub.dev/google/tfjs-model/imagenet/mobilenet_v3_small_100_224/feature_vector/5/default/1}
\showURL{%
\tempurl}


\bibitem[\protect\citeauthoryear{Gupta, Lin, Ji, Patel, and Vogel}{Gupta
  et~al\mbox{.}}{2020}]%
        {gupta2020replicate}
\bibfield{author}{\bibinfo{person}{Aakar Gupta}, \bibinfo{person}{Bo~Rui Lin},
  \bibinfo{person}{Siyi Ji}, \bibinfo{person}{Arjav Patel}, {and}
  \bibinfo{person}{Daniel Vogel}.} \bibinfo{year}{2020}\natexlab{}.
\newblock \showarticletitle{Replicate and reuse: Tangible interaction design
  for digitally-augmented physical media objects}. In
  \bibinfo{booktitle}{\emph{Proceedings of the 2020 CHI Conference on Human
  Factors in Computing Systems}}. \bibinfo{pages}{1--12}.
\newblock


\bibitem[\protect\citeauthoryear{He, Gkioxari, Doll{\'a}r, and Girshick}{He
  et~al\mbox{.}}{2017}]%
        {he2017mask}
\bibfield{author}{\bibinfo{person}{Kaiming He}, \bibinfo{person}{Georgia
  Gkioxari}, \bibinfo{person}{Piotr Doll{\'a}r}, {and} \bibinfo{person}{Ross
  Girshick}.} \bibinfo{year}{2017}\natexlab{}.
\newblock \showarticletitle{Mask r-cnn}. In
  \bibinfo{booktitle}{\emph{Proceedings of the IEEE international conference on
  computer vision}}. \bibinfo{pages}{2961--2969}.
\newblock


\bibitem[\protect\citeauthoryear{Held, Gupta, Curless, and Agrawala}{Held
  et~al\mbox{.}}{2012}]%
        {held20123d}
\bibfield{author}{\bibinfo{person}{Robert Held}, \bibinfo{person}{Ankit Gupta},
  \bibinfo{person}{Brian Curless}, {and} \bibinfo{person}{Maneesh Agrawala}.}
  \bibinfo{year}{2012}\natexlab{}.
\newblock \showarticletitle{3D puppetry: a kinect-based interface for 3D
  animation.}. In \bibinfo{booktitle}{\emph{UIST}}, Vol.~\bibinfo{volume}{12}.
  Citeseer, \bibinfo{pages}{423--434}.
\newblock


\bibitem[\protect\citeauthoryear{Hettiarachchi and Wigdor}{Hettiarachchi and
  Wigdor}{2016}]%
        {hettiarachchi2016annexing}
\bibfield{author}{\bibinfo{person}{Anuruddha Hettiarachchi} {and}
  \bibinfo{person}{Daniel Wigdor}.} \bibinfo{year}{2016}\natexlab{}.
\newblock \showarticletitle{Annexing reality: Enabling opportunistic use of
  everyday objects as tangible proxies in augmented reality}. In
  \bibinfo{booktitle}{\emph{Proceedings of the 2016 CHI Conference on Human
  Factors in Computing Systems}}. \bibinfo{pages}{1957--1967}.
\newblock


\bibitem[\protect\citeauthoryear{Heun, Hobin, and Maes}{Heun
  et~al\mbox{.}}{2013}]%
        {heun2013reality}
\bibfield{author}{\bibinfo{person}{Valentin Heun}, \bibinfo{person}{James
  Hobin}, {and} \bibinfo{person}{Pattie Maes}.}
  \bibinfo{year}{2013}\natexlab{}.
\newblock \showarticletitle{Reality editor: programming smarter objects}. In
  \bibinfo{booktitle}{\emph{Proceedings of the 2013 ACM conference on Pervasive
  and ubiquitous computing adjunct publication}}. \bibinfo{pages}{307--310}.
\newblock


\bibitem[\protect\citeauthoryear{Huo and Ramani}{Huo and Ramani}{2016}]%
        {huo2016window}
\bibfield{author}{\bibinfo{person}{Ke Huo} {and} \bibinfo{person}{Karthik
  Ramani}.} \bibinfo{year}{2016}\natexlab{}.
\newblock \showarticletitle{Window-Shaping: 3D Design Ideation in Mixed
  Reality}. In \bibinfo{booktitle}{\emph{Proceedings of the 2016 Symposium on
  Spatial User Interaction}}. \bibinfo{pages}{189--189}.
\newblock


\bibitem[\protect\citeauthoryear{Huo, Wang, Paredes, Villanueva, Cao, and
  Ramani}{Huo et~al\mbox{.}}{2018}]%
        {huo2018synchronizar}
\bibfield{author}{\bibinfo{person}{Ke Huo}, \bibinfo{person}{Tianyi Wang},
  \bibinfo{person}{Luis Paredes}, \bibinfo{person}{Ana~M Villanueva},
  \bibinfo{person}{Yuanzhi Cao}, {and} \bibinfo{person}{Karthik Ramani}.}
  \bibinfo{year}{2018}\natexlab{}.
\newblock \showarticletitle{Synchronizar: Instant synchronization for
  spontaneous and spatial collaborations in augmented reality}. In
  \bibinfo{booktitle}{\emph{Proceedings of the 31st Annual ACM Symposium on
  User Interface Software and Technology}}. \bibinfo{pages}{19--30}.
\newblock


\bibitem[\protect\citeauthoryear{Inc.}{Inc.}{2022a}]%
        {noauthor_aero_nodate}
\bibfield{author}{\bibinfo{person}{Adobe Inc.}}
  \bibinfo{year}{2022}\natexlab{a}.
\newblock \bibinfo{title}{Adobe Aero}.
\newblock
\newblock
\urldef\tempurl%
\url{https://www.adobe.com/in/products/aero.html}
\showURL{%
\tempurl}


\bibitem[\protect\citeauthoryear{Inc.}{Inc.}{2022b}]%
        {noauthor_XD_nodate}
\bibfield{author}{\bibinfo{person}{Adobe Inc.}}
  \bibinfo{year}{2022}\natexlab{b}.
\newblock \bibinfo{title}{Adobe XD}.
\newblock
\newblock
\urldef\tempurl%
\url{https://www.adobe.com/in/products/xd.html}
\showURL{%
\tempurl}


\bibitem[\protect\citeauthoryear{Inc.}{Inc.}{2022c}]%
        {inc_realitycomposer_nodate}
\bibfield{author}{\bibinfo{person}{Apple Inc.}}
  \bibinfo{year}{2022}\natexlab{c}.
\newblock \bibinfo{title}{Reality Composer}.
\newblock
\newblock
\urldef\tempurl%
\url{https://developer.apple.com/augmented-reality/reality-composer/}
\showURL{%
\tempurl}


\bibitem[\protect\citeauthoryear{Inc.}{Inc.}{2017}]%
        {noauthor_gravity_nodate}
\bibfield{author}{\bibinfo{person}{Gravity~Sketch Inc.}}
  \bibinfo{year}{2017}\natexlab{}.
\newblock \bibinfo{title}{Gravity Sketch}.
\newblock
\newblock
\urldef\tempurl%
\url{https://www.gravitysketch.com/}
\showURL{%
\tempurl}


\bibitem[\protect\citeauthoryear{Inc.}{Inc.}{2022d}]%
        {noauthor_insvision_nodate}
\bibfield{author}{\bibinfo{person}{InVision Inc.}}
  \bibinfo{year}{2022}\natexlab{d}.
\newblock \bibinfo{title}{InVision}.
\newblock
\newblock
\urldef\tempurl%
\url{https://www.invisionapp.com/}
\showURL{%
\tempurl}


\bibitem[\protect\citeauthoryear{Inc.}{Inc.}{2022e}]%
        {noauthor_sketch_nodate}
\bibfield{author}{\bibinfo{person}{Sketch Inc.}}
  \bibinfo{year}{2022}\natexlab{e}.
\newblock \bibinfo{title}{Sketch}.
\newblock
\newblock
\urldef\tempurl%
\url{https://www.sketch.com/}
\showURL{%
\tempurl}


\bibitem[\protect\citeauthoryear{Inc.}{Inc.}{2022f}]%
        {noauthor_sketchup_nodate}
\bibfield{author}{\bibinfo{person}{SketchUp Inc.}}
  \bibinfo{year}{2022}\natexlab{f}.
\newblock \bibinfo{title}{Sketchup}.
\newblock
\newblock
\urldef\tempurl%
\url{https://www.sketchup.com/page/homepage}
\showURL{%
\tempurl}


\bibitem[\protect\citeauthoryear{Inc.}{Inc.}{2022g}]%
        {noauthor_unreal_nodate}
\bibfield{author}{\bibinfo{person}{Unreal Inc.}}
  \bibinfo{year}{2022}\natexlab{g}.
\newblock \bibinfo{title}{Unreal Engine}.
\newblock
\newblock
\urldef\tempurl%
\url{https://www.unrealengine.com/en-US}
\showURL{%
\tempurl}


\bibitem[\protect\citeauthoryear{Ishii}{Ishii}{2004}]%
        {ishii2004bottles}
\bibfield{author}{\bibinfo{person}{Hiroshi Ishii}.}
  \bibinfo{year}{2004}\natexlab{}.
\newblock \showarticletitle{Bottles: A transparent interface as a tribute to
  mark weiser}.
\newblock \bibinfo{journal}{\emph{IEICE Transactions on information and
  systems}} \bibinfo{volume}{87}, \bibinfo{number}{6} (\bibinfo{year}{2004}),
  \bibinfo{pages}{1299--1311}.
\newblock


\bibitem[\protect\citeauthoryear{Ishii and Ullmer}{Ishii and Ullmer}{1997}]%
        {ishii1997tangible}
\bibfield{author}{\bibinfo{person}{Hiroshi Ishii} {and} \bibinfo{person}{Brygg
  Ullmer}.} \bibinfo{year}{1997}\natexlab{}.
\newblock \showarticletitle{Tangible bits: towards seamless interfaces between
  people, bits and atoms}. In \bibinfo{booktitle}{\emph{Proceedings of the ACM
  SIGCHI Conference on Human factors in computing systems}}.
  \bibinfo{pages}{234--241}.
\newblock


\bibitem[\protect\citeauthoryear{Jackson and Keefe}{Jackson and Keefe}{2016}]%
        {jackson2016lift}
\bibfield{author}{\bibinfo{person}{Bret Jackson} {and}
  \bibinfo{person}{Daniel~F Keefe}.} \bibinfo{year}{2016}\natexlab{}.
\newblock \showarticletitle{Lift-off: Using reference imagery and freehand
  sketching to create 3d models in vr}.
\newblock \bibinfo{journal}{\emph{IEEE transactions on visualization and
  computer graphics}} \bibinfo{volume}{22}, \bibinfo{number}{4}
  (\bibinfo{year}{2016}), \bibinfo{pages}{1442--1451}.
\newblock


\bibitem[\protect\citeauthoryear{Jones, Benko, Ofek, and Wilson}{Jones
  et~al\mbox{.}}{2013}]%
        {jones2013illumiroom}
\bibfield{author}{\bibinfo{person}{Brett~R Jones}, \bibinfo{person}{Hrvoje
  Benko}, \bibinfo{person}{Eyal Ofek}, {and} \bibinfo{person}{Andrew~D
  Wilson}.} \bibinfo{year}{2013}\natexlab{}.
\newblock \showarticletitle{IllumiRoom: peripheral projected illusions for
  interactive experiences}. In \bibinfo{booktitle}{\emph{Proceedings of the
  SIGCHI Conference on Human Factors in Computing Systems}}.
  \bibinfo{pages}{869--878}.
\newblock


\bibitem[\protect\citeauthoryear{Jordan, Devasia, Hong, Williams, and
  Breazeal}{Jordan et~al\mbox{.}}{2021}]%
        {jordan2021poseblocks}
\bibfield{author}{\bibinfo{person}{Brian Jordan}, \bibinfo{person}{Nisha
  Devasia}, \bibinfo{person}{Jenna Hong}, \bibinfo{person}{Randi Williams},
  {and} \bibinfo{person}{Cynthia Breazeal}.} \bibinfo{year}{2021}\natexlab{}.
\newblock \showarticletitle{PoseBlocks: A toolkit for creating (and dancing)
  with AI}. In \bibinfo{booktitle}{\emph{Proceedings of the AAAI Conference on
  Artificial Intelligence}}, Vol.~\bibinfo{volume}{35}.
  \bibinfo{pages}{15551--15559}.
\newblock


\bibitem[\protect\citeauthoryear{Kaimoto, Monteiro, Faridan, Li, Farajian,
  Kakehi, Nakagaki, and Suzuki}{Kaimoto et~al\mbox{.}}{2022}]%
        {kaimoto2022sketched}
\bibfield{author}{\bibinfo{person}{Hiroki Kaimoto}, \bibinfo{person}{Kyzyl
  Monteiro}, \bibinfo{person}{Mehrad Faridan}, \bibinfo{person}{Jiatong Li},
  \bibinfo{person}{Samin Farajian}, \bibinfo{person}{Yasuaki Kakehi},
  \bibinfo{person}{Ken Nakagaki}, {and} \bibinfo{person}{Ryo Suzuki}.}
  \bibinfo{year}{2022}\natexlab{}.
\newblock \showarticletitle{Sketched Reality: Sketching Bi-Directional
  Interactions Between Virtual and Physical Worlds with AR and Actuated
  Tangible UI}.
\newblock \bibinfo{journal}{\emph{arXiv preprint arXiv:2208.06341}}
  (\bibinfo{year}{2022}).
\newblock


\bibitem[\protect\citeauthoryear{Kang, Shokeen, Byrne, Norooz, Bonsignore,
  Williams-Pierce, and Froehlich}{Kang et~al\mbox{.}}{2020}]%
        {kang2020armath}
\bibfield{author}{\bibinfo{person}{Seokbin Kang}, \bibinfo{person}{Ekta
  Shokeen}, \bibinfo{person}{Virginia~L Byrne}, \bibinfo{person}{Leyla Norooz},
  \bibinfo{person}{Elizabeth Bonsignore}, \bibinfo{person}{Caro
  Williams-Pierce}, {and} \bibinfo{person}{Jon~E Froehlich}.}
  \bibinfo{year}{2020}\natexlab{}.
\newblock \showarticletitle{ARMath: augmenting everyday life with math
  learning}. In \bibinfo{booktitle}{\emph{Proceedings of the 2020 CHI
  Conference on Human Factors in Computing Systems}}. \bibinfo{pages}{1--15}.
\newblock


\bibitem[\protect\citeauthoryear{Kato and Billinghurst}{Kato and
  Billinghurst}{1999}]%
        {kato1999marker}
\bibfield{author}{\bibinfo{person}{Hirokazu Kato} {and} \bibinfo{person}{Mark
  Billinghurst}.} \bibinfo{year}{1999}\natexlab{}.
\newblock \showarticletitle{Marker tracking and hmd calibration for a
  video-based augmented reality conferencing system}. In
  \bibinfo{booktitle}{\emph{Proceedings 2nd IEEE and ACM International Workshop
  on Augmented Reality (IWAR'99)}}. IEEE, \bibinfo{pages}{85--94}.
\newblock


\bibitem[\protect\citeauthoryear{Kelly, Shapiro, de~Halleux, and Ball}{Kelly
  et~al\mbox{.}}{2018}]%
        {kelly2018arcadia}
\bibfield{author}{\bibinfo{person}{Annie Kelly}, \bibinfo{person}{R~Benjamin
  Shapiro}, \bibinfo{person}{Jonathan de Halleux}, {and}
  \bibinfo{person}{Thomas Ball}.} \bibinfo{year}{2018}\natexlab{}.
\newblock \showarticletitle{ARcadia: A rapid prototyping platform for real-time
  tangible interfaces}. In \bibinfo{booktitle}{\emph{Proceedings of the 2018
  CHI Conference on Human Factors in Computing Systems}}.
  \bibinfo{pages}{1--8}.
\newblock


\bibitem[\protect\citeauthoryear{Knibbe, Grossman, and Fitzmaurice}{Knibbe
  et~al\mbox{.}}{2015}]%
        {knibbe2015smart}
\bibfield{author}{\bibinfo{person}{Jarrod Knibbe}, \bibinfo{person}{Tovi
  Grossman}, {and} \bibinfo{person}{George Fitzmaurice}.}
  \bibinfo{year}{2015}\natexlab{}.
\newblock \showarticletitle{Smart makerspace: An immersive instructional space
  for physical tasks}. In \bibinfo{booktitle}{\emph{Proceedings of the 2015
  International Conference on Interactive Tabletops \& Surfaces}}.
  \bibinfo{pages}{83--92}.
\newblock


\bibitem[\protect\citeauthoryear{Krau{\ss}, Nebeling, Jasche, and
  Boden}{Krau{\ss} et~al\mbox{.}}{2022}]%
        {krauss2022elements}
\bibfield{author}{\bibinfo{person}{Veronika Krau{\ss}},
  \bibinfo{person}{Michael Nebeling}, \bibinfo{person}{Florian Jasche}, {and}
  \bibinfo{person}{Alexander Boden}.} \bibinfo{year}{2022}\natexlab{}.
\newblock \showarticletitle{Elements of XR Prototyping: Characterizing the Role
  and Use of Prototypes in Augmented and Virtual Reality Design}. In
  \bibinfo{booktitle}{\emph{CHI Conference on Human Factors in Computing
  Systems}}. \bibinfo{pages}{1--18}.
\newblock


\bibitem[\protect\citeauthoryear{Ledo, Houben, Vermeulen, Marquardt, Oehlberg,
  and Greenberg}{Ledo et~al\mbox{.}}{2018}]%
        {ledo2018evaluation}
\bibfield{author}{\bibinfo{person}{David Ledo}, \bibinfo{person}{Steven
  Houben}, \bibinfo{person}{Jo Vermeulen}, \bibinfo{person}{Nicolai Marquardt},
  \bibinfo{person}{Lora Oehlberg}, {and} \bibinfo{person}{Saul Greenberg}.}
  \bibinfo{year}{2018}\natexlab{}.
\newblock \showarticletitle{Evaluation strategies for HCI toolkit research}. In
  \bibinfo{booktitle}{\emph{Proceedings of the 2018 CHI Conference on Human
  Factors in Computing Systems}}. \bibinfo{pages}{1--17}.
\newblock


\bibitem[\protect\citeauthoryear{Lee, Kim, and Billinghurst}{Lee
  et~al\mbox{.}}{2005}]%
        {lee2005immersive}
\bibfield{author}{\bibinfo{person}{Gun~A Lee}, \bibinfo{person}{Gerard~J Kim},
  {and} \bibinfo{person}{Mark Billinghurst}.} \bibinfo{year}{2005}\natexlab{}.
\newblock \showarticletitle{Immersive authoring: What you experience is what
  you get (wyxiwyg)}.
\newblock \bibinfo{journal}{\emph{Commun. ACM}} \bibinfo{volume}{48},
  \bibinfo{number}{7} (\bibinfo{year}{2005}), \bibinfo{pages}{76--81}.
\newblock


\bibitem[\protect\citeauthoryear{Leiva and Beaudouin-Lafon}{Leiva and
  Beaudouin-Lafon}{2018}]%
        {leiva2018montage}
\bibfield{author}{\bibinfo{person}{Germ{\'a}n Leiva} {and}
  \bibinfo{person}{Michel Beaudouin-Lafon}.} \bibinfo{year}{2018}\natexlab{}.
\newblock \showarticletitle{Montage: a video prototyping system to reduce
  re-shooting and increase re-usability}. In
  \bibinfo{booktitle}{\emph{Proceedings of the 31st Annual ACM Symposium on
  User Interface Software and Technology}}. \bibinfo{pages}{675--682}.
\newblock


\bibitem[\protect\citeauthoryear{Leiva, Gr{\o}nb{\ae}k, Klokmose, Nguyen, Kazi,
  and Asente}{Leiva et~al\mbox{.}}{2021}]%
        {leiva2021rapido}
\bibfield{author}{\bibinfo{person}{Germ{\'a}n Leiva},
  \bibinfo{person}{Jens~Emil Gr{\o}nb{\ae}k},
  \bibinfo{person}{Clemens~Nylandsted Klokmose}, \bibinfo{person}{Cuong
  Nguyen}, \bibinfo{person}{Rubaiat~Habib Kazi}, {and} \bibinfo{person}{Paul
  Asente}.} \bibinfo{year}{2021}\natexlab{}.
\newblock \showarticletitle{Rapido: Prototyping Interactive AR Experiences
  through Programming by Demonstration}. In \bibinfo{booktitle}{\emph{The 34th
  Annual ACM Symposium on User Interface Software and Technology}}.
  \bibinfo{pages}{626--637}.
\newblock


\bibitem[\protect\citeauthoryear{Leiva, Nguyen, Kazi, and Asente}{Leiva
  et~al\mbox{.}}{2020}]%
        {leiva2020pronto}
\bibfield{author}{\bibinfo{person}{Germ{\'a}n Leiva}, \bibinfo{person}{Cuong
  Nguyen}, \bibinfo{person}{Rubaiat~Habib Kazi}, {and} \bibinfo{person}{Paul
  Asente}.} \bibinfo{year}{2020}\natexlab{}.
\newblock \showarticletitle{Pronto: Rapid augmented reality video prototyping
  using sketches and enaction}. In \bibinfo{booktitle}{\emph{Proceedings of the
  2020 CHI Conference on Human Factors in Computing Systems}}.
  \bibinfo{pages}{1--13}.
\newblock


\bibitem[\protect\citeauthoryear{Li, Annett, Hinckley, Singh, and Wigdor}{Li
  et~al\mbox{.}}{2019}]%
        {li2019holodoc}
\bibfield{author}{\bibinfo{person}{Zhen Li}, \bibinfo{person}{Michelle Annett},
  \bibinfo{person}{Ken Hinckley}, \bibinfo{person}{Karan Singh}, {and}
  \bibinfo{person}{Daniel Wigdor}.} \bibinfo{year}{2019}\natexlab{}.
\newblock \showarticletitle{Holodoc: Enabling mixed reality workspaces that
  harness physical and digital content}. In
  \bibinfo{booktitle}{\emph{Proceedings of the 2019 CHI Conference on Human
  Factors in Computing Systems}}. \bibinfo{pages}{1--14}.
\newblock


\bibitem[\protect\citeauthoryear{Liao, Karim, Jadon, Kazi, and Suzuki}{Liao
  et~al\mbox{.}}{2022}]%
        {liao2022realitytalk}
\bibfield{author}{\bibinfo{person}{Jian Liao}, \bibinfo{person}{Adnan Karim},
  \bibinfo{person}{Shivesh~Singh Jadon}, \bibinfo{person}{Rubaiat~Habib Kazi},
  {and} \bibinfo{person}{Ryo Suzuki}.} \bibinfo{year}{2022}\natexlab{}.
\newblock \showarticletitle{RealityTalk: Real-Time Speech-Driven Augmented
  Presentation for AR Live Storytelling}. In
  \bibinfo{booktitle}{\emph{Proceedings of the 35th Annual ACM Symposium on
  User Interface Software and Technology}}. \bibinfo{pages}{1--12}.
\newblock


\bibitem[\protect\citeauthoryear{Lindlbauer, Gr{\o}nb{\ae}k, Birk, Halskov,
  Alexa, and M{\"u}ller}{Lindlbauer et~al\mbox{.}}{2016}]%
        {lindlbauer2016combining}
\bibfield{author}{\bibinfo{person}{David Lindlbauer},
  \bibinfo{person}{Jens~Emil Gr{\o}nb{\ae}k}, \bibinfo{person}{Morten Birk},
  \bibinfo{person}{Kim Halskov}, \bibinfo{person}{Marc Alexa}, {and}
  \bibinfo{person}{J{\"o}rg M{\"u}ller}.} \bibinfo{year}{2016}\natexlab{}.
\newblock \showarticletitle{Combining shape-changing interfaces and spatial
  augmented reality enables extended object appearance}. In
  \bibinfo{booktitle}{\emph{Proceedings of the 2016 CHI Conference on Human
  Factors in Computing Systems}}. \bibinfo{pages}{791--802}.
\newblock


\bibitem[\protect\citeauthoryear{MacIntyre, Gandy, Dow, and Bolter}{MacIntyre
  et~al\mbox{.}}{2004}]%
        {macintyre2004dart}
\bibfield{author}{\bibinfo{person}{Blair MacIntyre}, \bibinfo{person}{Maribeth
  Gandy}, \bibinfo{person}{Steven Dow}, {and} \bibinfo{person}{Jay~David
  Bolter}.} \bibinfo{year}{2004}\natexlab{}.
\newblock \showarticletitle{DART: a toolkit for rapid design exploration of
  augmented reality experiences}. In \bibinfo{booktitle}{\emph{Proceedings of
  the 17th annual ACM symposium on User interface software and technology}}.
  \bibinfo{pages}{197--206}.
\newblock


\bibitem[\protect\citeauthoryear{Nebeling and Madier}{Nebeling and
  Madier}{2019}]%
        {nebeling2019360proto}
\bibfield{author}{\bibinfo{person}{Michael Nebeling} {and}
  \bibinfo{person}{Katy Madier}.} \bibinfo{year}{2019}\natexlab{}.
\newblock \showarticletitle{360proto: Making interactive virtual reality \&
  augmented reality prototypes from paper}. In
  \bibinfo{booktitle}{\emph{Proceedings of the 2019 CHI Conference on Human
  Factors in Computing Systems}}. \bibinfo{pages}{1--13}.
\newblock


\bibitem[\protect\citeauthoryear{Nebeling, Nebeling, Yu, and Rumble}{Nebeling
  et~al\mbox{.}}{2018}]%
        {nebeling2018protoar}
\bibfield{author}{\bibinfo{person}{Michael Nebeling}, \bibinfo{person}{Janet
  Nebeling}, \bibinfo{person}{Ao Yu}, {and} \bibinfo{person}{Rob Rumble}.}
  \bibinfo{year}{2018}\natexlab{}.
\newblock \showarticletitle{Protoar: Rapid physical-digital prototyping of
  mobile augmented reality applications}. In
  \bibinfo{booktitle}{\emph{Proceedings of the 2018 CHI Conference on Human
  Factors in Computing Systems}}. \bibinfo{pages}{1--12}.
\newblock


\bibitem[\protect\citeauthoryear{Nebeling and Speicher}{Nebeling and
  Speicher}{2018}]%
        {nebeling2018trouble}
\bibfield{author}{\bibinfo{person}{Michael Nebeling} {and}
  \bibinfo{person}{Maximilian Speicher}.} \bibinfo{year}{2018}\natexlab{}.
\newblock \showarticletitle{The trouble with augmented reality/virtual reality
  authoring tools}. In \bibinfo{booktitle}{\emph{2018 IEEE international
  symposium on mixed and augmented reality adjunct (ISMAR-Adjunct)}}. IEEE,
  \bibinfo{pages}{333--337}.
\newblock


\bibitem[\protect\citeauthoryear{Ng, Shin, Plopski, Sandor, and Saakes}{Ng
  et~al\mbox{.}}{2018}]%
        {ng2018situated}
\bibfield{author}{\bibinfo{person}{Gary Ng}, \bibinfo{person}{Joon~Gi Shin},
  \bibinfo{person}{Alexander Plopski}, \bibinfo{person}{Christian Sandor},
  {and} \bibinfo{person}{Daniel Saakes}.} \bibinfo{year}{2018}\natexlab{}.
\newblock \showarticletitle{Situated game level editing in augmented reality}.
  In \bibinfo{booktitle}{\emph{Proceedings of the Twelfth International
  Conference on Tangible, Embedded, and Embodied Interaction}}.
  \bibinfo{pages}{409--418}.
\newblock


\bibitem[\protect\citeauthoryear{Nintendo}{Nintendo}{2022a}]%
        {noauthor_mariokartlive_nodate}
\bibfield{author}{\bibinfo{person}{Nintendo}.}
  \bibinfo{year}{2022}\natexlab{a}.
\newblock \bibinfo{title}{Mario Kart Live}.
\newblock
\newblock
\urldef\tempurl%
\url{https://mklive.nintendo.com/}
\showURL{%
\tempurl}


\bibitem[\protect\citeauthoryear{Nintendo}{Nintendo}{2022b}]%
        {noauthor_nintendolabo_nodate}
\bibfield{author}{\bibinfo{person}{Nintendo}.}
  \bibinfo{year}{2022}\natexlab{b}.
\newblock \bibinfo{title}{Nintendo Labo}.
\newblock
\newblock
\urldef\tempurl%
\url{https://www.nintendo.co.uk/Nintendo-Labo/Nintendo-Labo-1328637.html}
\showURL{%
\tempurl}


\bibitem[\protect\citeauthoryear{OpenCV}{OpenCV}{2022}]%
        {noauthor_opencv_nodate}
\bibfield{author}{\bibinfo{person}{OpenCV}.} \bibinfo{year}{2022}\natexlab{}.
\newblock \bibinfo{title}{OpenCV}.
\newblock
\newblock
\urldef\tempurl%
\url{https://opencv.org/}
\showURL{%
\tempurl}


\bibitem[\protect\citeauthoryear{Park and Shin}{Park and Shin}{2021}]%
        {park2021tooee}
\bibfield{author}{\bibinfo{person}{Youngki Park} {and} \bibinfo{person}{Youhyun
  Shin}.} \bibinfo{year}{2021}\natexlab{}.
\newblock \showarticletitle{Tooee: A Novel Scratch Extension for K-12 Big Data
  and Artificial Intelligence Education Using Text-Based Visual Blocks}.
\newblock \bibinfo{journal}{\emph{IEEE Access}}  \bibinfo{volume}{9}
  (\bibinfo{year}{2021}), \bibinfo{pages}{149630--149646}.
\newblock


\bibitem[\protect\citeauthoryear{Rajaram and Nebeling}{Rajaram and
  Nebeling}{2022}]%
        {rajaram2022paper}
\bibfield{author}{\bibinfo{person}{Shwetha Rajaram} {and}
  \bibinfo{person}{Michael Nebeling}.} \bibinfo{year}{2022}\natexlab{}.
\newblock \showarticletitle{Paper Trail: An Immersive Authoring System for
  Augmented Reality Instructional Experiences}. In
  \bibinfo{booktitle}{\emph{CHI Conference on Human Factors in Computing
  Systems}}. \bibinfo{pages}{1--16}.
\newblock


\bibitem[\protect\citeauthoryear{Ramos, Meek, Simard, Suh, and Ghorashi}{Ramos
  et~al\mbox{.}}{2020}]%
        {ramos2020interactive}
\bibfield{author}{\bibinfo{person}{Gonzalo Ramos}, \bibinfo{person}{Christopher
  Meek}, \bibinfo{person}{Patrice Simard}, \bibinfo{person}{Jina Suh}, {and}
  \bibinfo{person}{Soroush Ghorashi}.} \bibinfo{year}{2020}\natexlab{}.
\newblock \showarticletitle{Interactive machine teaching: a human-centered
  approach to building machine-learned models}.
\newblock \bibinfo{journal}{\emph{Human--Computer Interaction}}
  \bibinfo{volume}{35}, \bibinfo{number}{5-6} (\bibinfo{year}{2020}),
  \bibinfo{pages}{413--451}.
\newblock


\bibitem[\protect\citeauthoryear{Redmon, Divvala, Girshick, and Farhadi}{Redmon
  et~al\mbox{.}}{2016}]%
        {redmon2016you}
\bibfield{author}{\bibinfo{person}{Joseph Redmon}, \bibinfo{person}{Santosh
  Divvala}, \bibinfo{person}{Ross Girshick}, {and} \bibinfo{person}{Ali
  Farhadi}.} \bibinfo{year}{2016}\natexlab{}.
\newblock \showarticletitle{You only look once: Unified, real-time object
  detection}. In \bibinfo{booktitle}{\emph{Proceedings of the IEEE conference
  on computer vision and pattern recognition}}. \bibinfo{pages}{779--788}.
\newblock


\bibitem[\protect\citeauthoryear{Reipschl{\"a}ger, Engert, and
  Dachselt}{Reipschl{\"a}ger et~al\mbox{.}}{2020}]%
        {reipschlager2020augmented}
\bibfield{author}{\bibinfo{person}{Patrick Reipschl{\"a}ger},
  \bibinfo{person}{Severin Engert}, {and} \bibinfo{person}{Raimund Dachselt}.}
  \bibinfo{year}{2020}\natexlab{}.
\newblock \showarticletitle{Augmented displays: Seamlessly extending
  interactive surfaces with head-mounted augmented reality}. In
  \bibinfo{booktitle}{\emph{Extended Abstracts of the 2020 CHI Conference on
  Human Factors in Computing Systems}}. \bibinfo{pages}{1--4}.
\newblock


\bibitem[\protect\citeauthoryear{Russakovsky, Deng, Su, Krause, Satheesh, Ma,
  Huang, Karpathy, Khosla, Bernstein, et~al\mbox{.}}{Russakovsky
  et~al\mbox{.}}{2015}]%
        {russakovsky2015imagenet}
\bibfield{author}{\bibinfo{person}{Olga Russakovsky}, \bibinfo{person}{Jia
  Deng}, \bibinfo{person}{Hao Su}, \bibinfo{person}{Jonathan Krause},
  \bibinfo{person}{Sanjeev Satheesh}, \bibinfo{person}{Sean Ma},
  \bibinfo{person}{Zhiheng Huang}, \bibinfo{person}{Andrej Karpathy},
  \bibinfo{person}{Aditya Khosla}, \bibinfo{person}{Michael Bernstein},
  {et~al\mbox{.}}} \bibinfo{year}{2015}\natexlab{}.
\newblock \showarticletitle{Imagenet large scale visual recognition challenge}.
\newblock \bibinfo{journal}{\emph{International journal of computer vision}}
  \bibinfo{volume}{115}, \bibinfo{number}{3} (\bibinfo{year}{2015}),
  \bibinfo{pages}{211--252}.
\newblock


\bibitem[\protect\citeauthoryear{Sabuncuoglu and Sezgin}{Sabuncuoglu and
  Sezgin}{2022}]%
        {sabuncuoglu2022prototyping}
\bibfield{author}{\bibinfo{person}{Alpay Sabuncuoglu} {and}
  \bibinfo{person}{T~Metin Sezgin}.} \bibinfo{year}{2022}\natexlab{}.
\newblock \showarticletitle{Prototyping Products using Web-based AI Tools:
  Designing a Tangible Programming Environment with Children}. In
  \bibinfo{booktitle}{\emph{6th FabLearn Europe/MakeEd Conference 2022}}.
  \bibinfo{pages}{1--6}.
\newblock


\bibitem[\protect\citeauthoryear{Saquib, Kazi, Wei, and Li}{Saquib
  et~al\mbox{.}}{2019}]%
        {saquib2019interactive}
\bibfield{author}{\bibinfo{person}{Nazmus Saquib},
  \bibinfo{person}{Rubaiat~Habib Kazi}, \bibinfo{person}{Li-Yi Wei}, {and}
  \bibinfo{person}{Wilmot Li}.} \bibinfo{year}{2019}\natexlab{}.
\newblock \showarticletitle{Interactive body-driven graphics for augmented
  video performance}. In \bibinfo{booktitle}{\emph{Proceedings of the 2019 CHI
  Conference on Human Factors in Computing Systems}}. \bibinfo{pages}{1--12}.
\newblock


\bibitem[\protect\citeauthoryear{Seichter, Looser, and Billinghurst}{Seichter
  et~al\mbox{.}}{2008}]%
        {seichter2008composar}
\bibfield{author}{\bibinfo{person}{Hartmut Seichter}, \bibinfo{person}{Julian
  Looser}, {and} \bibinfo{person}{Mark Billinghurst}.}
  \bibinfo{year}{2008}\natexlab{}.
\newblock \showarticletitle{ComposAR: An intuitive tool for authoring AR
  applications}. In \bibinfo{booktitle}{\emph{2008 7th IEEE/ACM International
  Symposium on Mixed and Augmented Reality}}. IEEE, \bibinfo{pages}{177--178}.
\newblock


\bibitem[\protect\citeauthoryear{Shorey and Girouard}{Shorey and
  Girouard}{2017}]%
        {shorey2017bendtroller}
\bibfield{author}{\bibinfo{person}{Paden Shorey} {and} \bibinfo{person}{Audrey
  Girouard}.} \bibinfo{year}{2017}\natexlab{}.
\newblock \showarticletitle{Bendtroller: An exploration of in-game action
  mappings with a deformable game controller}. In
  \bibinfo{booktitle}{\emph{Proceedings of the 2017 CHI Conference on Human
  Factors in Computing Systems}}. \bibinfo{pages}{1447--1458}.
\newblock


\bibitem[\protect\citeauthoryear{Speicher, Lewis, and Nebeling}{Speicher
  et~al\mbox{.}}{2021}]%
        {speicher2021designers}
\bibfield{author}{\bibinfo{person}{Maximilian Speicher}, \bibinfo{person}{Katy
  Lewis}, {and} \bibinfo{person}{Michael Nebeling}.}
  \bibinfo{year}{2021}\natexlab{}.
\newblock \showarticletitle{Designers, the stage is yours! medium-fidelity
  prototyping of augmented \& virtual reality interfaces with 360theater}.
\newblock \bibinfo{journal}{\emph{Proceedings of the ACM on Human-Computer
  Interaction}} \bibinfo{volume}{5}, \bibinfo{number}{EICS}
  (\bibinfo{year}{2021}), \bibinfo{pages}{1--25}.
\newblock


\bibitem[\protect\citeauthoryear{Steimle, Jordt, and Maes}{Steimle
  et~al\mbox{.}}{2013}]%
        {steimle2013flexpad}
\bibfield{author}{\bibinfo{person}{J{\"u}rgen Steimle},
  \bibinfo{person}{Andreas Jordt}, {and} \bibinfo{person}{Pattie Maes}.}
  \bibinfo{year}{2013}\natexlab{}.
\newblock \showarticletitle{Flexpad: highly flexible bending interactions for
  projected handheld displays}. In \bibinfo{booktitle}{\emph{Proceedings of the
  SIGCHI Conference on Human Factors in Computing Systems}}.
  \bibinfo{pages}{237--246}.
\newblock


\bibitem[\protect\citeauthoryear{Suzuki, Kazi, Wei, DiVerdi, Li, and
  Leithinger}{Suzuki et~al\mbox{.}}{2020}]%
        {suzuki2020realitysketch}
\bibfield{author}{\bibinfo{person}{Ryo Suzuki}, \bibinfo{person}{Rubaiat~Habib
  Kazi}, \bibinfo{person}{Li-Yi Wei}, \bibinfo{person}{Stephen DiVerdi},
  \bibinfo{person}{Wilmot Li}, {and} \bibinfo{person}{Daniel Leithinger}.}
  \bibinfo{year}{2020}\natexlab{}.
\newblock \showarticletitle{Realitysketch: Embedding responsive graphics and
  visualizations in AR through dynamic sketching}. In
  \bibinfo{booktitle}{\emph{Proceedings of the 33rd Annual ACM Symposium on
  User Interface Software and Technology}}. \bibinfo{pages}{166--181}.
\newblock


\bibitem[\protect\citeauthoryear{Three.js}{Three.js}{2022}]%
        {noauthor_threejs_nodate}
\bibfield{author}{\bibinfo{person}{Three.js}.} \bibinfo{year}{2022}\natexlab{}.
\newblock \bibinfo{title}{Ricardo Cabello}.
\newblock
\newblock
\urldef\tempurl%
\url{https://threejs.org/}
\showURL{%
\tempurl}


\bibitem[\protect\citeauthoryear{Tseng, Murai, Freed, Gelosi, Ta, and
  Kawahara}{Tseng et~al\mbox{.}}{2021}]%
        {tseng2021plushpal}
\bibfield{author}{\bibinfo{person}{Tiffany Tseng}, \bibinfo{person}{Yumiko
  Murai}, \bibinfo{person}{Natalie Freed}, \bibinfo{person}{Deanna Gelosi},
  \bibinfo{person}{Tung~D Ta}, {and} \bibinfo{person}{Yoshihiro Kawahara}.}
  \bibinfo{year}{2021}\natexlab{}.
\newblock \showarticletitle{PlushPal: Storytelling with interactive plush toys
  and machine learning}. In \bibinfo{booktitle}{\emph{Interaction design and
  children}}. \bibinfo{pages}{236--245}.
\newblock


\bibitem[\protect\citeauthoryear{Villar, Cletheroe, Saul, Holz, Regan,
  Salandin, Sra, Yeo, Field, and Zhang}{Villar et~al\mbox{.}}{2018}]%
        {villar2018project}
\bibfield{author}{\bibinfo{person}{Nicolas Villar}, \bibinfo{person}{Daniel
  Cletheroe}, \bibinfo{person}{Greg Saul}, \bibinfo{person}{Christian Holz},
  \bibinfo{person}{Tim Regan}, \bibinfo{person}{Oscar Salandin},
  \bibinfo{person}{Misha Sra}, \bibinfo{person}{Hui-Shyong Yeo},
  \bibinfo{person}{William Field}, {and} \bibinfo{person}{Haiyan Zhang}.}
  \bibinfo{year}{2018}\natexlab{}.
\newblock \showarticletitle{Project zanzibar: A portable and flexible tangible
  interaction platform}. In \bibinfo{booktitle}{\emph{Proceedings of the 2018
  CHI Conference on Human Factors in Computing Systems}}.
  \bibinfo{pages}{1--13}.
\newblock


\bibitem[\protect\citeauthoryear{Walsh, Von~Itzstein, and Thomas}{Walsh
  et~al\mbox{.}}{2013}]%
        {walsh2013tangible}
\bibfield{author}{\bibinfo{person}{James~A Walsh}, \bibinfo{person}{G~Stewart
  Von~Itzstein}, {and} \bibinfo{person}{Bruce~H Thomas}.}
  \bibinfo{year}{2013}\natexlab{}.
\newblock \showarticletitle{Tangible agile mapping: ad-hoc tangible user
  interaction definition}. In \bibinfo{booktitle}{\emph{AUIC}}. Citeseer,
  \bibinfo{pages}{3--12}.
\newblock


\bibitem[\protect\citeauthoryear{Walsh, Von~Itzstein, and Thomas}{Walsh
  et~al\mbox{.}}{2014}]%
        {walsh2014ephemeral}
\bibfield{author}{\bibinfo{person}{James~A Walsh}, \bibinfo{person}{Stewart
  Von~Itzstein}, {and} \bibinfo{person}{Bruce~H Thomas}.}
  \bibinfo{year}{2014}\natexlab{}.
\newblock \showarticletitle{Ephemeral interaction using everyday objects}. In
  \bibinfo{booktitle}{\emph{Proceedings of the Fifteenth Australasian User
  Interface Conference-Volume 150}}. \bibinfo{pages}{29--37}.
\newblock


\bibitem[\protect\citeauthoryear{Wang, Qian, He, Hu, Cao, and Ramani}{Wang
  et~al\mbox{.}}{2021b}]%
        {wang2021gesturar}
\bibfield{author}{\bibinfo{person}{Tianyi Wang}, \bibinfo{person}{Xun Qian},
  \bibinfo{person}{Fengming He}, \bibinfo{person}{Xiyun Hu},
  \bibinfo{person}{Yuanzhi Cao}, {and} \bibinfo{person}{Karthik Ramani}.}
  \bibinfo{year}{2021}\natexlab{b}.
\newblock \showarticletitle{GesturAR: An Authoring System for Creating Freehand
  Interactive Augmented Reality Applications}. In \bibinfo{booktitle}{\emph{The
  34th Annual ACM Symposium on User Interface Software and Technology}}.
  \bibinfo{pages}{552--567}.
\newblock


\bibitem[\protect\citeauthoryear{Wang, Qian, He, Hu, Huo, Cao, and Ramani}{Wang
  et~al\mbox{.}}{2020}]%
        {wang2020capturar}
\bibfield{author}{\bibinfo{person}{Tianyi Wang}, \bibinfo{person}{Xun Qian},
  \bibinfo{person}{Fengming He}, \bibinfo{person}{Xiyun Hu},
  \bibinfo{person}{Ke Huo}, \bibinfo{person}{Yuanzhi Cao}, {and}
  \bibinfo{person}{Karthik Ramani}.} \bibinfo{year}{2020}\natexlab{}.
\newblock \showarticletitle{CAPturAR: An augmented reality tool for authoring
  human-involved context-aware applications}. In
  \bibinfo{booktitle}{\emph{Proceedings of the 33rd Annual ACM Symposium on
  User Interface Software and Technology}}. \bibinfo{pages}{328--341}.
\newblock


\bibitem[\protect\citeauthoryear{Wang, Nguyen, Asente, and Dorsey}{Wang
  et~al\mbox{.}}{2021a}]%
        {wang2021distanciar}
\bibfield{author}{\bibinfo{person}{Zeyu Wang}, \bibinfo{person}{Cuong Nguyen},
  \bibinfo{person}{Paul Asente}, {and} \bibinfo{person}{Julie Dorsey}.}
  \bibinfo{year}{2021}\natexlab{a}.
\newblock \showarticletitle{Distanciar: Authoring site-specific augmented
  reality experiences for remote environments}. In
  \bibinfo{booktitle}{\emph{Proceedings of the 2021 CHI Conference on Human
  Factors in Computing Systems}}. \bibinfo{pages}{1--12}.
\newblock


\bibitem[\protect\citeauthoryear{Whitlock, Mitchell, Pfeufer, Arnot, Craig,
  Wilson, Chung, and Szafir}{Whitlock et~al\mbox{.}}{2020}]%
        {whitlock2020mrcat}
\bibfield{author}{\bibinfo{person}{Matt Whitlock}, \bibinfo{person}{Jake
  Mitchell}, \bibinfo{person}{Nick Pfeufer}, \bibinfo{person}{Brad Arnot},
  \bibinfo{person}{Ryan Craig}, \bibinfo{person}{Bryce Wilson},
  \bibinfo{person}{Brian Chung}, {and} \bibinfo{person}{Danielle~Albers
  Szafir}.} \bibinfo{year}{2020}\natexlab{}.
\newblock \showarticletitle{MRCAT: In situ prototyping of interactive AR
  environments}. In \bibinfo{booktitle}{\emph{International Conference on
  Human-Computer Interaction}}. Springer, \bibinfo{pages}{235--255}.
\newblock


\bibitem[\protect\citeauthoryear{Williams, Kaputsos, and Breazeal}{Williams
  et~al\mbox{.}}{2021}]%
        {williams2021teacher}
\bibfield{author}{\bibinfo{person}{Randi Williams}, \bibinfo{person}{Stephen~P
  Kaputsos}, {and} \bibinfo{person}{Cynthia Breazeal}.}
  \bibinfo{year}{2021}\natexlab{}.
\newblock \showarticletitle{Teacher Perspectives on How To Train Your Robot: A
  Middle School AI and Ethics Curriculum}. In
  \bibinfo{booktitle}{\emph{Proceedings of the AAAI Conference on Artificial
  Intelligence}}, Vol.~\bibinfo{volume}{35}. \bibinfo{pages}{15678--15686}.
\newblock


\bibitem[\protect\citeauthoryear{Xiao, Harrison, and Hudson}{Xiao
  et~al\mbox{.}}{2013}]%
        {xiao2013worldkit}
\bibfield{author}{\bibinfo{person}{Robert Xiao}, \bibinfo{person}{Chris
  Harrison}, {and} \bibinfo{person}{Scott~E Hudson}.}
  \bibinfo{year}{2013}\natexlab{}.
\newblock \showarticletitle{WorldKit: rapid and easy creation of ad-hoc
  interactive applications on everyday surfaces}. In
  \bibinfo{booktitle}{\emph{Proceedings of the SIGCHI Conference on Human
  Factors in Computing Systems}}. \bibinfo{pages}{879--888}.
\newblock


\bibitem[\protect\citeauthoryear{Ye and Fu}{Ye and Fu}{2022}]%
        {ye2022progesar}
\bibfield{author}{\bibinfo{person}{Hui Ye} {and} \bibinfo{person}{Hongbo Fu}.}
  \bibinfo{year}{2022}\natexlab{}.
\newblock \showarticletitle{ProGesAR: Mobile AR Prototyping for Proxemic and
  Gestural Interactions with Real-world IoT Enhanced Spaces}. In
  \bibinfo{booktitle}{\emph{CHI Conference on Human Factors in Computing
  Systems}}. \bibinfo{pages}{1--14}.
\newblock


\bibitem[\protect\citeauthoryear{Yue, Yang, Ren, and Wang}{Yue
  et~al\mbox{.}}{2017}]%
        {yue2017scenectrl}
\bibfield{author}{\bibinfo{person}{Ya-Ting Yue}, \bibinfo{person}{Yong-Liang
  Yang}, \bibinfo{person}{Gang Ren}, {and} \bibinfo{person}{Wenping Wang}.}
  \bibinfo{year}{2017}\natexlab{}.
\newblock \showarticletitle{SceneCtrl: Mixed reality enhancement via efficient
  scene editing}. In \bibinfo{booktitle}{\emph{Proceedings of the 30th annual
  ACM symposium on user interface software and technology}}.
  \bibinfo{pages}{427--436}.
\newblock


\bibitem[\protect\citeauthoryear{Zheng, Gyory, and Do}{Zheng
  et~al\mbox{.}}{2020}]%
        {zheng2020tangible}
\bibfield{author}{\bibinfo{person}{Clement Zheng}, \bibinfo{person}{Peter
  Gyory}, {and} \bibinfo{person}{Ellen Yi-Luen Do}.}
  \bibinfo{year}{2020}\natexlab{}.
\newblock \showarticletitle{Tangible interfaces with printed paper markers}. In
  \bibinfo{booktitle}{\emph{Proceedings of the 2020 ACM designing interactive
  systems conference}}. \bibinfo{pages}{909--923}.
\newblock


\bibitem[\protect\citeauthoryear{Zhou, Sykes, Fels, and Kin}{Zhou
  et~al\mbox{.}}{2020}]%
        {zhou2020gripmarks}
\bibfield{author}{\bibinfo{person}{Qian Zhou}, \bibinfo{person}{Sarah Sykes},
  \bibinfo{person}{Sidney Fels}, {and} \bibinfo{person}{Kenrick Kin}.}
  \bibinfo{year}{2020}\natexlab{}.
\newblock \showarticletitle{Gripmarks: Using Hand Grips to Transform In-Hand
  Objects into Mixed Reality Input}. In \bibinfo{booktitle}{\emph{Proceedings
  of the 2020 CHI Conference on Human Factors in Computing Systems}}.
  \bibinfo{pages}{1--11}.
\newblock


\bibitem[\protect\citeauthoryear{Zhou and Yatani}{Zhou and Yatani}{2022}]%
        {zhou2022gesture}
\bibfield{author}{\bibinfo{person}{Zhongyi Zhou} {and} \bibinfo{person}{Koji
  Yatani}.} \bibinfo{year}{2022}\natexlab{}.
\newblock \showarticletitle{Gesture-aware Interactive Machine Teaching with
  In-situ Object Annotations}.
\newblock \bibinfo{journal}{\emph{arXiv preprint arXiv:2208.01211}}
  (\bibinfo{year}{2022}).
\newblock


\bibitem[\protect\citeauthoryear{Zhu and Grossman}{Zhu and Grossman}{2020}]%
        {zhu2020bishare}
\bibfield{author}{\bibinfo{person}{Fengyuan Zhu} {and} \bibinfo{person}{Tovi
  Grossman}.} \bibinfo{year}{2020}\natexlab{}.
\newblock \showarticletitle{Bishare: Exploring bidirectional interactions
  between smartphones and head-mounted augmented reality}. In
  \bibinfo{booktitle}{\emph{Proceedings of the 2020 CHI Conference on Human
  Factors in Computing Systems}}. \bibinfo{pages}{1--14}.
\newblock


\end{thebibliography}

\end{document}